\documentclass[twocolumn]{aastex62}
\usepackage{amsmath}

\newcommand{\solarmass}{\ensuremath{\mathrm{M}_{\odot}}}
\newcommand{\solarlum}{\ensuremath{\mathrm{L}_{\odot}}}
\newcommand{\Msun}{\solarmass}
\newcommand{\Lsun}{\solarlum}

\newcommand{\LIR}{\ensuremath{L_{\mathrm{IR}}}}
\newcommand{\LCO}{\ensuremath{L'_{\mathrm{CO}}}}
\newcommand{\LCI}{\ensuremath{L_{\mathrm{\CI}}}}

\newcommand{\Ldust}{\ensuremath{L_{\mathrm{dust}}}}
\newcommand{\Mmol}{\ensuremath{M_{\mathrm{mol}}}}
\newcommand{\MH}{\ensuremath{M_{\mathrm{H}_{2}}}}
\newcommand{\MCI}{\ensuremath{M_{\mathrm{\CI}}}}
\newcommand{\SigmaSFR}{\ensuremath{\Sigma_{\mathrm{SFR}}}}

\newcommand{\HH}{\ensuremath{\mathrm{H}_{2}}}
\newcommand{\CIHH}{\ensuremath{\CI/[\HH]}}

\newcommand{\aco}{\ensuremath{\alpha_{\mathrm{CO}}}}

\newcommand{\Tkin}{\ensuremath{T_{\mathrm{kin}}}}
\newcommand{\Tex}{\ensuremath{T_{\mathrm{ex}}}}
\newcommand{\Tdust}{\ensuremath{T_{\mathrm{dust}}}}
\newcommand{\nHH}{\ensuremath{n_{\HH}}}
\newcommand{\rhoHH}{\ensuremath{\rho_{\HH}}}

\newcommand{\avg}[1]{\ensuremath{\langle #1 \rangle}}

\newcommand{\CI}{\ensuremath{\mathrm{[C\, \textsc{i}]}}}
\newcommand{\CII}{\ensuremath{\mathrm{[C\, \textsc{ii}]}}}

\newcommand{\avgCIabundCO}{\ensuremath{(1.9 \pm 0.4) \times 10^{-5}}}
\newcommand{\avgCIabundCOtherm}{\ensuremath{(2.2 \pm 0.4) \times 10^{-5}}}
\newcommand{\avgCIabundDust}{\ensuremath{(2.2 \pm 0.3) \times 10^{-5}}}

\newcommand{\rTwoOneLVGzLoOne}{\ensuremath{0.83 \pm 0.12}}
\newcommand{\rTwoOneLVGzLoTwo}{\ensuremath{0.75 \pm 0.11}}
\newcommand{\rThreeOneLVGzHiOne}{\ensuremath{0.92 \pm 0.17}}
\newcommand{\rThreeOneLVGzHiTwo}{\ensuremath{0.80 \pm 0.14}}

\newcommand{\rFourTwostack}{\ensuremath{0.33 \pm 0.04}}
\newcommand{\rFiveTwostack}{\ensuremath{0.23 \pm 0.02}}
\newcommand{\rThreeOnestack}{\ensuremath{0.77 \pm 0.14}}
\newcommand{\rSevenOnestack}{\ensuremath{0.19 \pm 0.04}}

\received{18 May 2020}
\revised{26 August 2020}
\accepted{31 August 2020}

\submitjournal{ApJ}

\shorttitle{ASPECS-LP: CO excitation, \CI\ and ISM conditions in galaxies at
  $z=1-3$}

\shortauthors{Boogaard et al.}

\begin{document}

\title{The ALMA Spectroscopic Survey in the HUDF: CO Excitation and Atomic
  Carbon in Star-Forming Galaxies at $z=1-3$}

\correspondingauthor{Leindert Boogaard}
\email{boogaard@strw.leidenuniv.nl}

\author[0000-0002-3952-8588]{Leindert A. Boogaard} \affil{Leiden Observatory,
  Leiden University, PO Box 9513, NL-2300 RA Leiden, The Netherlands}

\author[0000-0001-5434-5942]{Paul van der Werf} \affil{Leiden Observatory, Leiden University, PO Box
  9513, NL-2300 RA Leiden, The Netherlands}

\author[0000-0003-4678-3939]{Axel Wei\ss} \affil{Max-Planck-Institut f\"{u}r Radioastronomie, Auf dem
  H\"{u}gel 69, 53121 Bonn, Germany}

\author[0000-0003-1151-4659]{Gerg\"{o} Popping} \affil{European Southern
  Observatory, Karl-Schwarzschild-Str. 2, D-85748, Garching, Germany}

\author[0000-0002-2662-8803]{Roberto Decarli} \affil{INAF-Osservatorio di
  Astrofisica e Scienza dello Spazio, via Gobetti 93/3, I-40129, Bologna,
  Italy}

\author[0000-0003-4793-7880]{Fabian Walter} \affil{Max Planck Institute f\"{u}r
  Astronomie, K\"{o}nigstuhl 17, 69117 Heidelberg, Germany} \affil{National Radio
  Astronomy Observatory, Pete V. Domenici Array Science Center, P.O. Box O,
  Socorro, NM 87801, USA}

\author[0000-0002-6290-3198]{Manuel Aravena} \affil{N\'ucleo de Astronom\'ia de
  la Facultad de Ingenier\'ia y Ciencias, Universidad Diego Portales,
  Av. Ej\'ercito Libertador 441, Santiago, Chile}

\author[0000-0002-4989-2471]{Rychard Bouwens} \affil{Leiden Observatory, Leiden University, PO Box
  9513, NL-2300 RA Leiden, The Netherlands}

\author[0000-0001-9585-1462]{Dominik Riechers} \affil{Cornell University, 220
  Space Sciences Building, Ithaca, NY 14853, USA} \affil{Max Planck Institute
  f\"{u}r Astronomie, K\"{o}nigstuhl 17, 69117 Heidelberg, Germany}

\author[0000-0003-3926-1411]{Jorge Gonz\'alez-L\'opez} \affil{N\'ucleo de
  Astronom\'ia de la Facultad de Ingenier\'ia y Ciencias, Universidad Diego
  Portales, Av. Ej\'ercito Libertador 441, Santiago, Chile} \affil{Instituto de
  Astrof\'{\i}sica, Facultad de F\'{\i}sica, Pontificia Universidad Cat\'olica
  de Chile Av. Vicu\~na Mackenna 4860, 782-0436 Macul, Santiago, Chile}

\author[0000-0003-3037-257X]{Ian Smail} \affil{Centre for Extragalactic Astronomy, Department of
  Physics, Durham University, South Road, Durham, DH1 3LE, UK}

\author[0000-0001-6647-3861]{Chris Carilli} \affil{National Radio Astronomy Observatory, Pete
  V. Domenici Array Science Center, P.O. Box O, Socorro, NM 87801, USA}
\affil{Battcock Centre for Experimental Astrophysics, Cavendish Laboratory,
  Cambridge CB3 0HE, UK}

\author[0000-0002-1173-2579]{Melanie Kaasinen} \affil{Max Planck Institute f\"{u}r Astronomie,
  K\"{o}nigstuhl 17, 69117 Heidelberg, Germany} \affil{Universit\"{a}t
  Heidelberg, Zentrum f\"{u}r Astronomie, Institut f\"{u}r Theoretische
  Astrophysik, Albert-Ueberle-Stra\ss e 2, D-69120 Heidelberg, Germany}

\author[0000-0002-3331-9590]{Emanuele Daddi} \affil{Laboratoire AIM, CEA/DSM-CNRS-Universite Paris
  Diderot, Irfu/Service d'Astrophysique, CEA Saclay, Orme des Merisiers, 91191
  Gif-sur-Yvette cedex, France}

\author[0000-0003-2027-8221]{Pierre Cox} \affil{Institut d'astrophysique de
  Paris, Sorbonne Université, CNRS, UMR 7095, 98 bis bd Arago, 7014 Paris,
  France}

\author[0000-0003-0699-6083]{Tanio D\'iaz-Santos} \affil{N\'ucleo de Astronom\'ia de la Facultad de
  Ingenier\'ia y Ciencias, Universidad Diego Portales, Av. Ej\'ercito
  Libertador 441, Santiago, Chile} \affil{Chinese Academy of Sciences South
  America Center for Astronomy (CASSACA), National Astronomical Observatories,
  CAS, Beijing 100101, China} \affil{Institute of Astrophysics, Foundation for
  Research and Technology -- Hellas (FORTH), Heraklion, GR-70013, Greece}

\author[0000-0003-4268-0393]{Hanae Inami} \affil{Hiroshima Astrophysical Science Center, Hiroshima
  University, 1-3-1 Kagamiyama, Higashi-Hiroshima, Hiroshima 739-8526, Japan}

\author{Paulo C.~Cortes} \affil{Joint ALMA Observatory - ESO, Av. Alonso de
  C\'ordova, 3104, Santiago, Chile} \affil{National Radio Astronomy
  Observatory, 520 Edgemont Rd, Charlottesville, VA, 22903, USA}

\author{Jeff Wagg} \affil{SKA Organization, Lower Withington Macclesfield,
  Cheshire SK11 9DL, UK}

\begin{abstract}
  We investigate the CO excitation and interstellar medium (ISM) conditions in
  a cold gas mass-selected sample of 22 star-forming galaxies at $z=0.46-3.60$,
  observed as part of the ALMA Spectroscopic Survey in the Hubble Ultra Deep
  Field (ASPECS).  Combined with VLA follow-up observations, we detect a total
  of 34 CO $J \rightarrow J-1$ transitions with $J=1$ up to 8 (and an
  additional 21 upper limits, up to $J=10$) and six $\CI$
  ${^{3}P_{1}} \rightarrow {^{3}P_{0}}$ and
  ${^{3}P_{2}} \rightarrow {^{3}P_{1}}$ transitions (and 12 upper limits).  The
  CO(2--1) and CO(3--2)-selected galaxies, at $\avg{z}=1.2$ and 2.5,
  respectively, exhibit a range in excitation in their mid-$J=4,5$ and
  high-$J=7,8$ lines, on average lower than (\LIR-brighter) BzK-color- and
  submillimeter-selected galaxies at similar redshifts.  The former implies
  that a warm ISM component is not necessarily prevalent in gas mass-selected
  galaxies at $\avg{z}=1.2$.  We use stacking and Large Velocity Gradient
  models to measure and predict the average CO ladders at $z<2$ and $z\ge2$,
  finding $r_{21} = \rTwoOneLVGzLoTwo$ and $r_{31} = \rThreeOnestack$,
  respectively.  From the models, we infer that the galaxies at $z\ge2$ have
  intrinsically higher excitation than those at $z<2$.  This fits a picture in
  which the global excitation is driven by an increase in the star formation
  rate surface density of galaxies with redshift.  We derive a neutral atomic
  carbon abundance of \avgCIabundCO, comparable to the Milky Way and
  main-sequence galaxies at similar redshifts, and fairly high densities
  ($\ge 10^{4}$\,cm$^{-3}$), consistent with the low-$J$ CO excitation.  Our
  results imply a decrease in the cosmic molecular gas mass density at $z\ge2$
  compared to previous ASPECS measurements.
\end{abstract}

\section{Introduction}
\label{sec:intro}
Cold molecular gas is the fuel for star formation.  Characterizing the mass of
the cold interstellar medium (ISM) and the internal physical conditions
(temperature, density and radiation field) is therefore fundamental to our
understanding of the process of star formation \citep[see the reviews
by][]{McKee2007, Kennicutt2012, Carilli2013}.  The majority of the star
formation at intermediate redshifts ($z=1-3$) takes place in galaxies which
have an average star formation rate for their stellar mass.  These galaxies lie
on the `main sequence of star-forming galaxies'---the empirical correlation
between the stellar mass and star formation rate of galaxies across cosmic time
\citep[e.g.,][]{Noeske2007a, Elbaz2011, Whitaker2014, Schreiber2015,
  Boogaard2018}.  Although measurements of the molecular gas mass in these
galaxies are now more frequently conducted, the physical conditions in the cold
ISM of star-forming galaxies at $z>1$ are still poorly constrained.

The mass of the molecular ISM is dominated by \HH, which does not radiate under
typical conditions, and must therefore be traced by other species.  The most
common and direct tracer of the molecular gas mass is the first rotational
transition of carbon monoxide $^{12}$C$^{16}$O $J=1\rightarrow 0$, hereafter
CO(1--0) \citep[e.g.,][]{Dickman1986, Solomon1987, Bolatto2008}.  Alternative
tracers of the molecular gas mass include the dust emission
\cite[e.g.,][]{Hildebrand1983, Magdis2012, Scoville2014, Scoville2016,
  Magnelli2020} and lines from fainter optically thin species, such as neutral
atomic carbon \citep[\CI;][]{Papadopoulos2004, Weiss2005a, Walter2011}, now
more frequently observed in star-forming galaxies at $z>1$
\citep[e.g.,][]{Popping2017, Valentino2018, Bourne2019}.

Measurements of the molecular gas mass via CO at $z>1$ are limited to the
specific transitions that can be observed through the atmospheric windows from
Earth.  Constraints on the CO excitation are therefore crucial to convert
observations from higher-$J$ lines back to CO(1--0).  The higher rotational
levels of CO (with quantum number $J>1$) are populated both radiatively and
collisionally and the rotational ladder of CO is therefore a key probe of the
density, $\nHH$, and kinetic temperature, $\Tkin$, of the emitting medium.  The
excitation of CO can be driven by a number of processes, related to star
formation, (galactic) dynamics (including shocks/mechanical heating) and
potential activity from an active galactic nucleus (AGN).  In the local
universe, observations with the \emph{Herschel} satellite have shown that the
CO excitation in (U)LIRGS, (Ultra) Luminous Infrared Galaxies with
$\LIR \ge 10^{11}$ ($10^{12}$) \citep{Sanders1996}, can often be well modeled
by the combination of a cold component (containing most of the mass) and a warm
component, dominating the emission below and above $J \approx 4$ respectively,
while heating from an AGN is the dominant contributor to the line emission only
for the levels above $J \approx 10$ \citep[e.g.,][]{vanderWerf2010, Greve2014,
  Rosenberg2015, Kamenetzky2014, Kamenetzky2017, Lu2017}.  The CO excitation in
sources at higher redshift has been a field of intense study, yet, to date,
only limited constraints exist regarding the CO ladder in star-forming galaxies
at $z>1$.

At the time of the review by \cite{Carilli2013}, the main sources studied in
multiple CO transitions at $z>1$ were quasars (QSOs), radio galaxies and
submillimeter-selected galaxies (SMGs), with high $\LIR \gg 10^{12}$\,\Lsun.
Overall, these early results were indicative of decreasing excitation (i.e., a
lower \nHH\ and \Tkin) going from quasars to SMGs.  Since then, the average CO
excitation of SMGs has been studied by \cite{Bothwell2013}, who characterized a
sample of mostly unlensed SMGs at $z=2-4$, up to CO(7--6) (including CO(1--0)
observations from \citealt{Carilli2010, Riechers2010, Riechers2011a,
  Ivison2011}).  \cite{Spilker2014} used ALMA spectral scan observations of 22
lensed SMGs detected with the South Pole Telescope (SPT) at $z=2-6$
\citep{Weiss2013} to stack CO(3--2) up to CO(6--5).  More recently,
\cite{Yang2017} studied \emph{Herschel}-selected, strongly lensed SMGs at
$z=2-4$ up to CO(8--7).  These studies find that the CO ladders of SMGs can
continue to rise up to $J\sim7$, testifying to a warm and dense
($n \ge 10^{5.5}$\,cm$^{-3}$) ISM.  The differences between the (low-$J$) CO
excitation in SMGs and (mid-IR selected) AGN have not been found to be
statistically significant \citep{Sharon2016, Kirkpatrick2019}.

In contrast, observations of CO excitation in main-sequence star-forming
galaxies (SFGs) at $z>1$ have only recently become possible, with the advent of
the Northern Extended Millimeter Array (NOEMA) and the Atacama Large Millimeter
Array (ALMA).  The Plateau de Bure Interferometer HIgh-z Blue Sequence Survey
(PHIBSS) has observed CO(3--2) in a sample of massive, main sequence-selected
galaxies between $z=1-3$ \citep{Genzel2010, Genzel2015, Tacconi2010,
  Tacconi2013, Tacconi2018}, with multi-line CO excitation follow-up of only a
few sources \citep{Bolatto2015, Brisbin2019}.  A number of SFGs, selected by
their BzK-color \citep{Daddi2004} and having a detection at 24\,\micron\ and
1.4\,GHz \citep{Daddi2010}, have been observed in more than one CO transition
from CO(1--0) to CO(3--2) \citep{Dannerbauer2009, Daddi2010, Aravena2010,
  Aravena2014}.  The CO ladder of four of these `BzK-selected' galaxies at
$z\approx1.5$ was characterized comprehensively by \cite{Daddi2015}.  They
found all sources were significantly excited in their CO(5--4) transition,
compared to the lower $J$ transitions, indicating the presence of both a cold
and a denser, possibly warmer gas component.  Very recently,
\cite{Valentino2020b} expanded these results with observations of a larger
sample of similarly IR-bright star-forming galaxies at $z=1.25$.  However, all
these samples were preselected based on their star formation rate, and are
still among the most massive and IR luminous main-sequence galaxies at these
redshifts, with only specific sources selected for multi-line follow-up.
Therefore, it remains unclear whether the excitation conditions found in these
sources are representative of the general population of star-forming galaxies
at these redshifts, in particular at lower masses and star formation rates.

The ALMA Spectroscopic Survey in the Hubble Ultra Deep Field (ASPECS;
\citealt{Walter2016}) provides a unique avenue to study the CO excitation,
molecular gas content and physical conditions of the cold ISM of star-forming
galaxies at high-redshift.  ASPECS is a flux-limited survey, designed to detect
CO in galaxies without preselection.  It thereby provides the most complete
inventory of the cosmic molecular gas density, $\rhoHH(z)$, to date
\citep{Decarli2016a, Decarli2019, Decarli2020}.  The galaxies detected in CO by
ASPECS are found to lie on, above and below the main sequence at $z=1-3$, with
near-solar metallicities \citep{Aravena2019, Boogaard2019}.  The coverage of
ASPECS (Band 3 and Band 6) provides simultaneous constraints on multiple lines
from CO, \CI\ for most sources, depending on the redshift (as well as any other
species in the frequency range).  Furthermore, the multiple tunings scanning
through the entire ALMA frequency bands give a high continuum sensitivity,
providing a deep (9.3\,$\mu$Jy/beam, \autoref{sec:alma}), contiguous continuum
map at 1.2\,mm in the HUDF \citep{Gonzalez-Lopez2020, Aravena2020}.  Using
earlier data from the ASPECS-Pilot program on a smaller area of the sky,
\cite{Decarli2016b} studied a sample of seven galaxies at $z=1-3$ (a subset of
the sources studied in this paper), finding that the CO excitation conditions
were overall lower than those typically found in starbursts, SMGs and QSO
environments.

This paper studies the CO excitation, atomic carbon emission and ISM conditions
in a flux-limited sample of 22 CO and/or dust continuum detected galaxies at
$z=1-3$ from the ASPECS Large Program (LP), supplemented by follow-up CO(1--0)
observations from VLASPECS \citep{Riechers2020b}.  The paper is organized as
follows.  We first present the ALMA and VLA observations and the physical
properties of the galaxies in the sample (\autoref{sec:observations}).  All
line fluxes are measured homogeneously through simultaneous Gaussian fitting
(\autoref{sec:methods}) and presented in \autoref{sec:results}.  We discuss the
mid- and high-$J$ CO excitation in the individual CO(2--1)- and
CO(3--2)-selected sources at $\avg{z}=1.2$ and $\avg{z}=2.5$, respectively, in
\autoref{sec:co-excitation-indiv} and compute the average CO ladders through
stacking (including individually undetected lines;
\autoref{sec:average-co-ladder}).  We then use Large Velocity Gradient models
to characterize the average ladders at $z\le2$ and $z>2$
(\autoref{sec:lvg-modeling}).  We further analyze the low-$J$ CO excitation by
placing our galaxies on empirical relations with the rest-frame 850\,\micron\
dust luminosity (\autoref{sec:dust-cont-vers}).  We next turn to the neutral
atomic carbon, discuss its mass and abundance, and use PDR models to analyze
the average ISM conditions in our galaxies (\autoref{sec:atomic-carbon}).  The
implications of our measurements on the average low-, mid- and high-$J$ CO
excitation in star-forming galaxies at $z\ge1$ are discussed in
\autoref{sec:discussion}.  Finally, we conclude with the implications of our
results for the inference of the cosmic molecular gas density from ASPECS, as
these are the galaxies that directly inform that measurement
(\autoref{sec:impl-cosm-molec}).  Throughout this paper, we use a
\cite{Chabrier2003} initial mass function and a concordance flat $\Lambda$CDM
cosmology with $H_{0} = 70$\,km\,s$^{-1}$\,Mpc$^{-1}$, $\Omega_{m} = 0.3$ and
$\Omega_{\Lambda} = 0.7$, in good agreement with the results from
\cite{PlanckCollaboration2015}.

\begin{deluxetable*}{cccccccccc}
  \tablecaption{Physical properties of the ASPECS-LP sample considered in this paper \label{tab:sample}}
  \tablehead{
    \colhead{ID 1mm} & \colhead{ID 3mm} &
    \colhead{ID 9mm} &\colhead{$z$} &
    \colhead{$\log M_{*}$} & \colhead{$\log \mathrm{SFR}$} & \colhead{$\log
      \LIR $} & \colhead{$\log \SigmaSFR$} &  \colhead{$r_{e}$}&\colhead{X-ray} \\
    \colhead{} & \colhead{} & \colhead{} & \colhead{} &
    \colhead{(M$_{\odot}$)} & \colhead{(M$_{\odot}$\,yr$^{-1}$)} &
    \colhead{(L$_{\odot}$)} & \colhead{(M$_{\odot}$\,yr$^{-1}$\,kpc$^{-2}$)} &
    \colhead{(arcsec)} & \colhead{}
  }
  \colnumbers
  \startdata
1mm.C01             & 3mm.01   & 9mm.1   & 2.543 & $10.4 \pm 0.1$ & $2.37 \pm 0.10$ & $12.9 \pm 0.1$ & $1.07 \pm 0.19$  & $0.21 \pm 0.04$ & AGN \\
1mm.C03             & 3mm.04   & \nodata & 1.414 & $11.3 \pm 0.1$ & $1.72 \pm 0.13$ & $12.0 \pm 0.1$ & $-0.82 \pm 0.14$ & $0.88 \pm 0.04$ & \nodata  \\
1mm.C04             & 3mm.03   & 9mm.6   & 2.454 & $10.7 \pm 0.2$ & $1.78 \pm 0.21$ & $11.9 \pm 0.2$ & $-0.46 \pm 0.22$ & $0.63 \pm 0.04$ & \nodata  \\
1mm.C05             & 3mm.05   & \nodata & 1.551 & $11.5 \pm 0.1$ & $1.79 \pm 0.17$ & $12.0 \pm 0.2$ & $-0.83 \pm 0.18$ & $0.98 \pm 0.04$ & AGN \\
1mm.C06             & 3mm.07   & 9mm.3   & 2.696 & $11.1 \pm 0.1$ & $2.32 \pm 0.14$ & $12.4 \pm 0.1$ & $0.10 \pm 0.15$  & $0.61 \pm 0.04$ & \nodata  \\
1mm.C07             & \nodata  & 9mm.7   & 2.580 & $11.0 \pm 0.1$ & $1.65 \pm 0.14$ & $11.9 \pm 0.1$ & $0.48 \pm 0.24$  & $0.18 \pm 0.04$ & AGN \\
1mm.C09             & 3mm.13   & \nodata & 3.601 & $9.8 \pm 0.2$  & $1.58 \pm 0.21$ & $11.6 \pm 0.2$ & $0.06 \pm 0.25$  & $0.27 \pm 0.04$ & \nodata  \\
1mm.C10             & \nodata  & \nodata & 1.997 & $11.1 \pm 0.1$ & $2.04 \pm 0.10$ & $12.4 \pm 0.1$ & $-0.16 \pm 0.12$ & $0.60 \pm 0.04$ & X \\
1mm.C12             & 3mm.15   & \nodata & 1.096 & $9.5 \pm 0.1$  & $1.55 \pm 0.10$ & $11.7 \pm 0.1$ & $-0.82 \pm 0.11$ & $0.73 \pm 0.04$ & AGN \\
1mm.C13             & 3mm.10   & \nodata & 1.037 & $11.1 \pm 0.1$ & $1.27 \pm 0.10$ & $11.6 \pm 0.1$ & $-0.36 \pm 0.15$ & $0.31 \pm 0.04$ & \nodata  \\
1mm.C14a            & \nodata  & 9mm.5   & 1.999 & $10.8 \pm 0.1$ & $1.70 \pm 0.17$ & $11.9 \pm 0.2$ & $0.20 \pm 0.22$  & $0.27 \pm 0.04$ & \nodata  \\
1mm.C16             & 3mm.06   & \nodata & 1.095 & $10.6 \pm 0.1$ & $1.52 \pm 0.10$ & $11.5 \pm 0.1$ & $-0.76 \pm 0.11$ & $0.66 \pm 0.04$ & X \\
1mm.C15             & 3mm.02   & \nodata & 1.317 & $11.2 \pm 0.1$ & $1.05 \pm 0.12$ & $11.5 \pm 0.1$ & $-0.95 \pm 0.14$ & $0.48 \pm 0.04$ & \nodata  \\
1mm.C19             & 3mm.12   & 9mm.4   & 2.574 & $10.6 \pm 0.1$ & $1.54 \pm 0.20$ & $11.6 \pm 0.2$ & $-0.43 \pm 0.21$ & $0.46 \pm 0.04$ & AGN \\
1mm.C20             & \nodata  & \nodata & 1.093 & $10.9 \pm 0.1$ & $0.97 \pm 0.14$ & $11.2 \pm 0.1$ & $-1.01 \pm 0.16$ & $0.46 \pm 0.04$ & \nodata  \\
1mm.C25             & 3mm.14   & \nodata & 1.098 & $10.6 \pm 0.1$ & $1.35 \pm 0.11$ & $11.4 \pm 0.1$ & $-0.00 \pm 0.19$ & $0.22 \pm 0.04$ & \nodata  \\
1mm.C23             & 3mm.08   & \nodata & 1.382 & $10.7 \pm 0.1$ & $1.60 \pm 0.12$ & $11.7 \pm 0.1$ & $-1.03 \pm 0.12$ & $0.99 \pm 0.04$ & \nodata  \\
1mm.C30             & \nodata  & \nodata & 0.458 & $10.0 \pm 0.1$ & $1.12 \pm 0.10$ & $11.0 \pm 0.1$ & $-0.01 \pm 0.22$ & $0.17 \pm 0.04$ & X \\
\nodata             & 3mm.11   & \nodata & 1.096 & $10.2 \pm 0.1$ & $0.99 \pm 0.11$ & $11.0 \pm 0.1$ & $-0.70 \pm 0.15$ & $0.33 \pm 0.04$ & \nodata  \\
\nodata$^{\dagger}$ & 3mm.09   & 9mm.2   & 2.698 & $11.1 \pm 0.1$ & $2.54 \pm 0.10$ & $12.6 \pm 0.1$ & $2.05 \pm 0.42$  & $0.08 \pm 0.04$ & AGN \\
Faint.1mm.C20       & 3mm.16   & \nodata & 1.294 & $10.3 \pm 0.1$ & $1.06 \pm 0.14$ & $11.0 \pm 0.2$ & $-1.10 \pm 0.15$ & $0.57 \pm 0.04$ & \nodata  \\
\nodata             & MP.3mm.2 & \nodata & 1.087 & $10.4 \pm 0.1$ & $1.40 \pm0.10$ & $11.5 \pm 0.1$ & $-0.95 \pm 0.11$ & $0.71 \pm 0.04$ & X \\
\enddata
\tablenotemark{$\dagger$}{Object falls outside of the Band 6 mosaic, but is the
  brightest 1\,mm continuum source in the ASPECS field
  \citep[cf.][]{Dunlop2017}.}

\tablecomments{(1) ASPECS-LP continuum ID \citep{Gonzalez-Lopez2020,
    Aravena2020}. (2) ASPECS-LP line ID \citep{Boogaard2019}. (3) VLASPECS ID
  \citep{Riechers2020b} (4) Spectroscopic redshift. (5) Stellar mass. (6) Star
  formation rate. (7) Infrared luminosity; \LIR($3-1000$\,\micron). (8) Star
  formation rate surface density;
  $\Sigma_{\rm SFR} = \mathrm{SFR} / 2 \pi r_{e}^{2}$.  (9) \emph{HST}/F160W
  effective radius from \cite{vanderWel2012}, for which we adopt an 0\farcs04
  error floor. (10) X-ray classification as either hosting an active galactic
  nucleus (AGN) or another source of X-ray emission (X) \citep{Luo2017}.
  Columns (5)--(7) were derived with \textsc{Magphys} \citep{DaCunha2008,
    DaCunha2015}.  We conservatively fold in a 0.1\,dex error to the
  \textsc{Magphys} uncertainties, to account for underestimated and systematic
  uncertainties, and report the values as
  $p_{50} \pm \sqrt{(0.5 (p_{84} - p_{16}))^{2} + 0.1^{2}}$, where $p_{i}$ is
  the $i^{\mathrm{th}}$ percentile.}
\end{deluxetable*}
\newpage
\section{Observations and ancillary data}
\label{sec:observations}
\subsection{ALMA Spectroscopic Survey Data Reduction}
\label{sec:alma}
The ASPECS data consists of two spectral scan mosaics over the deepest part of
the Hubble Ultra Deep Field \citep[HUDF;][]{Illingworth2013, Koekemoer2013}.
The raw ASPECS data were processed with \textsc{casa} \citep{McMullin2007} as
described in \cite{Gonzalez-Lopez2019} for Band~3 and \cite{Decarli2020} for
Band~6.  The visibilities were imaged using the task \textsc{tclean}, adopting
natural weighting.  The complete mosaics cover an area of 4.6~arcmin$^{2}$
(Band~3) and 2.9~arcmin$^{2}$ (Band~6), measured as the region in which the
primary beam sensitivity is $\ge 50\%$ of the peak sensitivity (6.1 and
3.7~arcmin$^{2}$ when measured down to 20\%).

The Band~3 data cube ranges from $84 - 115$\,GHz, with a channel width of
7.813\,MHz, corresponding to velocity resolution of
$\Delta v \approx 23.5$\,km\,s$^{-1}$ at 99.5\,GHz.  The spatial resolution of
the naturally weighted cube is $\approx 1\farcs8 \times 1\farcs5$ (at
99.5\,GHz).  The sensitivity varies across the frequency range, reaching an
average root-mean-square (rms) sensitivity per channel of
$\approx 0.2$\,mJy\,beam$^{-1}$, varying across the frequency range \citep[see
][Fig.~3]{Gonzalez-Lopez2019}.  The Band~6 data cube spans from
$212 - 272$\,GHz, and was resampled at a channel width of 15.627\,MHz,
corresponding to $\Delta v \approx 19.4$\,km\,s$^{-1}$ at 242\,GHz.  The
naturally weighted cube has a beam size of $\approx 1\farcs5 \times 1\farcs1$
and reaches an average rms depth of $\approx 0.5$\,mJy\,beam$^{-1}$ per channel
\citep[see ][Fig. 1]{Decarli2020}.

To create continuum maps, we collapse both the Band 6 (1.2\,mm) and Band 3
(3.0\,mm) data cubes over their full frequency range.  The deepest parts of the
continuum reach $3.8\,\mu$Jy\,beam$^{-1}$ in Band~3, with a beam size of
$2\farcs8 \times 1\farcs7$, and $9.3\,\mu$Jy\,beam$^{-1}$ in Band~6, with a
beam size of $1\farcs5 \times 1\farcs1$ \citep{Gonzalez-Lopez2019,
  Gonzalez-Lopez2020}.  The absolute flux calibration is expected to be
reliable at the $\sim 10\%$ level.

\subsection{ASPECS Sample}
\label{sec:sample}
We search for line and continuum sources in the ASPECS data cubes, which is
described in \cite{Gonzalez-Lopez2019} and \cite{Gonzalez-Lopez2020}.  In the
Band 3 data we detect 16 CO emitters at high significance from the line search,
plus 2 additional CO emitters based on a MUSE redshift prior
\citep{Boogaard2019}.  For five of these sources we also detect the continuum
at 3\,mm \citep{Gonzalez-Lopez2019}.  From the Band 6 data we detect 35 sources
in 1.2\,mm dust continuum at high significance, 32 of which show counterparts
in the optical/NIR imaging \citep{Gonzalez-Lopez2020, Aravena2020}.  We conduct
a search for emission lines in the Band 6 cube following the same approach as
for the Band 3 data.  This reveals several CO (and \CI) emitters, all
coinciding with sources detected in the Band 6 continuum image, with one
exception: a narrow CO line in one of the CO-emitters also found in Band 3
(3mm.11; not detected in continuum at all).  Notably, we did not find any
high-significance lines in sources not already detected the dust continuum.
The Band 6 continuum sources furthermore encompass all Band 3 CO emitters
\citep{Aravena2020}, with four exceptions: The first two are the lowest mass
and SFR source of the main sample (3mm.11) and the faintest source in CO
(ASPECS-LP-MP.3mm.02).  The third is 3mm.16, which does however have a
dust-continuum counterpart in the supplementary catalog of 26 sources at lower
significance (the `Faint' sample), which were selected based on the presence of
a optical/near-IR counterpart \citep{Gonzalez-Lopez2020}.  The fourth is
3mm.09, which is the brightest source at 1.3\,mm in the field
\citep[UDF1;][]{Dunlop2017}.  This source was detected towards the edge of the
Band 3 mosaic (at $40$\% of the primary beam peak sensitivity; hereafter PB
response) and is at the extreme edge of the Band 6 mosaic.  The CO(7--6) and
\CI(2--1) lines lie at $6\%$ of the PB response at 218\,GHz and the source
falls outside the continuum map (below 10\% PB response).  We do include this
source in this paper, but note that the upper limits on the lines in Band 6 are
essentially unconstraining.  For the SED fitting
(\autoref{sec:multi-wavel-data}) we use the continuum measurement at 1.3\,mm.

We therefore consider all of the Band 3 and Band 6 continuum and line sources
that are detected in at least one line.  In total, the sample consists of 22
sources.  The majority of the sample is low-$J$ CO selected in Band 3 (17/22).
There are 5 exceptions, i.e., sources which are added based on the Band 6 data.
Three sources lack coverage of any CO lines in Band 3.  These include (1mm.C10
and C14a) at $z \approx 1.99$ and 1mm.C30 at $z=0.46$.  One source, 1mm.C20,
does not show CO lines in Band 3 nor 6, but is detected in \CI\ in Band 6.
Lastly, we report a new CO(3--2) detection for 1mm.C07 in Band 3.  This source
was not included in the original sample from \cite{Gonzalez-Lopez2019} because
the line is below their single-line fidelity threshold and the source lacks a
MUSE redshift.  However, this source is now confirmed through the detection of
the high-$J$ CO and \CI\ lines in band 6.  One Band 3 CO(2--1) emitter
(MP.3mm.1; based on a MUSE prior) is not included in this paper, because we
re-measure the integrated flux to be slightly below $3\sigma$.  This is likely
because we convolve both cubes to a slightly larger beam size, in order to
consistently measure the line ratios, at the cost of signal-to-noise (see
\autoref{sec:spectr-line-fitt}).

\begin{figure}[t]
\includegraphics[width=0.5\textwidth]{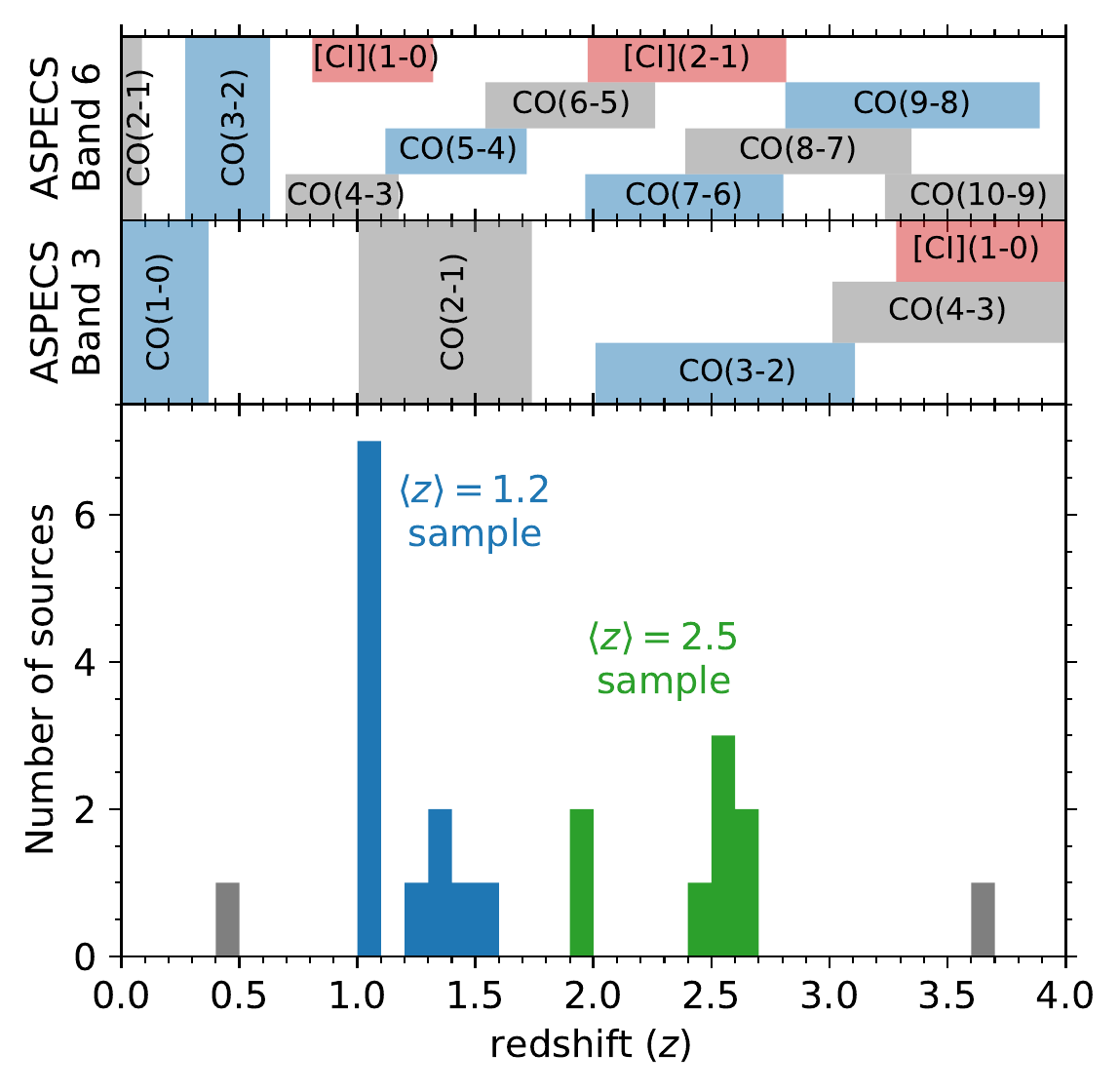}
\caption{Redshift distribution of the ASPECS sources discussed in this paper
  (bottom panel).  The top panels indicate the lines covered by ASPECS in Band
  3 and Band 6 at different redshift ranges (colored just to make them more
  easily distinguishable).  We highlight the samples at $z=1.0-1.6$ (blue) and
  $z=2.0-2.7$ (green), for which we have coverage of both a low-$J$ and a
  mid/high-$J$ CO line.  Additional VLA CO(1-0) follow-up is available for all
  but one source in the $\avg{z}=2.5$ sample (\autoref{sec:vla-data-from};
  \citealt{Riechers2020b}).\label{fig:hist}}
\end{figure}

The full sample is listed in \autoref{tab:sample}.  It spans redshifts from
$z=0.46 - 3.60$, with the majority of the sample being at $z=1-3$.  We show the
redshift distribution in \autoref{fig:hist}, highlighting the spectral lines
covered by ASPECS in the top panels.  The final redshifts are determined from
our fits of the CO and/or [CI] line(s), using the redshifts from the MUSE HUDF
survey and our literature compilation \citep[see][]{Boogaard2019, Decarli2019}
as prior information (\autoref{sec:spectr-line-fitt}).

\subsection{Very Large Array Observations (VLASPECS)}
\label{sec:vla-data-from}
The CO(1--0) transition in the ASPECS galaxies between $z=1.99 - 2.70$ was
observed with the Karl G. Jansky Very Large Array (VLA) as part of the VLASPECS
survey (\citealt{Riechers2020b}; VLA program ID: 19B-131; PI: Riechers).  Two
pointings were conducted with the D array in the Ka band, over a continuous
bandwidth of 30.593--38.662\,GHz at 2\,MHz spectral resolution, resulting in a
17\,km\,s$^{-1}$ resolution (at 35\,GHz).  The naturally-weighted cube has an
average rms noise level of $\approx 0.1$\,mJy\,beam$^{-1}$\,channel$^{-1}$
(increasing by about a factor of two from the low- to the high-frequency edge
of the bandpass, as expected) and a beam size of $4\farcs99 \times 1\farcs96$.
Given the recent flaring activity in the calibrator, the absolute flux is
conservatively considered to be reliable at the $\sim 15\%$ level.  The full
data reduction and presentation is part of \cite{Riechers2020b}.  In this paper,
we focus primarily on the CO excitation and analyze the data in concert with
the higher-$J$ CO lines.

\subsection{Multi-wavelength data \& SED fitting}
\label{sec:multi-wavel-data}
The wealth of multi-wavelength photometry available over the HUDF provides good
constraints on the spectral energy distribution (SED) of each of the ASPECS
galaxies.  By modeling the SEDs using the \textsc{Magphys}
\citep{DaCunha2008,DaCunha2015}, we derive stellar masses, star formation rates
(SFRs) and IR luminosities ($L_{\rm IR}$; $3 - 1000\,\mu$m).  We follow the
same procedure as described in \cite{Boogaard2019}, utilizing the UV --
24$\,\mu$m photometry from 3D-HST \citep{Skelton2014, Whitaker2014}, in
combination with the \emph{Herschel} $70 - 160$\,\micron\ data from
\cite{Elbaz2011} and the 3\,mm continuum from \cite{Gonzalez-Lopez2019}.
Superseding the earlier fits, we now include the updated 1.2\,mm flux
measurements from \cite{Gonzalez-Lopez2020}.  Furthermore, we include
$5 \sigma$ upper limits of 50\,$\mu$Jy and 20\,$\mu$Jy in the case of a
non-detection at 1.2\,mm and 3\,mm, respectively.  The fits for the full dust
continuum sample (including the sources not detected in CO) are presented in
\cite{Aravena2020}.  Following \cite{Aravena2020}, we conservatively fold in an
additional 0.1\,dex to the error bars to account for underestimated and
systematic uncertainties.  We derive average star formation rate surface
densities, $\SigmaSFR = \mathrm{SFR} / 2 \pi r_{e}^{2}$, using the
\emph{HST}/F160W half-light radii ($r_{e}$) from \cite{vanderWel2012}.  This is
a reasonable approximation for sources in which the radial extent of the star
formation follows the stellar disk, but should be considered as a lower limit
in the case of a more nuclear starburst.  The formal errors on the radii are of
order a few percent of the point spread function
($\mathrm{PSF} \sim 0\farcs16$), which we find to be very small.  Hence, we
conservatively adopt a floor on the error bar of $\mathrm{PSF}/4 = 0\farcs04$.
Lastly, the X-ray sources in the ASPECS sample are identified and classified
using the deep \emph{Chandra} 7\,Ms data from \cite{Luo2017} as described in
\cite{Boogaard2019}.

\section{Methods}
\label{sec:methods}
\subsection{Spectral line analysis}
\label{sec:spectr-line-fitt}
\begin{figure}[t]
\includegraphics[width=\columnwidth]{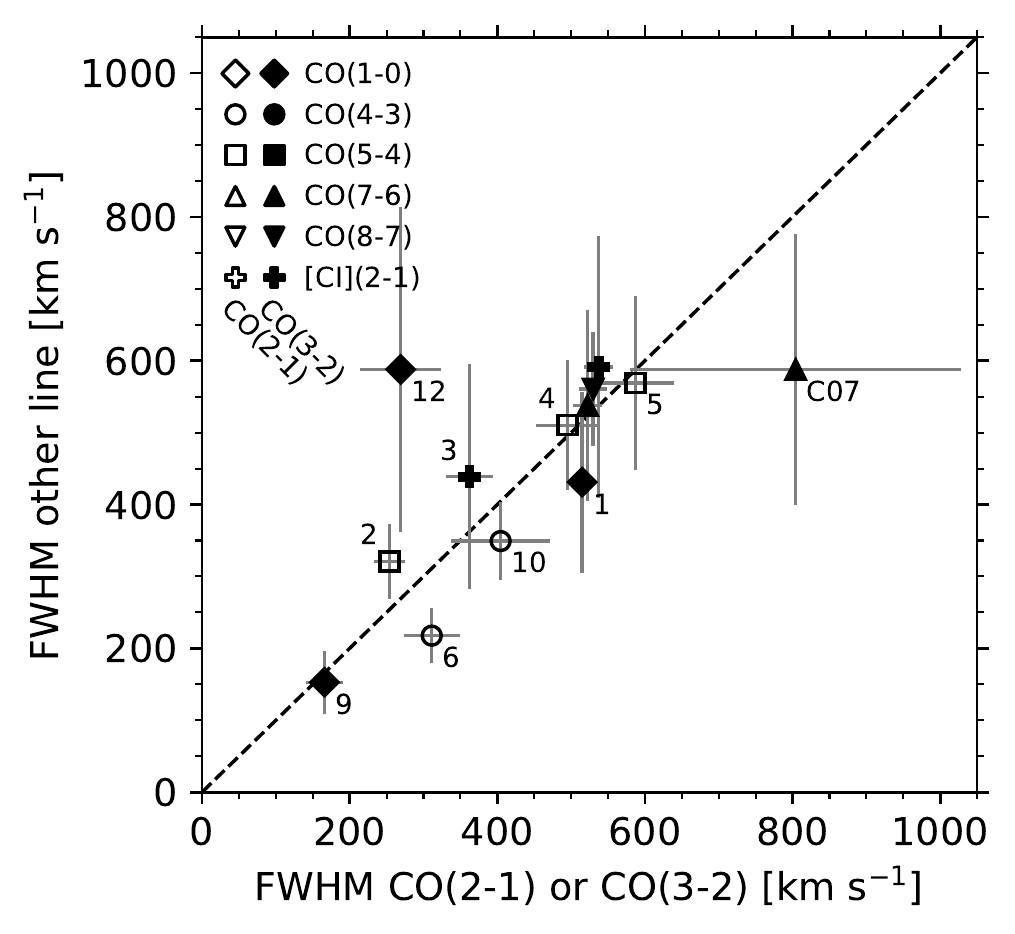}
\caption{Comparison of the line width (full-width at half-maximum; FWHM)
  between the CO(2--1) and CO(3--2) lines (by which the sample was selected)
  and the higher-$J$ CO ($J>4$) or \CI\ lines in individual ASPECS galaxies
  (extracted over the same 2\farcs2 aperture), as well as CO(1--0) from the
  VLA.  The sources were fit with a single redshift but allowing, \emph{for
    this figure only}, a varying line width for each transition.  Sources are
  identified by the 3mm.ID or else their 1mm.CID.  We only show sources where
  the relevant lines are detected with a $\mathrm{S/N} > 3$ in these fits.  We
  add a small positive offset to the multiple lines of 3mm.1, for readability.
  Overall, we find consistent line widths between the low-$J$ CO and higher-$J$
  CO/\CI\ lines.  Throughout the analysis presented in this work, we will
  therefore use a fixed line width to model the different transitions of a
  particular source, which is determined by fitting all the lines
  simultaneously (see \autoref{sec:spectr-line-fitt}).\label{fig:line-widths}}
\end{figure}
\begin{figure*}[t]
\includegraphics[width=\textwidth]{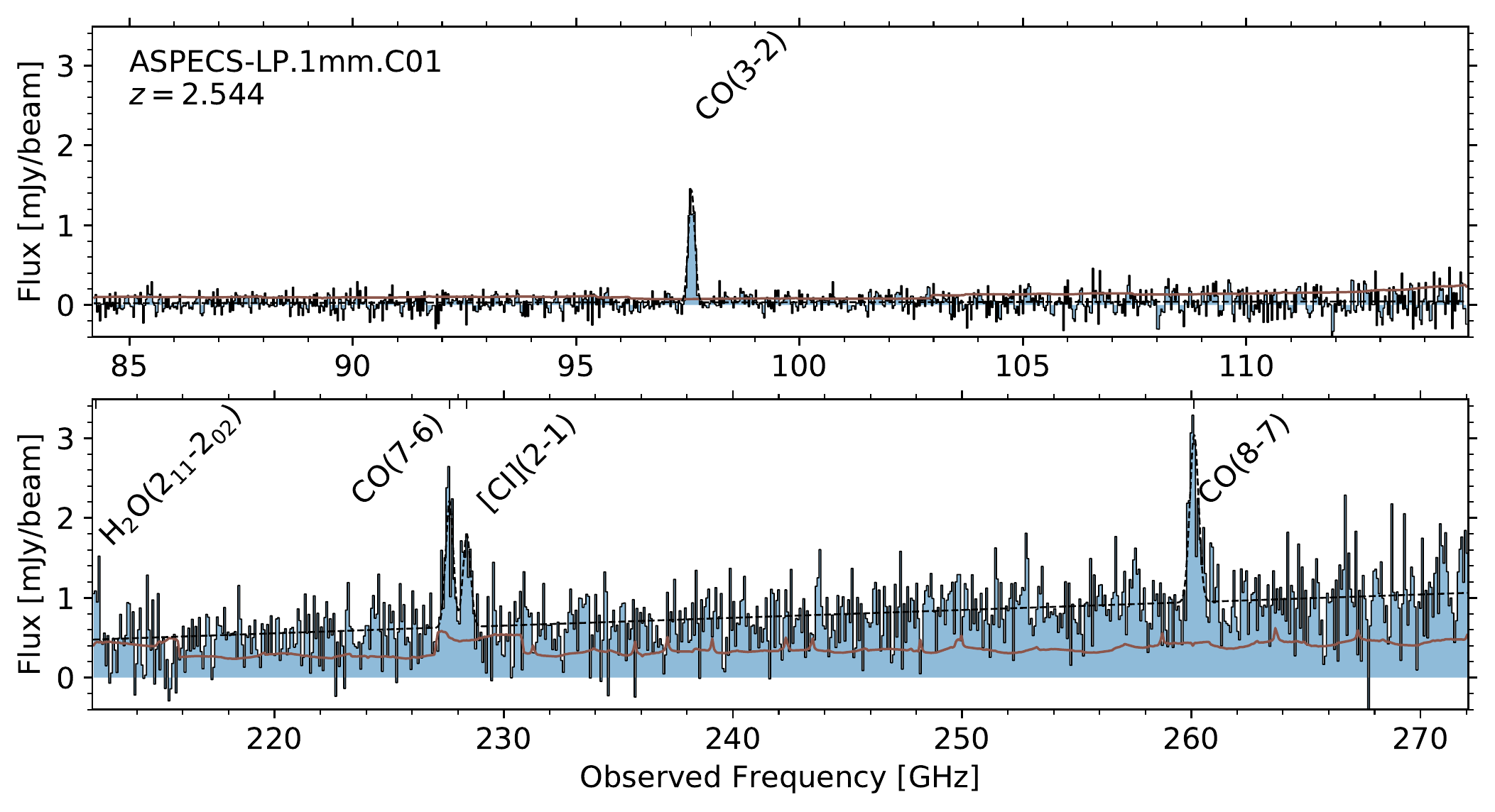}
\caption{Full spectrum in Band 3 (\textbf{Top}) and Band 6 (\textbf{Bottom}) of
  ASPECS-LP.1mm.C01, shown as an example of the ASPECS frequency coverage.  The
  brown line shows the root-mean-square noise in each of the channels.  For
  visualisation purposes, the spectra are averaged to a similar resolution of
  $95$\,km\,s$^{-1}$.  The black dashed line shows the best-fit model, which
  includes Gaussian line fits to the $^{12}$CO and atomic carbon (\CI) lines
  (constrained in redshift and line width by all the lines simultaneously) and
  a linear continuum (\autoref{sec:spectr-line-fitt}).  Note that in this
  particular source we also detect a water line at the edge of the Band 6, para
  H$_{2}$O($2_{11}\rightarrow2_{02}$), which is not included in the fitting
  (and not further discussed in the paper).\label{fig:aspecs1}}
\end{figure*}
We extract single pixel spectra from the naturally weighted Band 3 and Band 6
cubes, convolved to a common beam size of 2\farcs2.  In this way, we ensure
that the line fluxes are extracted over the same region of the galaxy, whilst
minimizing the impact of flux loss for sources that are more slightly extended
than the beam size of the naturally weighted cube.  We use the cubes at their
native spectral resolution in order to resolve even the narrowest lines
($\sim 50$\,km\,s$^{-1}$) into several resolution elements.  We adopt the
position of the dust continuum detection \citep{Gonzalez-Lopez2020,
  Aravena2020}, or, in the case of no dust continuum detection, the CO line
positions \citep{Gonzalez-Lopez2019, Boogaard2019}.  The beam size of the VLA
data is already larger than that of ASPECS ($\approx 5\farcs0 \times 2\farcs0$)
and only the brightest two sources are slightly resolved along the minor axis
by the VLA (similar to what is seen in ASPECS, which motivated the convolution
to 2\farcs2).  We therefore use the spectra extracted by \cite{Riechers2020b} in
order to measure the flux over as-similar regions as possible.

For each source, we simultaneously fit all CO and [CI] lines that are expected
to fall in Band 3, Band 6 and the VLA Ka band, based on the redshift from the
line-search, using the non-linear least square fitting code
\textsc{lmfit}\footnote{\url{https://lmfit.github.io/lmfit-py/}}
\citep{Newville2019}.  We first subtract the continuum in Band 6, which is
determined by fitting a first order polynomial to the median filtered spectrum.
All the lines in the continuum-free spectrum are modeled by Gaussian line
shapes, whose central frequencies are tied together by a single redshift.

Fitting the sources with the highest signal-to-noise (S/N) spectra, we find
that the widths of the different transitions are consistent in most cases.
This is illustrated in \autoref{fig:line-widths}, where we show the line width
measured in the CO(2--1) and CO(3--2) lines in Band 3 against the line width of
the other CO and \CI\ lines.  Here we only include sources with
$\mathrm{S/N} > 3$ in all the relevant lines in the free fit (which is more
conservative than for the fits where the line widths are tied together, because
of the additional degrees of freedom).  We therefore model all the lines with a
single line width.  Although this assumption is not strictly necessary, this
often improves the fitting of lines with lower signal-to-noise, where the line
width can be better constrained by the strongest lines.  The integrated line
fluxes are consistent within the uncertainties regardless of whether we force
the widths of the lines to match.  Furthermore, fitting the non-detected lines
simultaneously does not influence the fit of the detected lines within the
error (even in the most extreme case of a single detection and multiple upper
limits).  The observed line widths are likely governed by the global kinematics
of the source.  As such, the consistent line widths between the different
transitions suggest that the gas is not much more compact or extended in some
transitions compared to others, which supports our analysis of the global CO
excitation (see \autoref{sec:similar-line-widths} for further discussion).

As an illustration of the fitting procedure, we show the complete Band 3 and 6
spectrum of the brightest source, 1mm.C01, in \autoref{fig:aspecs1}, together
with the best fit model (lines and continuum).  This particular source is
detected in multiple lines as well as the dust continuum.

\subsection{Deriving line luminosities and molecular gas masses}
\label{sec:note-units-derived}
The line luminosities are commonly expressed in different units, useful for
different purposes, and we briefly review the relevant equations below
\citep[c.f.][]{Solomon2005, Obreschkow2009, Carilli2013}.  When expressed in
solar luminosities, the line luminosities indicate the total power emitted,
\begin{align}
  L  = 1.040 \times 10^{-3}\, S^{V} d_{L}^{2} \nu_{\rm obs} , \Lsun \label{Lline}.
\end{align}
Units of integrated brightness temperature are convenient to derive the line
excitation (notably, if the CO line emission originates in thermalized,
optically thick regions, \LCO\ is constant for all $J$ levels),
\begin{align}
  L' = 3.255 \times 10^{7}\, S^{V} d_{L}^{2} \nu_{\rm obs}^{-2} (1+z)^{-3} \, \mathrm{K\,km\,s}^{-1}\,\mathrm{pc}^{2} \label{Llineprime}.
\end{align}
In both equations, $S^{V} = \int S_{\nu} dv$ is the integrated line flux
($[S^{V}] = \mathrm{Jy\,km\,s}^{-1}$), $d_{L}$ is the luminosity distance
($[d_{L}] = \mathrm{Mpc}$) and $\nu_{\rm obs}$ is the observed line frequency
($[\nu_{\rm obs}] = \mathrm{GHz}$) \citep{Solomon1992}.  Note that the two
definitions are proportional, with
$\LCO = 3.130 \times 10^{-11} \nu_{\rm rest}^{-3} L_{\rm CO}$.

The CO excitation is typically reported as a brightness temperature ratio
between two transitions, which is computed from \LCO\ or $S^{V}$ as
\begin{align}
  r_{J_{2}\,J_{1}} = \frac{L'_{\mathrm{CO}\,J_{2}\rightarrow J_{2}-1}}{L'_{\mathrm{CO}\,J_{1}\rightarrow J_{1}-1}} = \frac{S^{V}_{\mathrm{CO}\,J_{2}\rightarrow J_{2}-1}}{S^{V}_{\mathrm{CO}\,J_{1} \rightarrow J_{1}-1}} \left(\frac{J_{1}}{J_{2}}\right)^{2} \label{eq:rJ1}
\end{align}

The relationship between the molecular gas mass ($M_{\rm mol}$) and the CO
luminosity (\LCO) is expressed as
\begin{align}
  M_{\rm mol} = \aco \frac{L'_{\mathrm{CO}\,J\rightarrow J-1}}{r_{J1}}\label{eq:Mmol},
\end{align}
where \aco\ is the conversion factor between CO luminosity and the total
molecular gas mass (including a factor of 1.36 to account for heavy elements,
primarily Helium; see \citealt{Bolatto2013} for a recent review).  We adopt an
$\aco=3.6$\,\Msun(K\,km\,s$^{-1}$\,pc$^{2}$)$^{-1}$ \citep{Daddi2010} where
needed (following the discussion in \citealt{Boogaard2019} and consistent with
the other ASPECS studies, as well as COLDz; \citealt{Riechers2019}).

\section{Results}
\label{sec:results}

\subsection{Observed emission lines from CO and \CI}
\label{sec:observ-emiss-lines}
We detect emission lines from CO and/or \CI\ in 22 distinct galaxies in the
ASPECS field, between redshifts $z=0.46 - 3.60$.  For the CO
$J \rightarrow J-1$ lines we measure 34 detections plus 21 upper limits, with
rotational quantum numbers between $J=1$ and 10.  We only probe the frequency
range for the CO(9--8) and CO(10-9) transitions in a single source at $z=3.60$
but neither is detected.  Therefore, we focus on the transitions up to
CO(8--7).  For atomic carbon we report six line detections plus 12 upper limits
in the $^{3}P_{1} \rightarrow {^{3}P_{0}}$ and
$^{3}P_{2} \rightarrow {^{3}P_{1}}$ transitions, hereafter \CI(1--0) and
\CI(2--1).

We measure the integrated line fluxes as described in
\autoref{sec:spectr-line-fitt} and show the individual line fits for all
sources in \autoref{fig:slfit} in \autoref{sec:spectral-line-fits}.  The
resulting redshifts, line widths (full width at half maximum; FWHM), central
frequencies and line fluxes for all sources can be found in
\autoref{tab:line-fluxes}.  In the remainder of this paper, we will treat
tentative lines with an integrated line flux smaller than $3\sigma$ in the VLA
and ALMA data as upper limits.  Here $\sigma$ is the uncertainty on the
Gaussian fit, measured over the same line width as the detected lines which
they are tied to.  As not all lines are perfectly described by single
Gaussians, we also compute the line fluxes by integrating the channels within
$1.4\times$ the FWHM and confirm these are consistent with the Gaussian fits to
within error.

Our method forces all lines for a source to a common line width, which may
result in different error bars for some lines than found based on a
signal-to-noise optimized extraction of each individual line
\citep{Gonzalez-Lopez2019, Riechers2020b}.  This more conservative treatment,
which is chosen to minimize biases for the specific analysis carried out in
this work, differs from the way they are used in other works in ASPECS focused
on studies of the global gas density evolution \citep{Decarli2019, Decarli2020,
  Riechers2020b}.  Comparing to the previous ASPECS papers, we find that our
fluxes in Band 3 are on average 20\% lower than those from
\cite{Gonzalez-Lopez2019}, but consistent with \citet[][for a small sub-set of
the sources]{Decarli2016b}.

The CO(1--0) observations cover all ASPECS sources between $z=1.99 - 2.70$,
except 1mm.C10, which lies outside of the VLA pointings
(\autoref{sec:vla-data-from}).  The CO(1--0) fluxes measured here are
consistent with \cite{Riechers2020b}, who measured the flux from the moment 0
maps collapsed over the channels in which emission was seen, while we obtain
larger uncertainties compared to the optimized extractions.  As all the lines
are relatively faint (due to the apparently high $r_{31}$, see
\autoref{sec:co-excitation}), this pushes the significance of some lines from
$>3\sigma$ into the $2.5-3\sigma$ range (and are therefore not shown in
\autoref{fig:line-widths}).  For 3mm.7, the CO(1--0) line-shape is consistent
with the CO(3--2), although the line is formally at $2.97\sigma$ in our fit.
In other cases, the line width of the feature at the frequency of CO(1--0)
appears different from the higher-$J$ lines (e.g., 3mm.3, 3mm.12), which could
be driven by the low signal-to-noise \citep[see][]{Riechers2020b}.  An
interesting case is 1mm.C14a, where the apparent CO(1--0) line appears offset
both spatially and in velocity by $\sim 200$\,km\,s$^{-1}$, compared to the
combined CO(6--5), CO(7--6) and [CI](2--1) lines.  For this source we will use
the fit results tied to the (formally undetected) CO(1--0) line for
consistency, but note that if we only fit the other lines we find a slightly
lower redshift solution ($z=1.9963$) and higher S/N, such that the \CI(2--1)
line is also at $>3\sigma$.

\section{CO excitation}
\label{sec:co-excitation}

\subsection{Individual sources}
\label{sec:co-excitation-indiv}
\begin{figure}[t]
\includegraphics[width=0.5\textwidth]{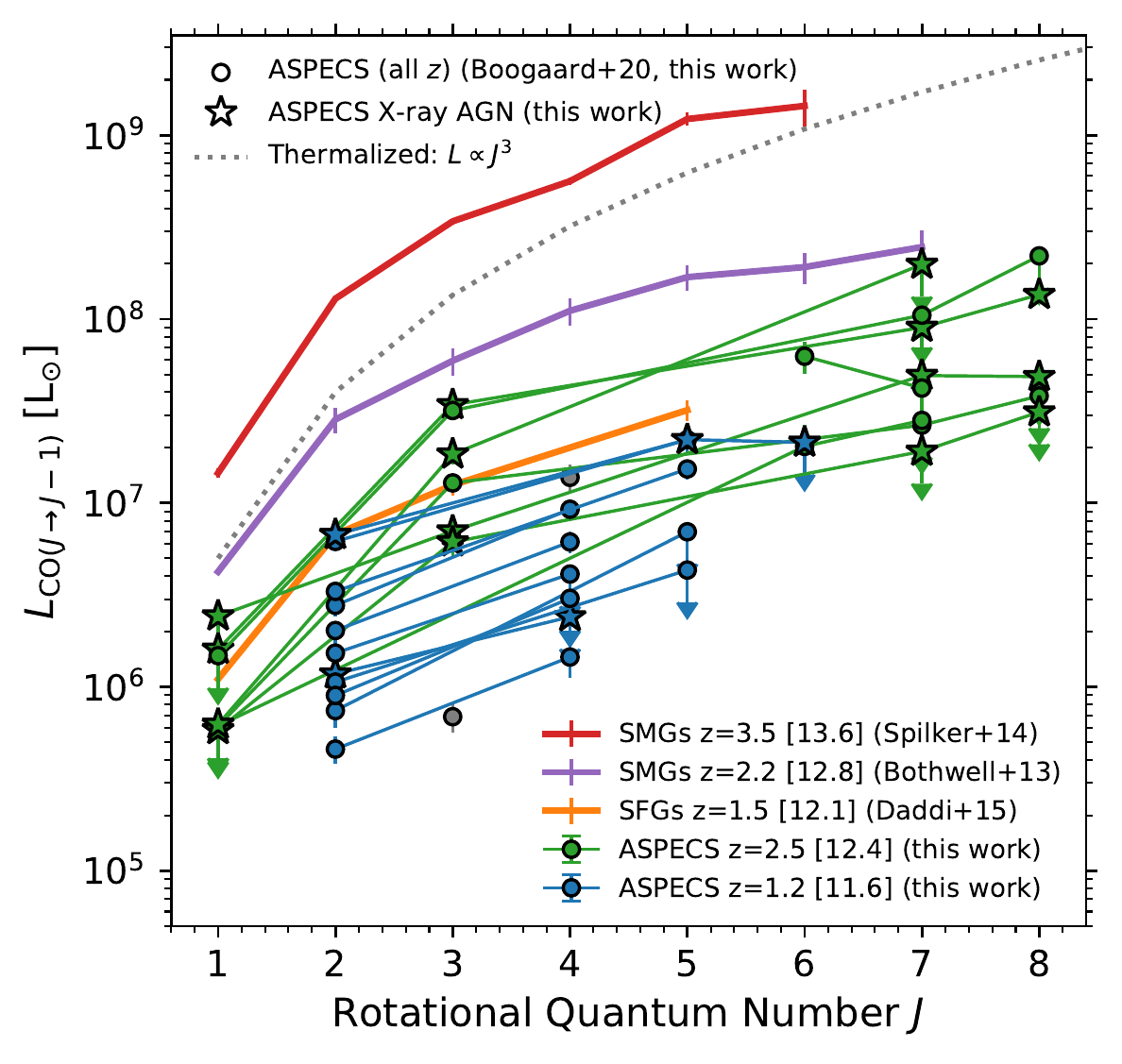}
\caption{CO line luminosities (in units of \Lsun) of the ASPECS galaxies
  (colored circles).  Downward pointing arrows indicate 3$\sigma$ upper limits.
  Stars indicate X-ray sources classified as AGN \citep{Luo2017}.  For
  comparison, we show the average CO ladders of $\avg{z}=1.5$ star-forming
  galaxies \citep{Daddi2015} and sub-millimeter galaxies at $\avg{z}=2.2$
  \citep{Bothwell2013} and $\avg{z}=3.5$ \citep{Spilker2014}, and a thermalized
  ladder (arbitrarily scaled to $5\times10^{6}$~\Lsun).  The average infrared
  luminosity ($\log L_{\rm IR}[\mathrm{L}_{\odot}]$) of the different samples
  is indicated between brackets in the legend.  Overall, the ASPECS galaxies
  probe lower infrared luminosities than typical samples at their respective
  redshifts.
  \label{fig:co_individual_lum}}
\end{figure}

\begin{figure*}[t]
\includegraphics[width=0.5\textwidth]{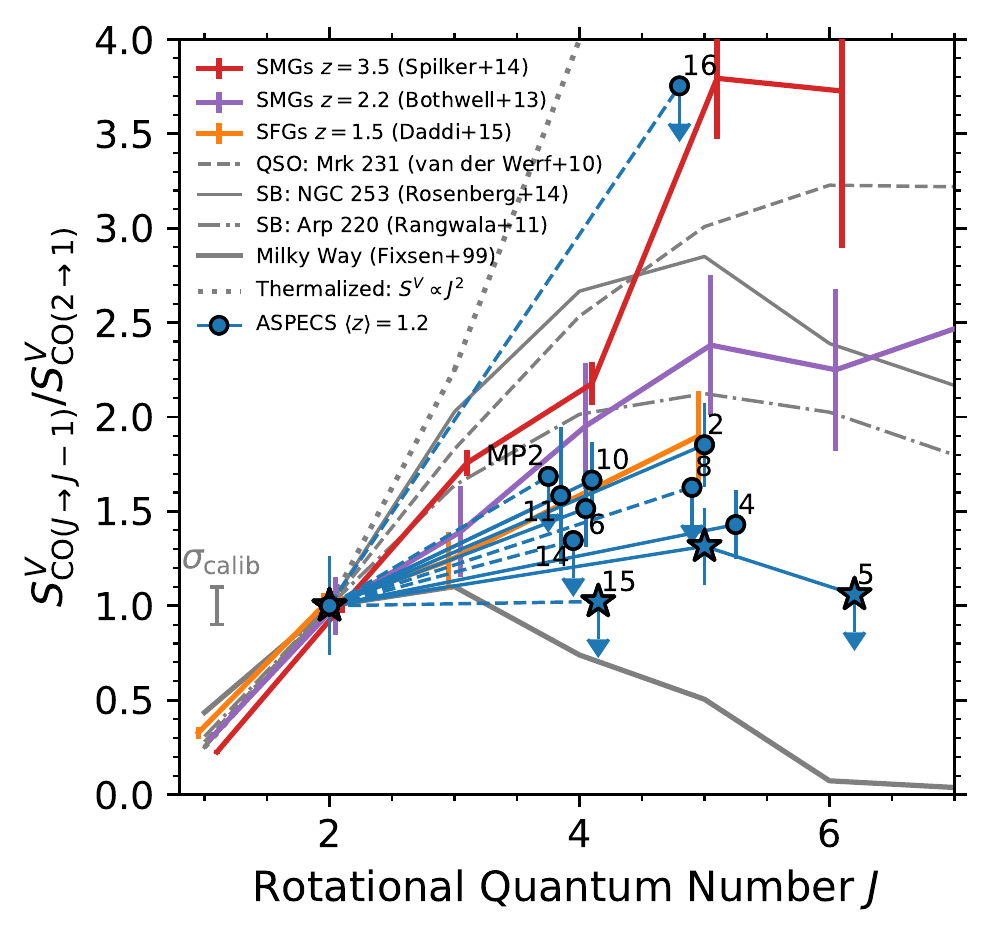}
\includegraphics[width=0.5\textwidth]{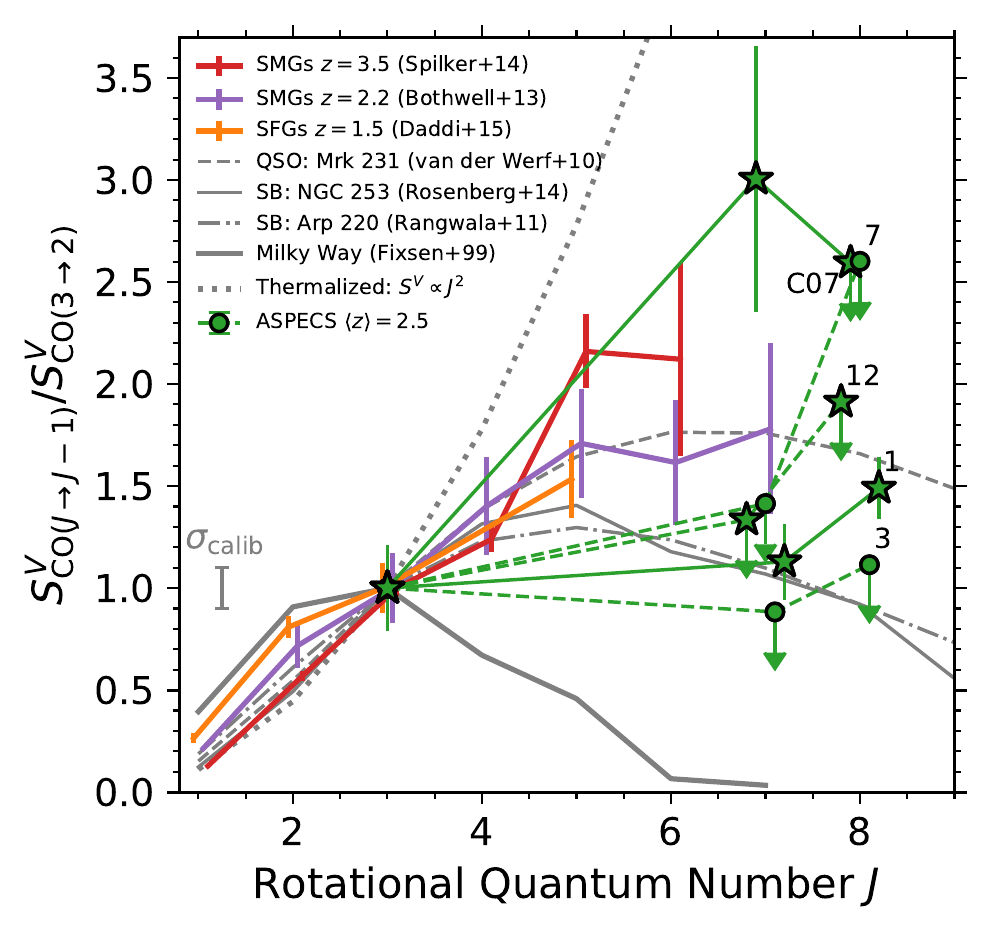}
\caption{CO ladders for the ASPECS galaxies at $\avg{z}=1.2$ (\textbf{left})
  and $\avg{z} = 2.5$ (\textbf{right}), normalized to CO(2--1) and CO(3--2)
  respectively, in units of integrated line flux
  ($[S^{V}]=\mathrm{Jy\,km\,s}^{-1}$).  We include all sources with coverage of
  at least two lines and a detection in the low-$J$ line (except 3mm.09, which
  has a weakly constraining upper limit putting CO(7--6) just below the
  thermalized value).  Downward pointing arrows indicate 3$\sigma$ upper limits
  on the mid/high-$J$ transition(s) and are connected to the lower-$J$
  transition with a dotted line.  Stars indicate X-ray sources classified as
  AGN \citep{Luo2017}.  The grey errorbar indicates the calibration
  uncertainty.  The galaxies at $\avg{z}=1.2$, show excitation in their mid-$J$
  lines, CO(4--3) and CO(5--4), that is consistent with, or lower than, what is
  found in the BzK-selected star-forming galaxies \citep{Daddi2015}.  The range
  in excitation suggests that an additional, warmer, component is present some,
  but not all, sources.  At $\avg{z} = 2.5$, the excitation in the high-$J$
  lines, CO(7--6) and CO(8--7), is comparable to what is found in local
  starbursts \citep[e.g.,][]{Rangwala2011, Rosenberg2014}, but appears lower
  than the average sub-mm galaxy
  \citep{Bothwell2013}. \label{fig:co_individual_flux}}
\end{figure*}
\begin{figure}[t]
  \includegraphics[width=0.5\textwidth]{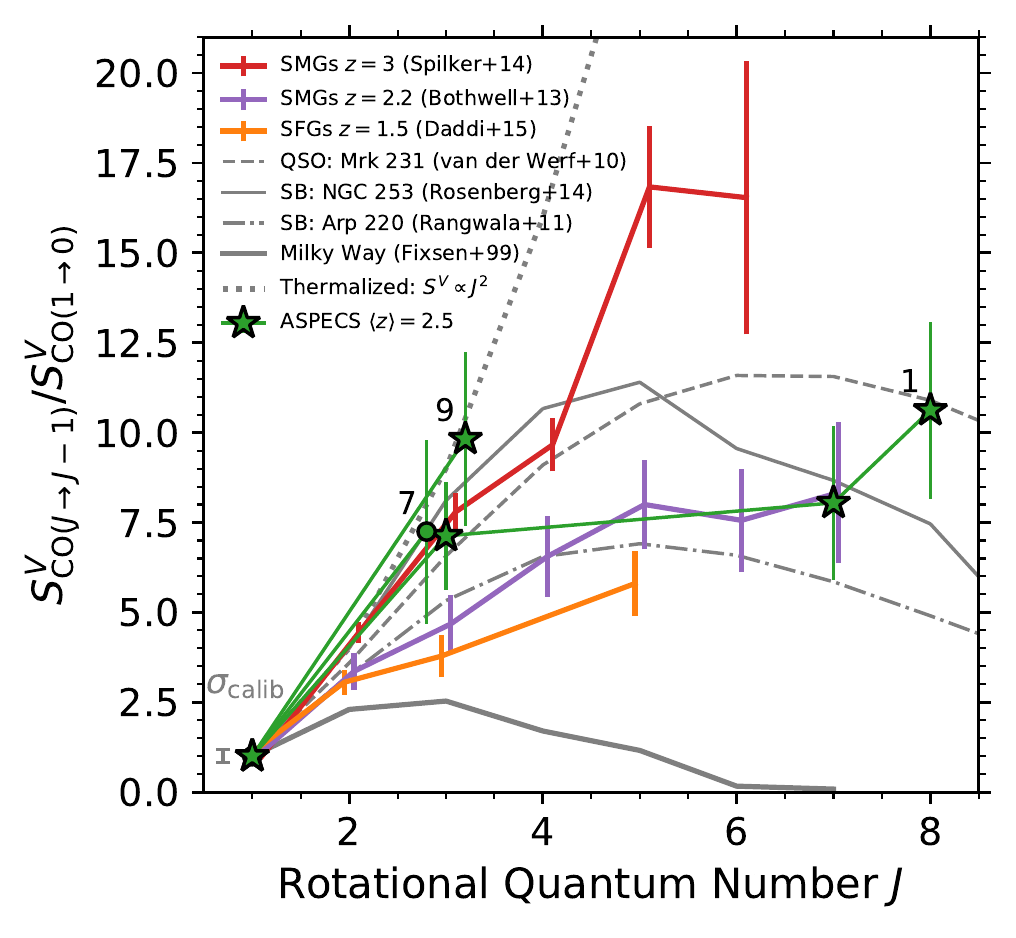}
  \caption{CO ladders of the ASPECS galaxies detected in CO(1--0) (green
    markers and lines), in units of integrated line flux
    ($[S^{V}]=\mathrm{Jy\,km\,s}^{-1}$), normalized to CO(1--0).  Stars
    indicate X-ray sources classified as AGN \citep{Luo2017}.  The grey
    errorbar indicates the combined calibration uncertainty on the ALMA and VLA
    data.  The literature sample shown here is the same as in
    \autoref{fig:co_individual_flux}.  For all ladders, we propagate the
    uncertainty on the transition to which the ladders are normalized to the
    higher-$J$ lines.  We add slight offsets in the horizontal direction for
    clarity. \label{fig:co10_vs_stack}}
\end{figure}

The CO line luminosities of all sources are shown in
\autoref{fig:co_individual_lum} (in units of \Lsun) including both detections
and $3\sigma$ upper limits.  The ASPECS observations naturally divide the
sample into different redshift bins, through the different low-, mid- and
high-$J$ CO lines that are covered in Band 3 and Band 6 at different redshifts
(\autoref{fig:hist}).  For the galaxies from $z=1.0 - 1.6$ ($\avg{z}=1.2$), we
measure the CO(2--1) line in Band 3 and either CO(4--3), CO(5--4), CO(6--5)
and/or \CI(1-0) in Band 6, depending on the exact redshift.  We cover both the
CO(6--5) and CO(7--6) lines in the two sources at $z\approx 1.997$, but just
miss the low-$J$ CO(3--2) line in Band 3.  For the higher redshift galaxies at
$z=2.4-2.7$, we cover CO(3--2) as well as CO(7--6), CO(8--7) and \CI(2-1).  The
VLA observations add constraints on CO(1--0) for all but one source at $z\ge2$.
Outside of these redshift bins we only have 1mm.C30 at $z=0.46$ observed
CO(3--2) in Band 6, for which we do not cover any other CO transition with
ASPECS, and 3mm.13 at $z=3.60$ for which we cover, but do not detect, CO(9--8)
or CO(10--9).

We compare our observations to the average CO ladders from different samples in
the literature: The BzK-selected SFGs at $\avg{z}=1.5$ from \cite{Daddi2015},
the SMGs at $\avg{z}=2.2$ from \cite{Bothwell2013} and the stacked CO ladder
for SPT-selected (lensed) SMGs at $\avg{z} = 3.5$ from \cite{Spilker2014}.  The
ASPECS galaxies at $\avg{z}=1.2$ are less massive and have a lower average
infrared luminosity, $\avg{\LIR} = 10^{11.6}$\,\Lsun, than the BzK galaxies at
$\avg{z}=1.5$ ($ 10^{12.1}$\,\Lsun).  This is also clearly reflected in their
overall lower CO luminosity.  The ASPECS galaxies at $\avg{z} = 2.5$ also have
a lower $\avg{\LIR} = 10^{{12.4}}$\,\Lsun\ compared to the SMGs at similar
redshifts.

We show the CO excitation ladders for the ASPECS galaxies at $\avg{z}=1.2$
(left) and $\avg{z}=2.5$ (right), relative to the low-$J$ CO(2--1) and CO(3--2)
transitions by which they were selected, respectively, in
\autoref{fig:co_individual_flux} (now as line flux ratios).  In addition to the
$z>1$ samples mentioned earlier, we also add the observed CO ladders for
several local sources: the Milky Way \citep[MW;][Inner Disk]{Fixsen1999} and
starburst NGC\,253 \citep{Rosenberg2014}, as well as the CO ladders for
Arp\,220 \citep{Rangwala2011} and the nearest known quasar, Mrk\,231
\citep{vanderWerf2010}, as modeled by the LVG models of \cite{Weiss2007}.  The
dotted line indicates a thermalized CO ladder (i.e., $S^{V} \propto J^{2}$).

The eleven CO(2--1)-selected galaxies at $\avg{z}=1.2$ (left panel) span a
range in excitation in their CO(4--3) and CO(5--4) lines.  Only one source (and
one weak upper limit) show excitation in the CO(5--4) line that is comparable
to the average of the BzK-selected SFGs \citep{Daddi2015}, while the other
measurements and limits are consistent with lower excitation.  We also add
direct measurements of the CO(4--3) transition to this picture (which was not
directly measured for the BzK galaxies).  This ratio is similar to the
(interpolated) value in the BzK galaxies for the three detected sources.  At
the same time we also infer upper limits consistent with lower excitation,
although none of the sources have limits strong enough to put them confidently
in the low-excitation regime of the Milky Way.  In all cases, the excitation is
significantly lower compared to SMGs at higher redshift and clearly not as high
as seen in the centers of the prototypical local starbursts Arp\,220 and
NGC\,253, nor Mrk\,231.

For the five CO(3--2) selected galaxies at $\avg{z}=2.5$ (right panel), we
probe the CO(7--6) and CO(8--7) lines.  Here, we find the brightest galaxy of
the survey (1mm.C01), which is an X-ray identified AGN with detections in all
three lines (cf.\ \autoref{fig:aspecs1}).  This source exhibits significant
excitation, out to $J=8$, at the level comparable to the local starbursts and
Mrk\,231, though still somewhat below the $\avg{z}=2.2$ SMGs at CO(7--6).
There is one other source detected in CO(7--6), 1mm.C07, which is also an X-ray
AGN.  This source shows the highest $r_{73}$ ratio of all sources, although we
caution that the line flux is uncertain for both lines (cf.\
\autoref{fig:line-widths}) and the CO(8--7) transition is undetected.  The
remaining sources at these redshifts are not detected in their high-$J$ lines.
At the sensitivity limit of ASPECS, this constrains their high-$J$ excitation
to be well below thermalized and comparable to the level of the local starburst
and somewhat below the \cite{Bothwell2013} SMGs.

We show the CO ladder normalized to CO(1--0) in \autoref{fig:co10_vs_stack},
for the sources with $\mathrm{S/N}>3$ in CO(1--0) in our joint fit.  In
contrast to the $r_{73}$ ratio, the $r_{31}$ ratio is typically higher than
that of the \cite{Bothwell2013} SMGs.

The X-ray radiation from an AGN can drive the emission of the high-$J$ CO lines
\citep[e.g.,][]{Meijerink2007, vanderWerf2010, Vallini2019}.  The stars in
\autoref{fig:co_individual_flux} and \autoref{fig:co10_vs_stack} indicate X-ray
identified AGN (1mm.C05, C12 at $\avg{z}=1.2$ and 1mm.C01, C07, C19 and 3mm.09
at $\avg{z} = 2.5$).  It is interesting to note that both sources detected in
the high-$J$ CO(7--6) line are X-ray AGN.  However, the upper limits on the
remaining galaxies do not distinguish them clearly from the detected sources.
At $\avg{z}=1.2$, the X-ray AGN lie at the low-excitation end of the sample,
which is consistent with the AGN not strongly driving the mid-$J$ lines.  Based
on the low number of sources, we are unable to draw strong conclusions here.
However, the results are consistent with recent work that did not find
statistically different excitation, up to CO(7--6), between galaxies with and
without an active nucleus \citep[e.g.,][]{Sharon2016, Kirkpatrick2019}.

\subsection{Stacked line fluxes}
\label{sec:average-co-ladder}
\begin{deluxetable}{cccc}
  \tablecaption{Average line fluxes from stacking
    \label{tab:stacking}}

  \tablehead{\colhead{Line} & \colhead{$N$} &
    \colhead{$S^{V}/S^{V}_{J=J_{\rm ref}}$} & $r_{J J_{l}}$}

  \colnumbers

  \startdata
  \multicolumn{4}{c}{$\avg{z}=1.2$; $J_{\rm ref} = J_{l} = 2$}\\
  \hline
  CO(2--1)       & 11      & $1.00 \pm 0.04$ & \nodata \\
  CO(4--3)       & 6       & $1.33 \pm 0.18$ & $0.33 \pm 0.04$ \\
  \CI(1--0)      & 8       & $0.33 \pm 0.18$ & \nodata         \\
  CO(5--4)       & 5       & $1.41 \pm 0.15$ & $0.23 \pm 0.02$ \\
  \hline
  \multicolumn{4}{c}{$\avg{z}=2.5$; $J_{\rm ref} = 3$; $J_{l}=1$}\\
  \hline
  CO(1--0)  & 6 & $0.14 \pm 0.03$ & \nodata         \\
  CO(3--2)  & 6 & $1.00 \pm 0.03$ & $0.77 \pm 0.14$ \\
  CO(7--6)  & 5 & $1.32 \pm 0.18$ & $0.19 \pm 0.04$ \\
  \CI(2--1) & 5 & $0.93 \pm 0.18$ & \nodata         \\
  CO(8--7)  & 5 & $1.10 \pm 0.20$ & $0.12 \pm 0.03$ \\
  \enddata
  \tablecomments{The lines fluxes are obtained through $1/\sigma$-weighted
    stacking, scaled to the reference transition in the stack, with propagated
    errors.  (1) Stacked transition (2) Number of objects in the stack of each
    transition. (3) Mean integrated line flux, normalized to the reference
    CO($J\rightarrow J-1$) transition in the stack, with $J=J_{\rm ref}$. (4)
    CO brightness temperature ratio with the lowest transition in the stack,
    $r_{J J_{l}} = L'_{\mathrm{CO}(J \rightarrow J-1)} /
    L'_{\mathrm{CO}(J_{l}-J_{l} - 1)}$.}
\end{deluxetable}

\begin{figure*}[t]
  \includegraphics[width=0.485\textwidth]{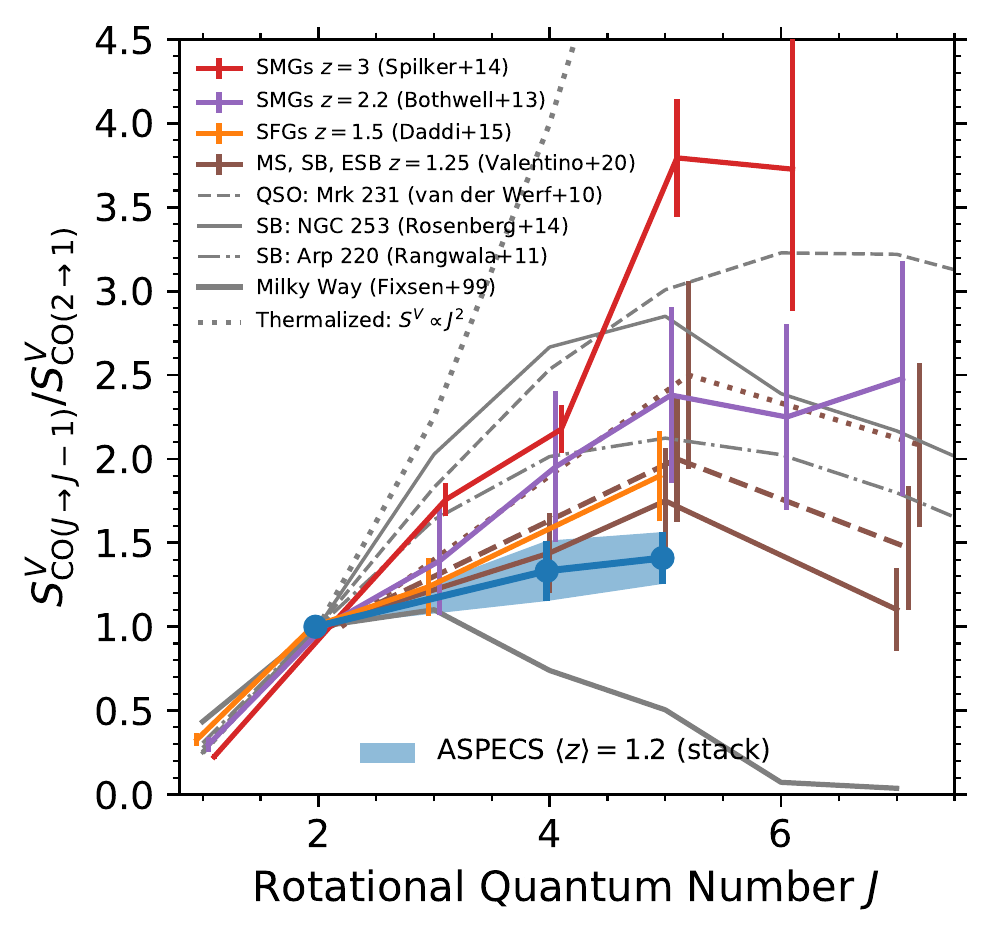}
  \includegraphics[width=0.5\textwidth]{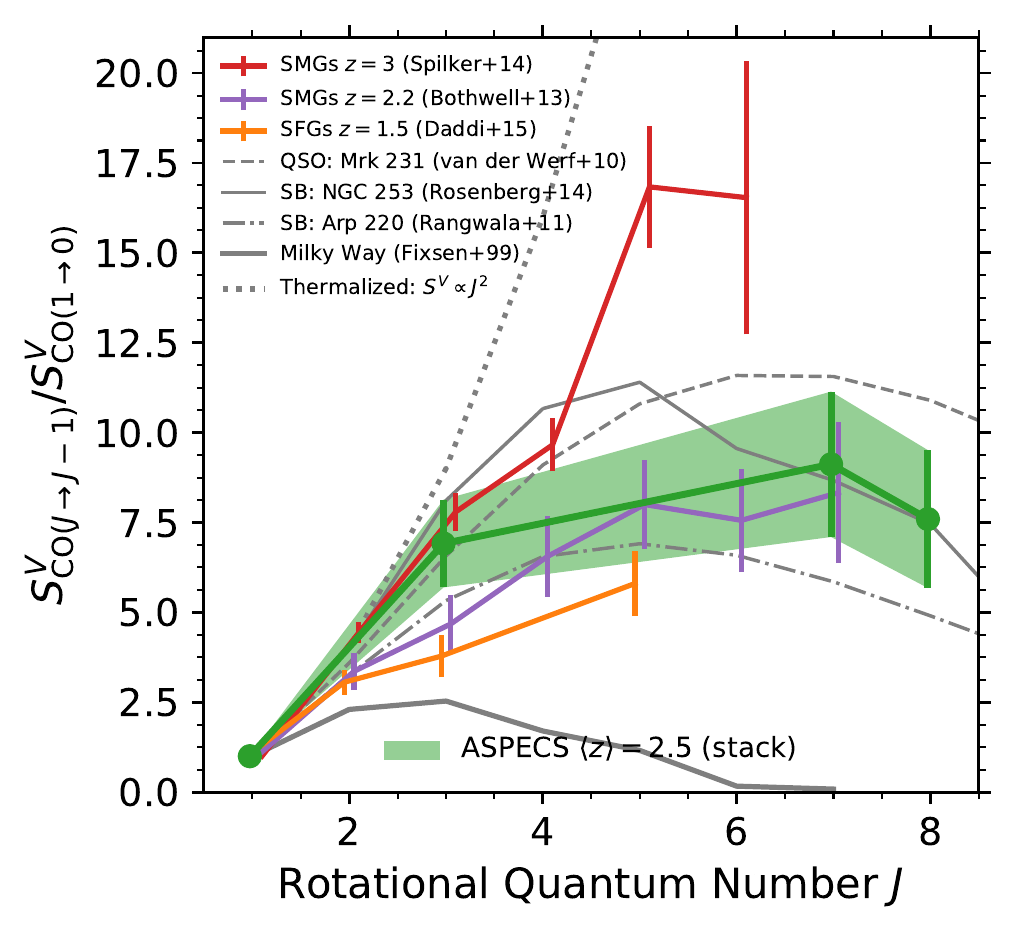}
  \caption{Average CO ladders for the ASPECS galaxies at $\avg{z}=1.2$
    (\textbf{left}) and $\avg{z}=2.5$ (\textbf{right}), in units of integrated
    line flux ($[S^{V}]=\mathrm{Jy\,km\,s}^{-1}$), obtained through
    $(1/\sigma)$-weighted mean stacking after scaling to a common CO(2--1) and
    CO(3--2) flux, respectively.  The solid line (and shaded region) show the
    mean stack of all sources.  The literature sample shown here is the same as
    in \autoref{fig:co_individual_flux}, with the addition of the recently
    observed main sequence and (extreme) starburst galaxies from
    \cite{Valentino2020b}, shown by the brown, solid and (dotted) dashed lines.
    For all ladders, we propagate the uncertainty on the transition to which
    the ladders are normalized to the higher-$J$ lines.  We add slight offsets
    in the horizontal direction for clarity. \label{fig:co_stack}}
\end{figure*}
We construct an average CO ladder in each of the two redshift bins by stacking
the CO lines in each transition.  The advantage of stacking (compared to taking
the average of the measured line fluxes) is that we can straightforwardly take
all sources into account in a non-parametric way, regardless of whether they
are detected in a specific transition or not.  Before stacking, we first take
out the intrinsic brightness variations in the sample by dividing their spectra
by the integrated flux in the CO(2--1) or CO(3--2) transition (by which they
were selected, depending on the redshift), as measured from the Gaussian fits.
In this way we determine the average excitation of the other lines in the
sample relative to CO(2--1) or CO(3--2) (including CO(1--0) and \CI).

Because we are stacking sources with different line widths, care must be taken
not to lose flux, while keeping an optimal signal-to-noise in the stack.
Therefore, we stack each transition individually in velocity space, such that
all the flux ends up in a single channel in the final stack
\citep[cf.][]{Spilker2014}.  We first create a grid of velocities centered
around zero.  We take a channel width of 700\,km\,s$^{-1}$ for the sources at
$\avg{z}=1.2$ and 800\,km\,s$^{-1}$ for the sources at $\avg{z}=2.5$, motivated
by the width of the broadest lines in our sample ($\mathrm{FWHM}\approx 590$
and 660\,km\,s$^{-1}$ in each redshift bin, respectively).  The average line
width of the sample is $\avg{\mathrm{FWHM}} = 330$\,km\,s$^{-1}$.  At this
channel width the CO(7--6) and [CI](2--1) lines, with a peak separation of
$1000$\,km\,s$^{-1}$, are not blended in the stack.  We find the results are
robust to modifying the channel width by $\pm 100$\,km\,s$^{-1}$.  After
subtracting the continuum from the Band 6 spectra (as in
\autoref{sec:spectr-line-fitt}), we convert each spectrum to velocity-space,
centered around the line.  We then bin the spectra onto the velocity grid and
stack them by taking the $1/\sigma$-weighted mean flux in each velocity bin
(where $\sigma$ is the root-mean-square error on the spectrum).  Likewise, we
determine the error spectrum of the stack by propagating the errors from
individual spectra.  We then measure the flux and error in the zero-velocity
bin, which is centered on the line.  We use a $1/\sigma$-weighting to avoid
strongly weighting towards the detected lines, while at the same time not
sacrificing too much signal-to-noise by not down-weighting very noisy spectra
(as in an unweighted stack).  Note this is different from \cite{Spilker2014},
who use a $1/\sigma^{2}$-weighted stack to obtain the highest possible
signal-to-noise ratio.

The resulting line fluxes, normalized to the reference transition in the stack
($J_{\rm ref}$), are provided in \autoref{tab:stacking}, where we also report
the line brightness temperature ratios (\autoref{eq:rJ1}) to the lowest-$J$
transition ($J_{l}$; note for the individual galaxies these are reported in
\autoref{tab:line-fluxes}).  We show the average ladders, normalised to
$J_{l}$, in \autoref{fig:co_stack}.

The stacks in the two redshift bins reinforce our results from
\autoref{sec:co-excitation-indiv}.  For the galaxies at $\avg{z}=1.2$,
excitation in the mid-$J$ lines, compared to CO(2--1) is
$r_{42} = \rFourTwostack$ and $r_{52} = \rFiveTwostack$.  This is on average
lower than BzK-selected galaxies, in particular in CO(5--4) transition
($r_{52} = 0.30\pm0.06$; \citealt{Daddi2015})\footnote{\cite{Daddi2015} did not
  measure the excitation in CO(4--3), but interpolating their CO ladder yields
  $r_{42}=0.41 \pm 0.09$ \citep[see][for details]{Decarli2016b}.}.  We now also
add the recently published CO ladders for star-forming galaxies at $z=1.25$
from \cite{Valentino2020b}, who separate their sample in main sequence galaxies
and (extreme) starbursts (the latter being defined as lying a factor
$\mathrm{SFR}/\mathrm{SFR}_{\mathrm{MS}} \ge 3.5 \times$ and $\ge 7\times$
above the main sequence of \citealt{Sargent2014}).  Their main sequence
galaxies show excitation intermediate between the BzK galaxies and ASPECS, with
$r_{42} = 0.36 \pm 0.06$ and $r_{52} = 0.28 \pm 0.05$.

At $\avg{z}=2.5$, we measure an average $r_{31} = \rThreeOnestack$ from the
stack of all sources.  For comparison, when considering the non-detections as
lower limits, the median of the individual measurements is $0.79 \pm 0.17$ (for
1mm.C01---fully consistent with $0.84 \pm 0.18$ as measured by
\citealt{Riechers2020b}).  The stacked $r_{31}$ value is higher than that found
for SMGs by \cite[][$r_{31} = 0.52 \pm 0.09$]{Bothwell2013}.  At the same time,
the high-$J$ excitation, compared to $J=3$ ($r_{73}$), is lower in our sample
compared to \cite{Bothwell2013}, as also seen in
\autoref{fig:co_individual_flux}.  The mean $r_{71} = \rSevenOnestack$ is
similar to that of 1mm.C01 alone, and comparable to the SMGs
($r_{71}=0.18 \pm 0.04$; \citealt{Bothwell2013}) and the local starburst
NGC\,253, while it lies below the local quasar Mrk\,231
(cf. \autoref{sec:co-excitation-indiv}).  Overall, the average ladder appears
similar to that found in local starburst galaxies, such as NGC\,253.

In addition to stacking all sources selected in a certain transition, we also
explored splitting the sample based on the presence of an AGN, or whether a
line was individually (un)detected.  We find marginal evidence of an overall
lower excitation in the galaxies without an X-ray detected AGN at $\avg{z}=2.5$
(in particular for the high-$J$ lines), but the limited numbers in the stack
prohibit firm conclusions.

\subsection{LVG modeling}
\label{sec:lvg-modeling}

\begin{deluxetable}{@{\extracolsep{5pt}}ccccc}
  \tablecaption{LVG modeling results
    \label{tab:lvg_model}}

  \tablehead{
     & \multicolumn{2}{c}{1-component} & \multicolumn{2}{c}{2-component$^{\dagger}$}\\
    \cline{2-3} \cline{4-5}
    \colhead{$J$} &
    \colhead{$S^{V}/S^{V}_{J=1}$} & \colhead{$r_{J 1}$} &
    \colhead{$S^{V}/S^{V}_{J=1}$} & \colhead{$r_{J 1}$}}

  \colnumbers

  \startdata
  \multicolumn{5}{c}{$z=1.0 - 1.6$ (12 galaxies)}\\
  \hline
  1 & $1.00 \pm 0.00$ & $1.00 \pm 0.00$ & $ 1.00 \pm 0.00$ & $1.00 \pm 0.00$ \\
  2 & $3.33 \pm 0.48$ & $0.83 \pm 0.12$ & $ 3.01 \pm 0.43$ & $0.75 \pm 0.11$ \\
  3 & $5.20 \pm 0.91$ & $0.58 \pm 0.10$ & $ 4.12 \pm 0.80$ & $0.46 \pm 0.09$ \\
  4 & $4.76 \pm 1.26$ & $0.30 \pm 0.08$ & $ 4.01 \pm 1.14$ & $0.25 \pm 0.07$ \\
  5 & $2.70 \pm 1.33$ & $0.11 \pm 0.05$ & $ 2.99 \pm 1.41$ & $0.12 \pm 0.06$ \\
  6 & $0.53 \pm 1.27$ & $0.01 \pm 0.04$ & $ 1.37 \pm 1.69$ & $0.04 \pm 0.05$ \\
  \hline
  \multicolumn{5}{c}{$z=2.0 - 2.7$ (8 galaxies)}\\
  \hline
  1 & $ 1.00 \pm 0.00$ & $1.00 \pm 0.00$ & $ 1.00 \pm 0.00$ & $1.00 \pm 0.00$ \\
  2 & $ 4.09 \pm 0.72$ & $1.02 \pm 0.18$ & $ 3.88 \pm 0.62$ & $0.97 \pm 0.15$ \\
  3 & $ 8.24 \pm 1.50$ & $0.92 \pm 0.17$ & $ 7.17 \pm 1.24$ & $0.80 \pm 0.14$ \\
  4 & $12.21 \pm 2.49$ & $0.76 \pm 0.16$ & $ 9.80 \pm 2.01$ & $0.61 \pm 0.13$ \\
  5 & $14.68 \pm 3.62$ & $0.59 \pm 0.14$ & $10.95 \pm 2.84$ & $0.44 \pm 0.11$ \\
  6 & $13.86 \pm 4.48$ & $0.39 \pm 0.12$ & $10.17 \pm 3.39$ & $0.28 \pm 0.09$ \\
  7 & $ 9.33 \pm 4.57$ & $0.19 \pm 0.09$ & $ 8.28 \pm 3.67$ & $0.17 \pm 0.07$ \\
  8 & $ 4.26 \pm 4.13$ & $0.07 \pm 0.06$ & $ 5.55 \pm 3.87$ & $0.09 \pm 0.06$ \\
  \enddata

  \tablenotemark{$\dagger$}{We adopt the two-component models throughout this
    paper.}

  \tablecomments{The average line ratios are computed by taking the
    $1/\sigma$-weighted mean of the LVG models of the individual sources in
    each redshift bin.  (1) CO($J \rightarrow J-1$) rotational quantum number
    $J$. (2) Single-component LVG model line flux, normalized to $J=1$. (3)
    Single-component LVG model CO brightness temperature ratio,
    $r_{J 1} = L'_{\mathrm{CO}(J \rightarrow J-1)} /
    L'_{\mathrm{CO}(1-0)}$. (4) Two-component LVG model line flux, normalized
    to $J=1$. (5) Two-component LVG model CO brightness temperature ratio.}
\end{deluxetable}

\begin{figure*}[t]
  \includegraphics[width=\columnwidth]{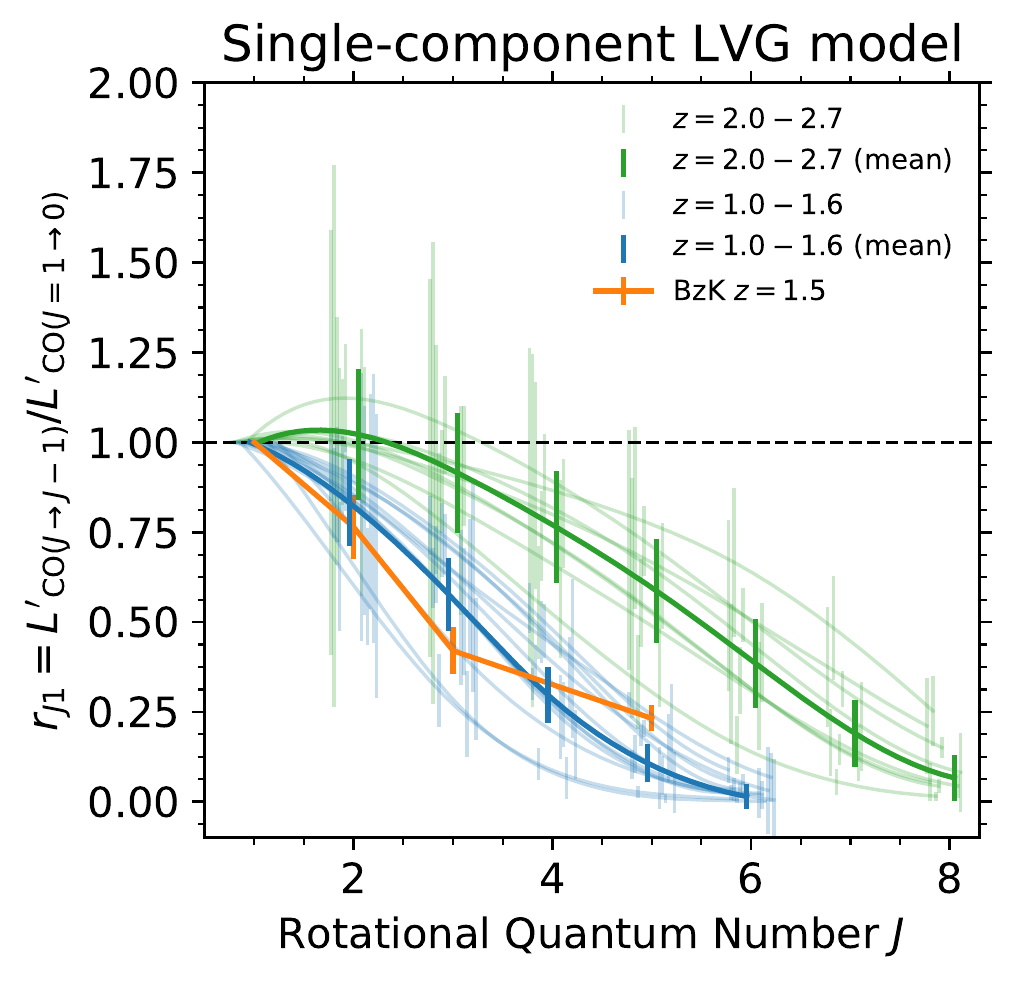}
  \includegraphics[width=\columnwidth]{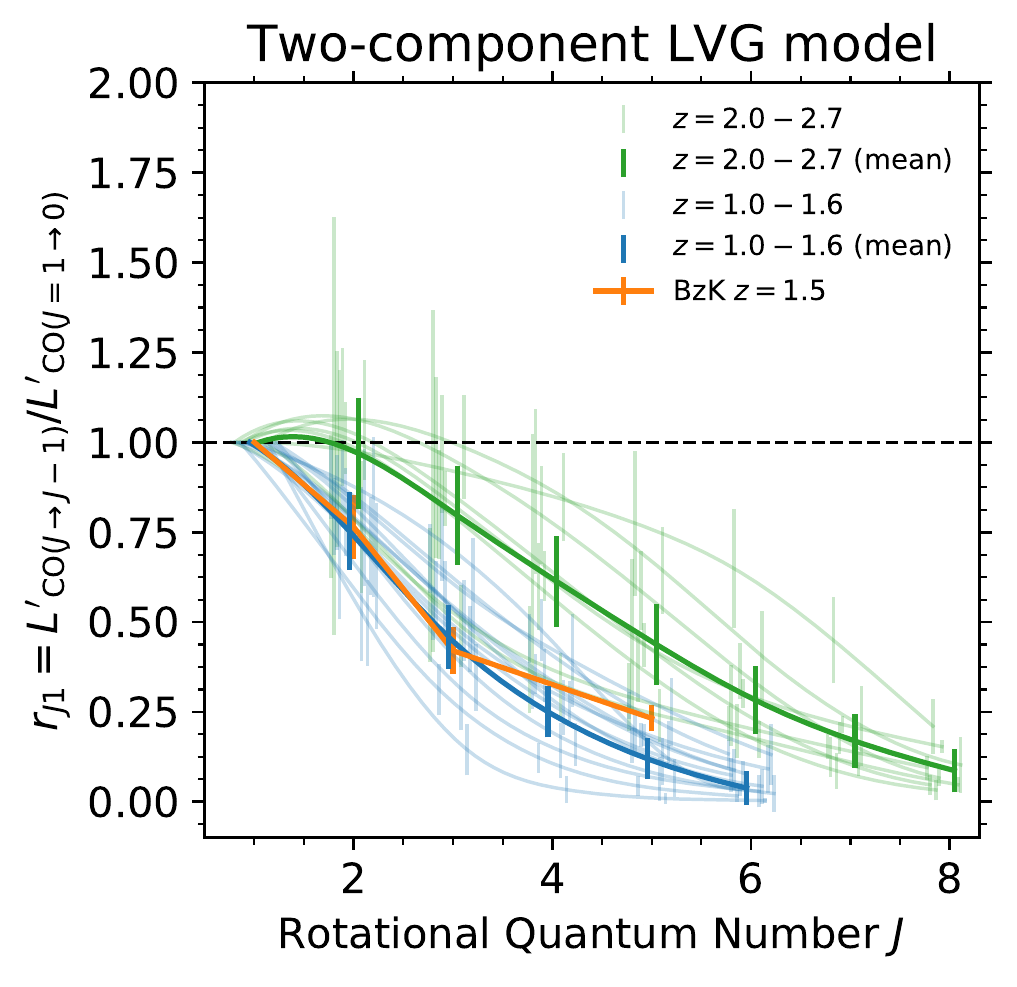}
  \caption{Predicted CO line luminosities ($L'_{{\rm CO}(J \rightarrow J-1)}$)
    for the ASPECS galaxies at $z=1.0 - 1.6$ and $z=2.0 - 2.7$, normalised to
    $L'_{{\rm CO}(1 \rightarrow 0)}$ (so the values on the ordinate are
    equivalent to $r_{J1}$). The CO line luminosities for the individual
    sources are predicted from the best-fit LVG model, assuming a single
    density and temperature component (left panel) as well as a two-component
    model (right panel).  The light-colored lines show the individual fits,
    while the strong-colored line shows the $1/\sigma$-weighted mean of the
    individual ladders.  While the temperature and density are degenerate in
    the fit, the emerging line luminosities are reasonably well
    constrained.  We show the BzK-selected galaxies from \cite{Daddi2015} for
    comparison and add horizontal offsets for clarity.  In both the single- and
    two-component models the CO(3--2) selected galaxies at $z=2.0-2.7$ show on-average
    higher excitation than the CO(2--1) selected galaxies at $z=1.0-1.6$.
    \label{fig:CO-axel}}
\end{figure*}

To further investigate the CO excitation, we study the CO ladder of all sources
at $z=1.0 - 1.6$ and $z=2.0 - 2.7$ in more detail by using a spherical,
isothermal large velocity gradient model (LVG), following \cite{Weiss2007}.
Because we only observe up to four CO lines, we cannot accurately constrain the
model parameters for individual sources.  Rather, we use the model to predict
the CO line luminosity of the neighboring, unobserved CO lines.  The background
to this approach is, that CO ladders cannot have arbitrary shapes and in this
sense our procedure can be viewed as the molecular line correspondence of
interpolating a sparsely sampled dust continuum SED.

In practice, we fit the observed CO line luminosities using a one- and a
two-component LVG model employing a Monte Carlo Bee algorithm \citep{Pham2009}
which samples randomly the parameter space and gives finer sampling for good
solutions (evaluated from a $\chi^{2}$ analysis for each model).  The model
predicted CO line luminosities, $L'_{\mathrm{CO}(J\rightarrow J-1)}$, and their
uncertainties are calculated using the probability-weighted mean of all
solutions and their standard deviations.  For the redshift $z=1.0 - 1.6$
sample, where we detect transitions up to $J=5$, we report the model-predicted
CO ladders up to $J=6$.  For the $z=2.0-2.7$ sample we report transitions up to
$J=8$, because the observations also cover higher transitions.  Typically, we
investigate on the order of $10^6$ models per galaxy.  The free parameters are
the \HH\ volume density, the kinetic temperature, the CO abundance per velocity
gradient and the source solid angle (expressed as the equivalent radius of the
emitting region, see \citealt{Weiss2007}).  We include an additional prior that
discards solutions where the peak of the CO ladder lies beyond the CO(7--6)
line.  This is motivated by our average ladder and there being only very few
extreme local ULIRGs and $z=2-3$ QSOs/SMGs where this is the case
\citep[cf.][]{Weiss2007proc, Carilli2013}.  Limits to the parameter space are:
$\log_{10}(\nHH) =1.0-7.0$\,cm$^{-3}$, $\Tkin =10-200$\,K,
[CO]/[\HH]\,($\Delta v/\Delta
r)^{-1}=10^{-3}-10^{-7}$\,(km\,s$^{-1}$\,pc$^{-1}$)$^{-1}$ and
$r_{\rm eff} = 1-10\,000$\,pc.

The CO ladders of the individual objects, derived from our single- and
two-component LVG fitting, are shown in \autoref{fig:CO-axel}, normalized to
the predicted CO(1--0) line luminosity.  We split the sample in two redshift
bins, based on the observed lines (similar to
\autoref{sec:co-excitation-indiv}).  We also compute the average ladder in each
redshift bin by computing the $1/\sigma$-weighted mean of the $L'_{\rm CO}$ for
each of the lines, after first rescaling to a common $L'_{\mathrm{CO(1-0)}}$
(to take out intrinsic variations in the luminosity).  The resulting average
ladders are provided in \autoref{tab:lvg_model}.

In general, the galaxies at $z \geq 2$ show more excited CO ladders than the
galaxies at $z<2$.  This could partially be a selection effect in the case of
the single component models, if the fit overpredicts the $J=3$ line luminosity
in an attempt to fit $J>6$, as suggested by the $r_{31} = \rThreeOneLVGzHiOne$
being slightly higher than the stacked value ($r_{31} = \rThreeOnestack$).
However, the two-component model at $z\geq2$ is still higher in $J=2$ and 3,
compared to the single-component fit at $z<2$ (i.e., the `maximal' value at
$z<2$), whereas the average $r_{31} = \rThreeOneLVGzHiTwo$ is fully consistent
with the stacked value.  This strongly suggests that there is a true, intrinsic
difference in excitation in the CO(2--1)-selected sample at $z<2$ compared to
the CO(3--2)-selected sample at $z\ge2$.  As we constrain two low/mid-$J$ lines
at both redshifts ($J=1$ and 3 at $z\ge2$, and $J=2$ and 4 at $z<2$), these
conclusions appears robust against the fact that we also probe higher-$J$ lines
at $z\ge2$.

At $z=1.0 - 1.6$, the single and two component models give formally consistent
results, whereas the mean of the low-$J$ lines is slightly higher for the
single component models ($r_{21} = \rTwoOneLVGzLoOne$).  The mean ladder of the
two-component model is similar to the result from \cite{Daddi2015} for $J=2$
and $J=3$ ($r_{21} = \rTwoOneLVGzLoTwo$), while yielding a lower $J=4$ and 5
(consistent with the stack).  Although some individual sources show ladders
consistent with thermalized $r_{21} = 1.0$ at these redshifts, the average is
subthermal.

In general, we note that the single component fits would over-predict the
low-$J$ excitation if the low-$J$ CO line luminosities have a significant
contribution from strongly subthermally excited gas.  This is particularly
significant at $z=2.0-2.7$, as the $J>6$ and $J\le3$ may not stem from the same
component.  However, this can also be important at $z=1.0-1.6$, if the CO
excitation is similar to the sources in \cite{Daddi2015} where the elevated
$J=5$ line luminosity is best described by a second, higher excitation
component.  This motivates the use of the two-component fit.  In contrast, the
observed CO transitions have little weight to constrain a two-component fit, in
particular at $z=1.0-1.6$, where we mostly only observe two CO transitions.

\begin{figure*}[t]
  \includegraphics[width=\columnwidth]{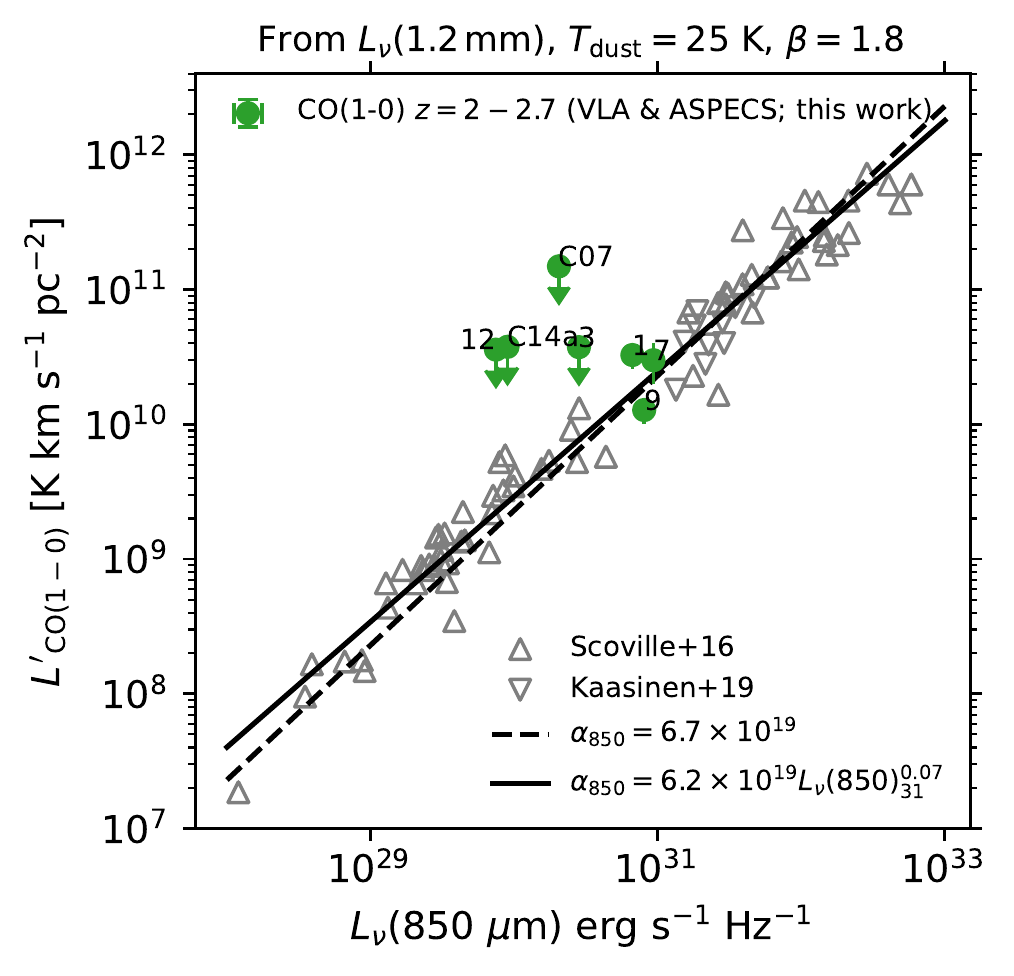}
  \includegraphics[width=\columnwidth]{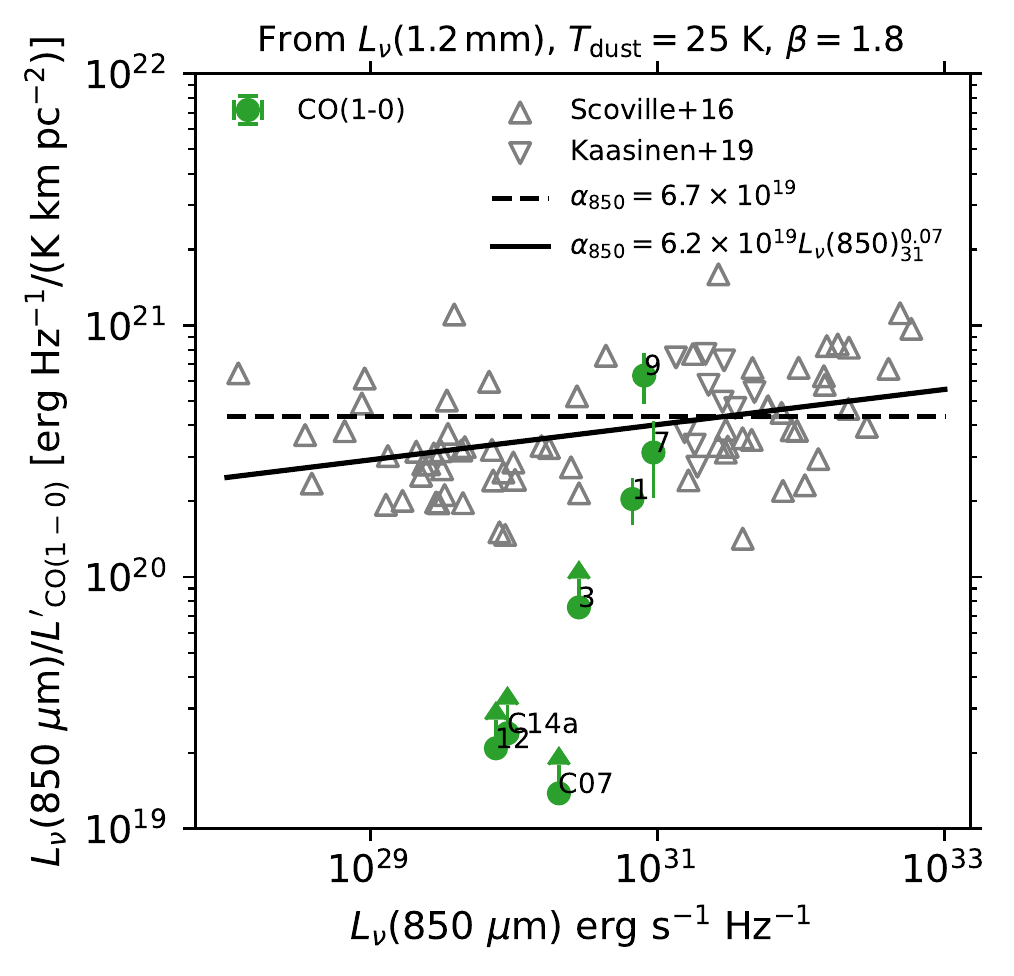}
  \caption{Rest frame luminosity at 850\,\micron\ compared to the CO(1--0)
    luminosity (\textbf{left}) and the ratio of
    $L_{\nu}(850\mu {\rm m})/L'_{{\rm CO}(1-0)}$ (\textbf{right}).  The
    CO(1--0) observations were taken with the VLA \citep{Riechers2020b} and are
    re-analysed in this paper.  Sources are indicated by the 3mm.ID (except
    1mm.C07 and 1mm.C14a).  The black lines show the best fit empirical
    relations from \citet[][assuming both a constant and dust luminosity
    dependent dust-to-gas conversion factor]{Scoville2016}, while the gray
    triangles show their calibration sample as well as more recent observations from
    \cite{Kaasinen2019}.
    \label{fig:L850_LCO10}}
\end{figure*}
\begin{figure*}[t]
  \includegraphics[width=\columnwidth]{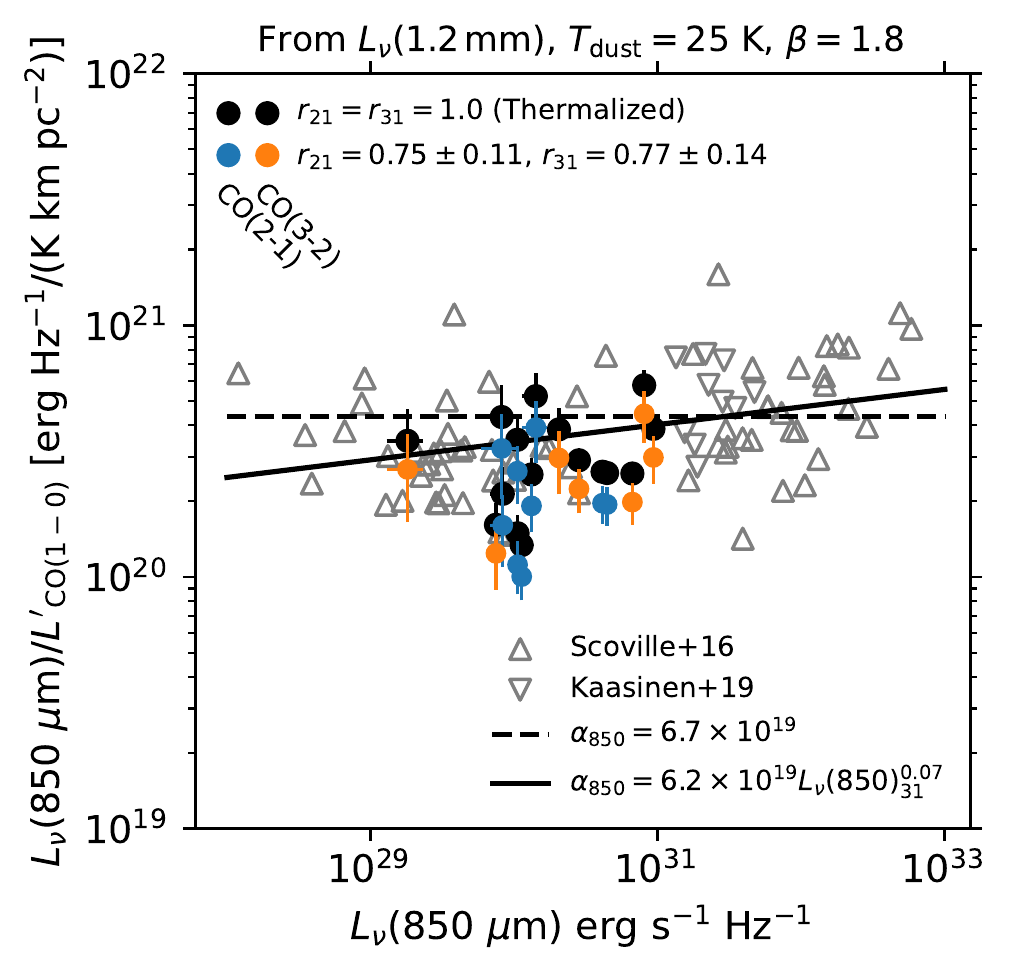}
  \includegraphics[width=\columnwidth]{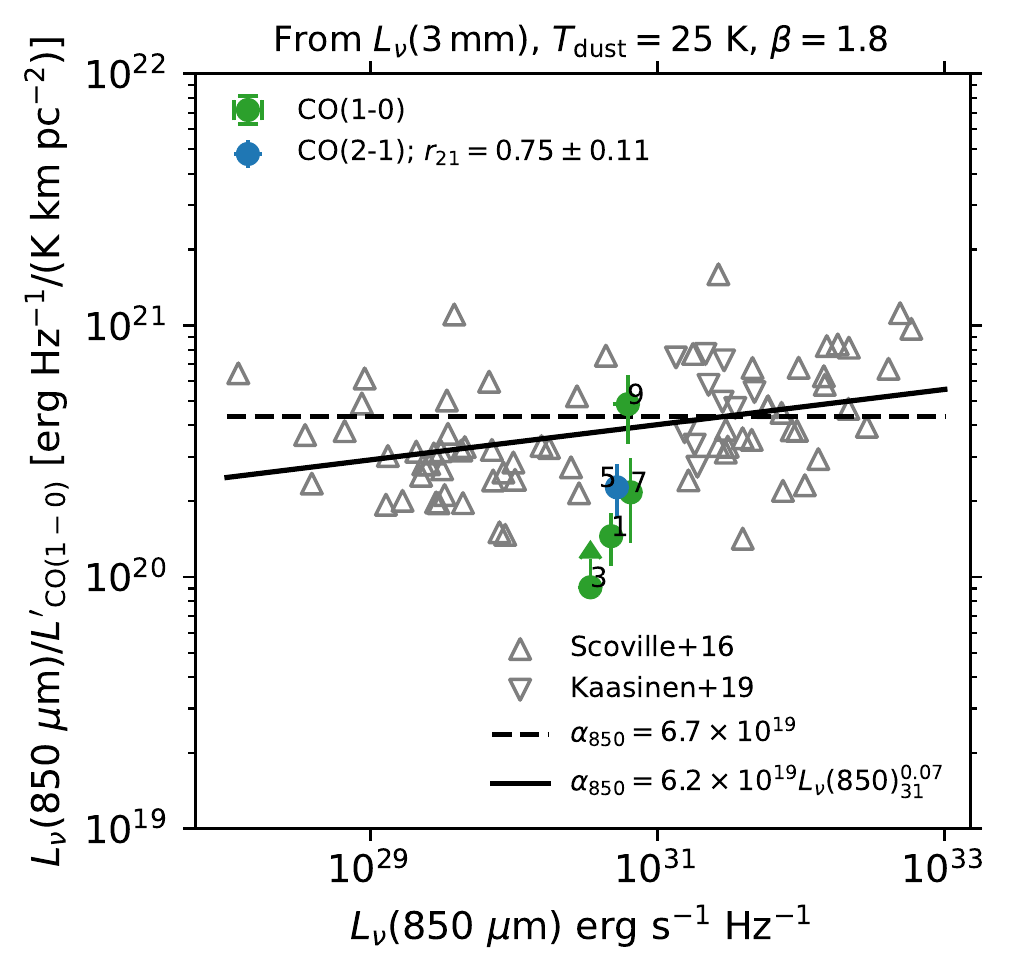}
  \caption{\textbf{Left:} The same as \autoref{fig:L850_LCO10} (right), but now
    using the inferred measurements of CO(1-0) from the low-$J$ CO(2--1) and
    CO(3--2) lines, using the excitation corrections from
    \autoref{sec:average-co-ladder}.  \textbf{Right} The same as
    \autoref{fig:L850_LCO10} (right), but now the $L_{\nu}(850\mu {\rm m})$ is
    inferred from the observed 3.0\,mm continuum instead (when detected).  The
    3.0\,mm continuum probes further down the Rayleigh-Jeans tail and is
    therefore less sensitive to the extrapolation to rest-frame 850\,\micron.
    The latter yields a slightly lower $L_{\nu}(850\mu {\rm m})$, but overall
    both methods give very consistent results.
    \label{fig:L850_LCO}}
\end{figure*}
\newpage
\subsection{Dust continuum versus low-$J$ CO}
\label{sec:dust-cont-vers}
\begin{deluxetable*}{ccccccc}
  \tablecaption{ASPECS-LP dust continuum data \label{tab:L850}}
  \tablehead{
    \colhead{ID 1mm} & \colhead{ID 3mm} & \colhead{$z$} &
    \colhead{$S_{\nu}(1.2\,\mathrm{mm})$} & \colhead{$S_{\nu}(3\,\mathrm{mm})$}
    & \colhead{$L_{\nu}(850\,\mu\mathrm{m,\,rest})^{a}$} &
    \colhead{$L_{\nu}(850\,\mu\mathrm{m,\,rest})^{b}$} \\
    \colhead{} & \colhead{} & \colhead{} & \colhead{$\mu$Jy} &
    \colhead{$\mu$Jy} & \colhead{($10^{29}$\,erg\,s$^{-1}$\,Hz$^{-1}$)} &
    \colhead{($10^{29}$\,erg\,s$^{-1}$\,Hz$^{-1}$)}
  }
  \colnumbers

  \startdata
  1mm.C01             & 3mm.01   & 2.543 & $752 \pm 24$  & $32.5 \pm 3.8$ & $66.6 \pm 2.1$ & $47.4 \pm 5.5$ \\
  1mm.C03             & 3mm.04   & 1.414 & $429 \pm 23$  & $\le 20$       & $41.3 \pm 2.2$ & \nodata \\
  1mm.C04             & 3mm.03   & 2.454 & $316 \pm 12$  & $22.7 \pm 4.2$ & $28.3 \pm 1.1$ & $34.0 \pm 6.3$ \\
  1mm.C05             & 3mm.05   & 1.551 & $461 \pm 28$  & $27.4 \pm 4.6$ & $44.4 \pm 2.7$ & $52.0 \pm 8.7$ \\
  1mm.C06             & 3mm.07   & 2.696 & $1071 \pm 47$ & $46.5 \pm 7.1$ & $93.3 \pm 4.1$ & $64.8 \pm 9.9$ \\
  1mm.C07             & \nodata  & 2.58  & $233 \pm 12$  & $\le 20$       & $20.5 \pm 1.0$ & \nodata \\
  1mm.C09             & 3mm.13   & 3.601 & $155 \pm 10$  & $\le 20$       & $12.4 \pm 0.8$ & \nodata \\
  1mm.C10             & \nodata  & 1.997 & $342 \pm 34$  & $\le 20$       & $32.1 \pm 3.2$ & \nodata \\
  1mm.C12             & 3mm.15   & 1.096 & $114 \pm 11$  & $\le 20$       & $10.5 \pm 1.0$ & \nodata \\
  1mm.C13             & 3mm.10   & 1.037 & $116 \pm 16$  & $\le 20$       & $10.6 \pm 1.4$ & \nodata \\
  1mm.C14a            & \nodata  & 1.999 & $96 \pm 10$   & $\le 20$       & $9.0 \pm 0.9$  & \nodata \\
  1mm.C16             & 3mm.06   & 1.095 & $143 \pm 18$  & $\le 20$       & $13.2 \pm 1.6$ & \nodata \\
  1mm.C15             & 3mm.02   & 1.317 & $118 \pm 13$  & $\le 20$       & $11.3 \pm 1.3$ & \nodata \\
  1mm.C19             & 3mm.12   & 2.574 & $85 \pm 12$   & $\le 20$       & $7.5 \pm 1.1$  & \nodata \\
  1mm.C20             & \nodata  & 1.093 & $94 \pm 16$   & $\le 20$       & $8.7 \pm 1.5$  & \nodata \\
  1mm.C25             & 3mm.14   & 1.098 & $90 \pm 17$   & $\le 20$       & $8.3 \pm 1.6$  & \nodata \\
  1mm.C23             & 3mm.08   & 1.382 & $148 \pm 30$  & $\le 20$       & $14.2 \pm 2.9$ & \nodata \\
  1mm.C30             & \nodata  & 0.458 & $34 \pm 10$   & $\le 20$       & $1.8 \pm 0.5$  & \nodata \\
  \nodata             & 3mm.11   & 1.096 & $\le 50$      & $\le 20$       & \nodata        & \nodata \\
  \nodata$^{\dagger}$ & 3mm.09   & 2.698 & $924 \pm 76$  & $44.5 \pm 9.7$ & $80.5 \pm 6.6$ & $62.0 \pm 13.5$ \\
  Faint.1mm.C20       & 3mm.16   & 1.294 & $86 \pm 24$   & $\le 20$       & $8.2 \pm 2.3$  & \nodata \\
  \nodata             & MP.3mm.2 & 1.087 & $\le 50$      & $\le 20$       & \nodata        & \nodata \\
  \enddata

  \tablenotemark{$\dagger$}{Object falls outside of the Band 6 mosaic, but is
    the brightest 1\,mm continuum source in the ASPECS field.  We adopt the
    $S_{\nu}(1.3\,\mathrm{mm})$ from \citep[][]{Dunlop2017}.}

  \tablenotemark{$a$}{Derived from $S_{\nu}$(1.2\,mm).}
  \tablenotemark{$b$}{Derived from $S_{\nu}$(3\,mm).}

\tablecomments{(1) ASPECS-LP continuum ID \citep{Gonzalez-Lopez2020,
    Aravena2020}. (2) ASPECS-LP line ID \citep{Boogaard2019}. (3) Redshift. (4)
  Flux density at 1.2\,mm \citep{Gonzalez-Lopez2020, Aravena2020}. (5) Flux
  density at 3\,mm \citep{Gonzalez-Lopez2019}. (6) Rest-frame 850~\micron\
  luminosity density inferred from $S_{\nu}$(1.2\,mm), assuming $\Tdust=25$\,K
  and $\beta=1.8$. (7) Rest-frame 850\,\micron\ luminosity density inferred
  from $S_{\nu}$(3\,mm), assuming $\Tdust=25$\,K and $\beta=1.8$.}
\end{deluxetable*}
The 1.2\,mm dust continuum emission provides an alternative way of measuring
the molecular gas mass, which is typically traced by the CO(1--0) emission
\citep[see][for an early reference]{Hildebrand1983}.  Because the Rayleigh
Jeans tail of the dust emission is nearly always optically thin, the dust
emission at long wavelengths is a direct probe of the total dust mass and
therefore the molecular gas mass, under the assumption that the dust emissivity
per unit dust mass and dust-to-gas ratio can be constrained
\citep{Scoville2014, Scoville2016}.  Motivating a mass-weighted cold dust
temperature $\Tdust = 25$\,K (which, in contrast to the light-weighted \Tdust,
is much less sensitive to the radiation field) and a dust emissivity index
$\beta = 1.8$, \cite{Scoville2016} show that the observed ratio between the
(inferred) dust luminosity at rest frame 850\,\micron,
$L_{\nu} (850\,\micron)$, and $L'_{\rm CO(1-0)}$ is relatively constant under
the wide range of conditions found in local star-forming galaxies, (U)LIRGS and
(mostly lensed) SMGs.  Recently, this has been further confirmed for a sample
of $z\sim2$ SFGs \citep{Kaasinen2019} as well as simulations \citep{Liang2018,
  Privon2018}.\footnote{Motivated by their observed correlation between
  $L_{\nu}(850\,\micron)$ and $L'_{\rm CO(1-0)}$, \cite{Scoville2016} then
  empirically calibrate the \Ldust-to-\Mmol\ ratio, assuming a CO-to-H$_{2}$
  mass conversion factor of
  $\aco = 6.5$\,\Msun(K\,km\,s$^{-1}$\,pc$^{2}$)$^{-1}$ (incl. He).  Note that,
  therefore, this estimate cannot be used to derive \aco\ independently.}

We can thus investigate whether our galaxies (that are observed in
$L'_{\rm CO(1-0)}$) follow the empirical relation with $L_{\nu}(850\,\micron)$
by \cite{Scoville2016}, by directly comparing to their calibration sample.  We
then use it to place constraints on the excitation for the sources only
observed in higher low-$J$ lines.  The advantage of this approach (rather than
comparing inferred gas masses) is that it is independent of $\aco$ and only
depends on the assumed excitation correction (\autoref{eq:rJ1}).  Furthermore,
we need not assume a gas-to-dust ratio, as this is implicit in the empirical
correlation (but it does depend on the assumptions for \Tdust\ and $\beta$,
mentioned above).  We stress that we cannot infer individual excitation
corrections in this manner, since the calibration only holds on average and has
a certain degree of intrinsic scatter\footnote{Using the data from
  \cite{Scoville2016}, we measure a scatter around the relation of about
  0.2~dex.  However, this includes the scatter due to measurement and
  extrapolation errors (which are not provided in the paper), therefore the
  intrinsic scatter is potentially smaller.}.

We estimate the rest frame $L_{\nu}(850\,\micron)$ for our sources from the
1.2\,mm continuum emission, assuming $\Tdust = 25$\,K and $\beta = 1.8$
(\autoref{tab:L850}).  While a $\Tdust = 25$\,K is arguably a good assumption
for the cold dust that traces the cold gas mass \citep{Scoville2016}, we note
that the observed SED, which should be used to scale the flux density to
rest-frame 850\,\micron, is dominated by the luminosity-weighted dust
temperature, which is likely higher.  However, we adopt $\Tdust = 25$\,K in
order to remain consistent with the calibration sample of \cite{Scoville2016}.
We show the $L_{\nu}(850\,\micron)$ against the CO(1--0) luminosity in
\autoref{fig:L850_LCO10}.  The ASPECS galaxies probe fainter dust luminosities
than the calibration sample(s) at high-$z$.  For the sources observed in
CO(1--0), we find that the three detections (including 3mm.07) and the upper
limits are consistent with the \citep{Scoville2016} relation.  In
\autoref{fig:L850_LCO} (left panel), we show the same ratio, but with the \LCO\
inferred from the low-$J$ CO lines.  Using $r_{21} = \rTwoOneLVGzLoTwo$
(\autoref{sec:lvg-modeling}), we find that the sources detected in CO(2--1) at
$\avg{z}=1.2$ on average lie relatively low compared to the Scoville relation,
although several individual sources follow it well.  Using the mean
$r_{31} = 0.77 \pm 0.14$ (measured from stacking,
\autoref{sec:average-co-ladder}) for the galaxies at $z=2.0-2.7$, we find that
most sources are consistent with the relation, including the galaxies not
individually detected in CO(1--0), although the sample average is slightly
below the relation.  Assuming that the rest-frame 850\,\micron\ and CO
luminosities are tightly correlated, this would suggest that the excitation
values we adopt are too low on average, in particular for CO(2--1).  For
comparison, we also show the case in which the low-$J$ lines are thermalized on
average ($r_{21} = r_{31} = 1.0$; black points).  We find an overall better
agreement assuming the lines are thermalized on average.  Although we cannot
constrain the $L'_{\mathrm{CO}(1-0)}$ for individual sources via the
$L_{\nu}(850\,\micron)$ calibration, the comparison implies that, on average,
the $r_{21}$ and $r_{31}$ may not be much lower than $\sim0.75$, on average, at
$\avg{z}=1.2$ and 2.5, respectively (consistent with the stacking and LVG
modeling).  Note that to make the CO and dust fully consistent with the
empirical relation, based on CO excitation alone, would imply suprathermalized
CO in some cases, which is not expected to occur under normal conditions in the
ISM, where the CO is optically thick (but $r_{J1}>1$ is possible if the CO is
optically thin).

An alternative explanation for the low
$L_{\nu}(850\mu {\rm m})/L'_{\mathrm{CO}}$ ratios is a bias due to the
CO-selection.  Comparing the primary, flux-limited samples (see
\autoref{sec:sample}) of both the CO and dust continuum-selected sources with a
redshift at which we can detect CO, we find that there are two CO(2-1)-selected
sources without dust continuum and potentially\footnote{The precise number is
  dependent on the accuracy of the redshift measurement available for the dust
  continuum sources.} two vice versa.  At the same time, all CO(3--2) emitters
are detected in dust-continuum, while there are potentially four dust-selected
sources at $z=2-3$ without CO(3--2).  While the number of galaxies under
consideration is modest, this argues against a strong selection effect, at
least for the CO(2--1)-selected sources, in which case we would expect a larger
number of dust-selected sources with CO emission (filling in the scatter above
the relation).  Because the CO detection limit increases relative to that of
the dust continuum (as the latter experiences a strong negative k-correction,
e.g., \citealt{Blain2002}), a selection effect is expected to be stronger for
the CO(3--2)-selected sources, as is indeed suggested by the above comparison.
However, the latter galaxies do not show systematically lower ratios, compared
to the CO(2--1) selected sources, and direct observations of CO(1--0) for a few
of the sources do not suggest a strong systematic offset.  Overall, we
therefore conclude that, while we cannot fully exclude the impact of selection,
it does not appear to play a dominant role at least for the CO(2--1)-selected
sources.

Finally, to investigate the influence of the Rayleigh Jeans correction on the
results (in particular for the higher redshift sources), we also infer
$L_{\nu}(850\mu\mathrm{m})$ from the 3.0\,mm continuum data, that has been
detected in four of the galaxies at $z\approx2.6$ and 3mm.05 at $z=1.55$
(\autoref{fig:L850_LCO10}, right panel).  The rest-frame
$L_{\nu}(850\mu\mathrm{m})$ luminosities inferred from 3.0\,mm are on average
$\approx 10\%$ lower than those from the 1.2\,mm, but overall we come to the
same conclusions.
\begin{figure*}[t]
  \includegraphics[width=\columnwidth]{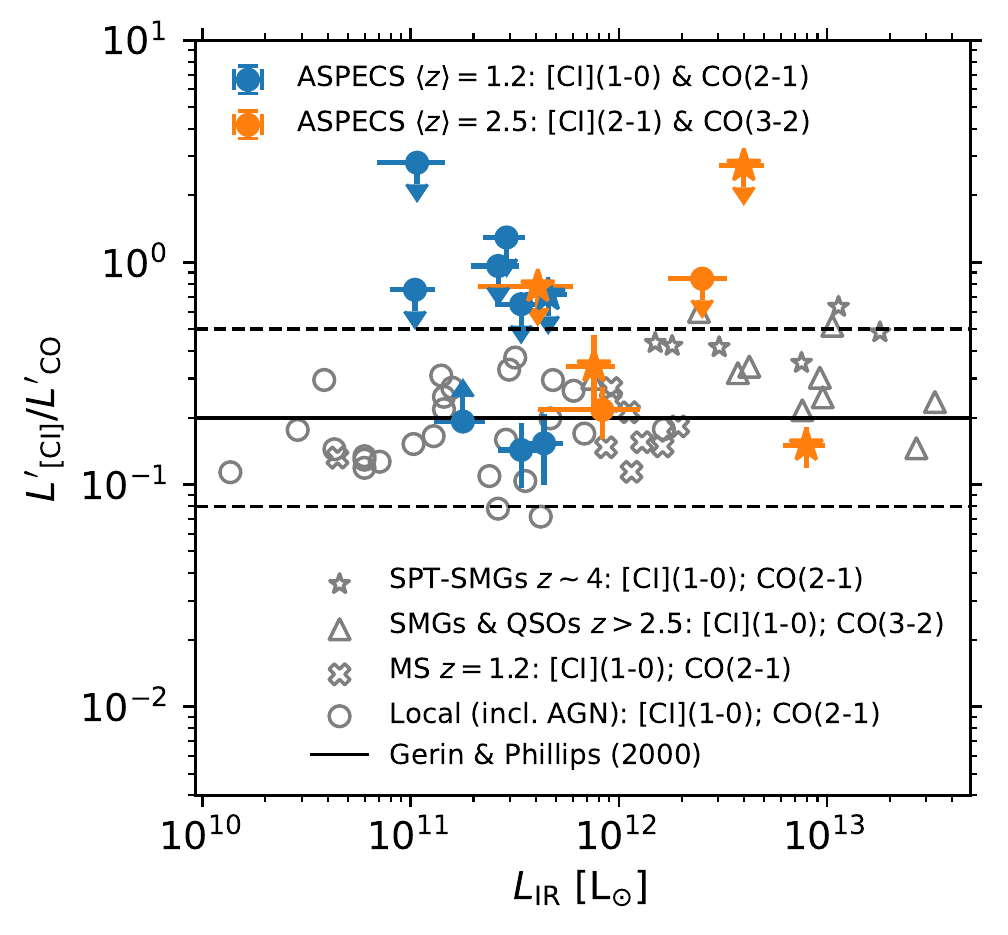}
  \includegraphics[width=\columnwidth]{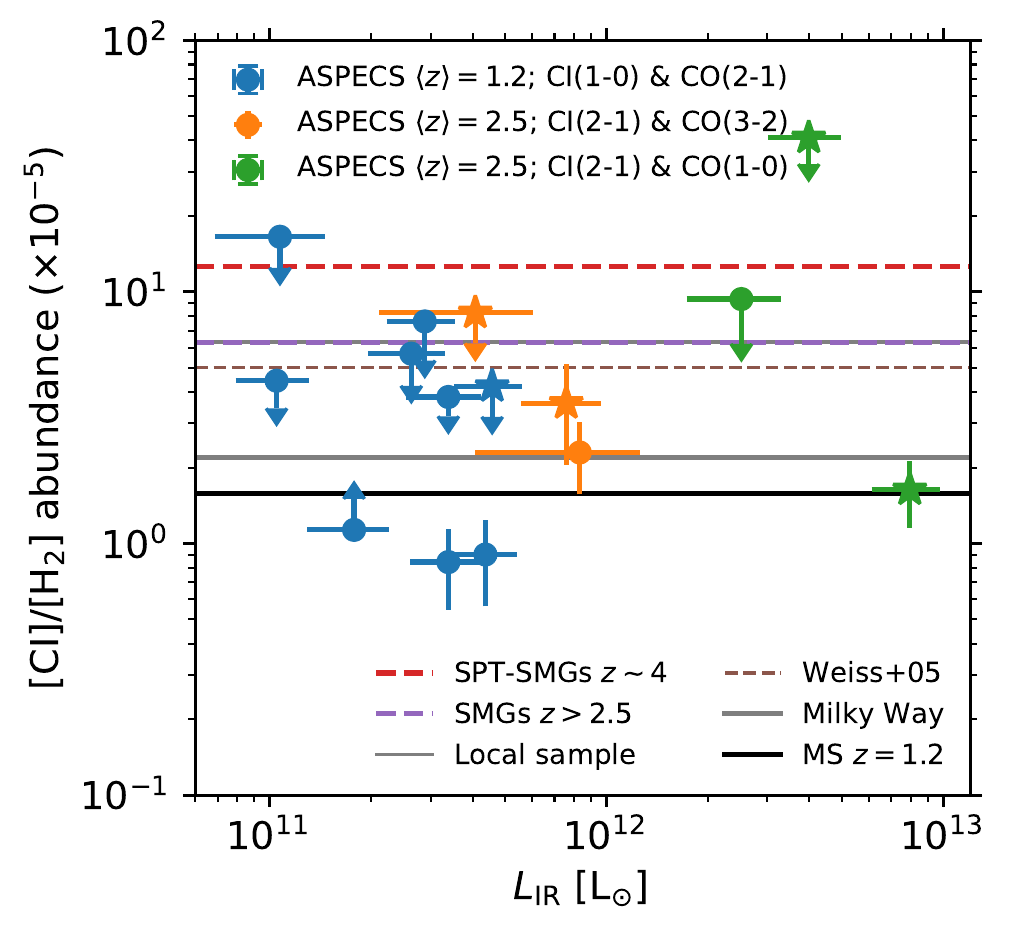}
  \caption{\textbf{Left:} \LCI/\LCO\ ratio for
    \CI(${^{3}P_{1}}\rightarrow {^{3}P_{0}}$) over CO(2--1) (blue) and
    \CI(${^{3}P_{2}}\rightarrow {^{3}P_{1}}$) over CO(3--2) (orange), where
    stars indicate X-ray AGN.  We compare the observed ratios to SPT-SMGs at
    $z=4$ \citep{Bothwell2017}, SMGs at $z\ge2.5$ \citep{Walter2011,
      Alaghband-Zadeh2013}, main-sequence galaxies at $z=1.2$ and local
    galaxies, as compiled by \cite{Valentino2018}, and the average ratio and
    scatter in the local sample from \cite{Gerin2000}.  Overall, the ratios
    broadly agree with the spread found for previous samples of star-forming
    galaxies. \textbf{Right:} Atomic carbon abundance in the ASPECS galaxies.
    The H$_{2}$ mass was derived from CO(2--1) (assuming
    $r_{21} = \rTwoOneLVGzLoTwo$) and CO(1--0) or CO(3--2) (assuming
    $r_{31} = \rThreeOnestack$), with
    $\aco=3.6$\,\Msun(K\,km\,s$^{-1}$\,pc$^{2}$)$^{-1}$, for the galaxies
    detected in \CI(1--0) and \CI(2--1) respectively.  We compare our
    measurements to the abundances for different galaxy types (excluding active
    galaxies), converted to a common \aco\ by \cite{Valentino2018}.  On average
    we find \CI\ abundances similar to the Milky Way \citep{Frerking1989} and
    the star-forming galaxies from \cite{Valentino2018} (who assumes a
    galaxy-specific \aco, which is 3.0 on average, and $r_{21} = 0.84$), with
    higher abundances at $\avg{z} = 2.5$ compared to $\avg{z} = 1.2$.  Note
    that the higher abundances in the sub-millimeter galaxies from literature
    are partly driven by the assumed lower \aco\ in these
    systems. \label{fig:abundance_CO}}
\end{figure*}
\newpage
\section{Atomic carbon}
\label{sec:atomic-carbon}
\subsection{Atomic carbon abundances}
\label{sec:atom-carb-abund}
\begin{deluxetable*}{ccccccc}
  \tablecaption{Masses from different tracers and neutral atomic carbon
    abundances for the \CI\ detected galaxies.\label{tab:ci-table}}
  \tablehead{\colhead{ID} & \colhead{$z$} &  \colhead{$M_{\rm mol,\,RJ}$} & \colhead{$M_{\rm mol,\,CO}$} &  \colhead{$M_{\CI}$} & \colhead{(\CIHH)$_{\rm RJ}$} & \colhead{(\CIHH)$_{\rm CO}$}\\
    \colhead{} & & \colhead{($\times 10^{10}$\,\Msun)} &
    \colhead{($\times 10^{10}$\,\Msun)} & \colhead{($\times 10^{6}$\,\Msun)} &
    \colhead{($\times 10^{-5}$)} & \colhead{($\times 10^{-5}$)} } \colnumbers
  \startdata
  1mm.C12             & 1.09 & $1.6 \pm 0.2$  & $1.4\pm0.3$   & $\le 2.7$    & $\le3.8$     & $\le4.2$  \\
  1mm.C13             & 1.03 & $1.6 \pm 0.2$  & $3.4\pm0.7$   & $1.4\pm 0.4$ & $1.9\pm 0.7$ & $0.9\pm 0.3$  \\
  1mm.C16             & 1.09 & $2.0 \pm 0.2$  & $2.5\pm0.4$   & $0.9\pm 0.3$ & $1.1\pm 0.4$ & $0.8\pm 0.3$  \\
  1mm.C15             & 1.31 & $1.7 \pm 0.2$  & $4.0\pm0.7$   & $\le 6.8$    & $\le9.1$     & $\le3.8$  \\
  1mm.C20             & 1.09 & $1.3 \pm 0.2$  & $\le3.3$      & $1.6\pm 0.5$ & $2.9\pm 1.0$ & $\ge1.1$  \\
  1mm.C25             & 1.09 & $1.3 \pm 0.2$  & $1.9\pm0.5$   & $\le 4.7$    & $\le8.5$     & $\le5.7$  \\
  3mm.11              & 1.09 & $\le 0.4$      & $0.6\pm0.1$   & $\le 1.1$    & \nodata      & $\le4.4$  \\
  3mm.16              & 1.29 & $1.2 \pm 0.3$  & $0.9\pm0.2$   & $\le 6.6$    & $\le12.3$    & $\le16.5$ \\
  MP.3mm.2            & 1.09 & $\le 0.4$      & $1.1\pm0.3$   & $\le 3.7$    & $\le20.2$    & $\le7.6$\\
  \hline
  1mm.C01             & 2.54 & $10.0 \pm 0.3$ & $11.7\pm 2.4$ & $8.5\pm 1.8$ & $1.9\pm 0.4$ & $1.6\pm 0.5$  \\
  1mm.C04$^{\dagger}$ & 2.45 & $4.2  \pm 0.2$ & $4.5\pm 0.9$  & $4.6\pm 1.1$ & $2.5\pm 0.6$ & $2.3\pm 0.7$  \\
  1mm.C06             & 2.69 & $14.0 \pm 0.6$ & $10.7\pm 3.6$ & $\le44.3$    & $\le7.2$     & $\le9.4$  \\
  1mm.C07$^{\dagger}$ & 2.58 & $3.1  \pm 0.2$ & $2.5\pm 0.7$  & $3.9\pm 1.3$ & $2.9\pm 1.0$ & $3.6\pm 1.6$  \\
  1mm.C19$^{\dagger}$ & 2.57 & $1.1  \pm 0.2$ & $2.2\pm 0.5$  & $\le7.3$     & $\le16.0$    & $\le8.2$  \\
  3mm.9 & 2.70 & $12.0\pm1.0$ & $4.6\pm1.0$ & $\le83.1$ & $\le 15.6$ &
  $\le 40.9$ \enddata \tablenotemark{$^{\dagger}$}{CO related properties
    derived from CO(3--2) assuming $r_{31}=\rThreeOnestack$}
  \tablecomments{Properties derived for \CI(1--0) and CO(2--1) at $1\le z<2$,
    assuming $r_{21} = \rTwoOneLVGzLoTwo$ (\textbf{top rows}) and from
    \CI(2--1) and CO(1--0), or CO(3--2) assuming $r_{31} = \rThreeOnestack$, at
    $2 \le z < 3$ (\textbf{bottom rows}).  In the case of a non-detection we
    report a $3\sigma$ upper limit. (1) ASPECS-LP ID (see
    \autoref{tab:sample}). (2) Redshift. (3) Molecular gas mass determined via
    the 1.2\,mm dust continuum emission on the Rayleigh Jeans tail
    (\autoref{sec:dust-cont-vers}; cf.\ \autoref{tab:L850}). (4) Molecular gas
    emission determined from the CO line luminosity emission assuming
    $\aco=3.6$\,\Msun\,(K\,km\,s$^{-1}$\,pc$^{2}$)$^{-1}$ (\autoref{eq:Mmol}).
    (5) Atomic carbon mass derived from \CI(1--0) and \CI(2--1) via
    \autoref{eq:MCI10} and \autoref{eq:MCI21}. (6) Neutral atomic carbon
    abundance computed with $M_{\rm mol, RJ}$ (\autoref{eq:ci-abund}). (7)
    Neutral atomic carbon abundance, computed with $M_{\rm mol, CO}$.}
\end{deluxetable*}

Atomic carbon has been suggested as a good alternative tracer of the molecular
gas mass.  This is motivated by the fact that the emission from atomic carbon
(\CI) has been found to be closely associated with CO emission in a range of
different environments in the Milky Way \citep{Stutzki1997, Ojha2001,
  Ikeda2002, Schneider2003} and in local galaxies \citep[e.g.,][]{Gerin2000,
  Israel2015, Jiao2019}.  There has been some debate to whether \CI\ can be
used to trace the total molecular gas mass, because the \CI\ emission was
originally predicted to arise only from a narrow \CII/\CI/CO transition zone in
molecular clouds on the basis of early theoretical work (\citealt{Tielens1985a,
  Tielens1985b}; cf.\ \citealt{Israel2015}).  However, more recent models have
supported the picture in which CO and \CI\ coexist over a wide range of
conditions \citep[see, e.g., ][]{Papadopoulos2004, Glover2015, Bisbas2015,
  Bisbas2017}.

The \CI\ lines are typically found to be optically thin \citep{Ojha2001,
  Ikeda2002, Weiss2003}.  As a result, the \CI\ column density in the upper
levels of the ${^{3}P_{2}} \rightarrow {^{3}P_{1}}$
($\nu_{\rm rest} = 809.342$\,GHz) and ${^{3}P_{1}} \rightarrow {^{3}P_{0}}$
($\nu_{\rm rest} = 492.161$\,GHz) transitions is directly related to the line
intensity, and depends only on the excitation temperature, \Tex\
\citep[e.g.,][given their low critical densities,
$< 10^{3}$\,cm$^{-3}$]{Frerking1989, Stutzki1997, Weiss2003}.  This means that
the atomic carbon mass (\MCI) can be directly inferred from the line
luminosity:
\begin{align}
\MCI &= 5.706 \times 10^{-4} Q(\Tex)\frac{e^{T_1 / \Tex}}{3}L'_{\CI(1-0)} \label{eq:MCI10}\\
\MCI &= 4.556 \times 10^{-4} Q(\Tex)\frac{e^{T_2 / \Tex}}{5}L'_{\CI(2-1)}. \label{eq:MCI21}
\end{align}
Here $T_1 = 23.6$ K and $T_2 = 62.5$ K are the energies of the ${^{3}P_{2}}$
and ${^{3}P_{1}}$ levels and
\(Q(\Tex) = 1 + 3 e^{-T_1/\Tex} + 5 e^{-T_2/\Tex}\) is the partition function
in the three-level system approximation \citep{Weiss2003, Weiss2005a}.

The excitation temperature itself can be measured directly from the ratio of
the integrated line intensities,
\begin{align}
  \label{eq:CI_Tex}
  \Tex = \frac{\mathrm{38.8\, K}}{\ln(2.11 / R)},
\end{align}
where $R = L'_{\CI(2-1)}/L'_{\CI(1-0)}$ \citep{Stutzki1997}.  \cite{Walter2011}
measured an excitation temperature of $\langle \Tex \rangle = 29.1 \pm 6.3$\,K
in a sample of $\langle z \rangle = 2.5$ SMGs.  As we never observe both \CI\
transitions in the same source, we assume a typical value of $\Tex = 30$\,K
\citep[cf.][corresponding to $R = 0.58$]{Weiss2005a, Bothwell2017,
  Popping2017, Valentino2018, Brisbin2019}.  Note from \autoref{eq:MCI10} and
\autoref{eq:MCI21} that the neutral atomic carbon mass is not a strong function
of the assumed excitation temperatures above $\approx 20$\,K for \CI(1--0) and
$\approx 40$\,K for \CI(2--1) \citep[as pointed out by][]{Weiss2005a}.

Before turning to the masses inferred from \CI\ and CO, we compare the line
luminosities directly, as a function of \LIR, in the left panel of
\autoref{fig:abundance_CO}.  In particular at $\avg{z}=1.2$, we probe \CI\ in
galaxies at lower \LIR\ than previous studies of similar sources.  Overall, the
ratios are comparable to those in the main-sequence galaxies from
\cite{Valentino2018} and the average ratio in a variety of local galaxies from
\cite{Gerin2000}.

We derive \CI\ masses of a few $\times 10^{6}$\,\Msun (\autoref{tab:ci-table}).
From the \CI\ masses, we derive the galaxy average, luminosity weighted,
neutral atomic carbon abundances,
\begin{align}
  \frac{\CI}{[\mathrm{H}_{2}]} = \frac{\MCI}{6\MH}, \label{eq:ci-abund}
\end{align}
where $\MH = \Mmol / 1.36$, not including He.  We use the CO-derived H$_{2}$
masses, adopting $r_{21}=\rTwoOneLVGzLoTwo$ and $r_{31}=\rThreeOnestack$ for
the sources without CO(1--0), and
$\aco=3.6$\,\Msun(K\,km\,s$^{-1}$\,pc$^{2}$)$^{-1}$.  The abundances are shown
as a function \LIR\ in \autoref{fig:abundance_CO}.  We find an average
abundance of $(\CIHH)_{\rm CO} = \avgCIabundCO$ (ignoring limits)\footnote{If
  we instead assume thermalized CO for all sources without CO(1--0), we derive
  $(\CIHH)_{\rm CO} = \avgCIabundCOtherm$.}.  Overall, the abundances are
broadly similar to those in the Milky Way
\citep[$2.2\times10^{-5}$,][]{Frerking1989} and in $z \approx 1.2$
main-sequence galaxies
\citep[$(1.6 \pm 0.8) \times 10^{-5}$,][]{Valentino2018}, but lower than in
high redshift SMGs \citep{Walter2011, Alaghband-Zadeh2013, Bothwell2017}.
However, as pointed out by \cite{Valentino2018}, these differences could also
be driven by the difference in adopted \aco, as their derived abundances assume
an $\aco=0.8$\,\Msun(K\,km\,s$^{-1}$\,pc$^{2}$)$^{-1}$, which is the typical
value assumed for these systems.  Finally, we come to similar conclusions if we
use the dust continuum based molecular gas masses instead,
$(\CIHH)_{\rm RJ} = \avgCIabundDust$, assuming the luminosity independent
calibration from \cite{Scoville2016} (\autoref{sec:dust-cont-vers}).

\subsection{PDR modeling}
\label{sec:pdr-modeling}
\begin{figure}[t]
  \includegraphics[width=\columnwidth]{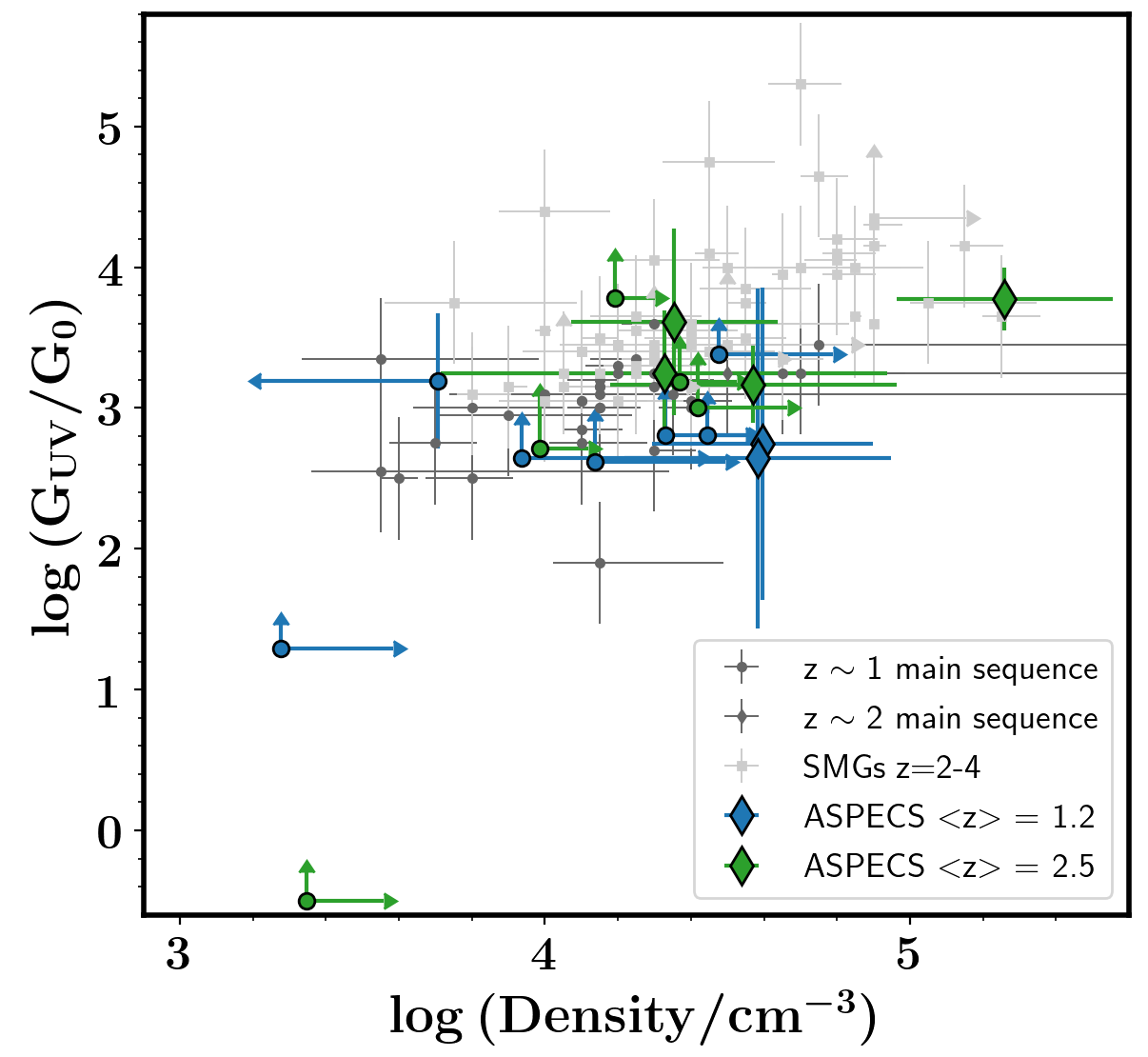}
  \caption{The ISM density and UV radiation field strength ($G_{\rm UV}$,
    relative to the local galactic interstellar radiation field, $G_{0}$,
    \citealt{Habing1968}) as inferred from PDR modeling.  The ASPECS galaxies
    are shown as diamonds or, in the case of a limit on either parameter,
    circles.  These galaxies are compared to main-sequence galaxies at $z\sim1$
    \citep{Valentino2018, Bourne2019, Valentino2020a} and $z\sim2$
    \citep{Popping2017, Talia2018} and $z=2-4$ SMGs \citep{Walter2011,
      Alaghband-Zadeh2013, Bothwell2017, Yang2017, Canameras2018, Nesvadba2019,
      Harrington2018, Andreani2018, Dannerbauer2019, Jin2019} for which
    \citet{Valentino2020a} re-derived the density and UV radiation field (using
    similar lines and model assumptions as in this paper).\label{fig:PDR}}
\end{figure}
We use the combination of \CI, CO and far-infrared dust emission (\LIR) to
explore the ISM properties of the galaxies in our sample using
photodissociation regions (PDR) models.  To this end, we use the results from
the \textsc{PDRToolbox} \citep{Kaufman2006, Pound2008}.  The
\textsc{PDRToolbox} is based on the one-dimensional models from
\citep{Kaufman2006} and solves for the chemistry, thermal balance and radiative
transfer, assuming metal, dust and polycyclic aromatic hydrocarbon (PAH)
abundances and a gas microturbulent velocity.  Every PDR is described by a
fixed number density of H nuclei and intensity of the impinging UV radiation
field, $G_{\rm UV}$, in units of the local galactic interstellar radiation
field, $G_0$ \citep{Habing1968}.  The \textsc{PDRToolbox} then provides the
line ratio of [CI], CO and \LIR\ as a function of the density and UV radiation
field of a PDR.

We estimate the ISM density and UV radiation field by using a combination of
the \CI\ and CO emission lines for each galaxy.  We specifically focus on CO
emission from rotational transitions equal to or lower than CO(4--3), unless
these are not available, as higher order CO emission originates from
significantly denser ISM than \CI\ \citep[cf.][]{Valentino2020a}.  We adopt a
numerical approach where we bootstrap the observed flux ratios within their
error a 1000 times and solve for the ISM density and UV radiation field for
each instance.  As the final density and radiation field we take the median of
these values.  The 68\% confidence interval is taken as the error on the
derived values.  For the non-detected lines we run the models using $3\sigma$
upper limits on the line fluxes and interpret the results as lower or upper
limits accordingly.  Similar analyses have been performed in, for instance,
\citet{Alaghband-Zadeh2013, Bothwell2017, Popping2017, Canameras2018,
  Brisbin2019} and \citet{Valentino2020a}.

The results of the PDR modeling are shown in \autoref{fig:PDR}.  In general, we
find that the PDR models predict fairly high densities,
$\ge 10^{4}$\,cm$^{-3}$, for all sources.  In the PDR model, this is
constrained by the ratio of \CI\ (with low critical density) over CO.  The UV
radiation field strength is primarily determined by the ratio of the lines over
the dust continuum and found to be $\ge 3\times10^{2}\,\mathrm{G}_{0}$ in most
cases.  The median G$_{\mathrm{UV}}$ of the detections appears to be larger at
$\avg{z}=2.5$ compared to $\avg{z}=1.2$ in our sample, though this difference
is not statistically significant.  Overall, the galaxies occupy the same
parameter space as the main-sequence galaxies from \citep{Valentino2018,
  Valentino2020a}, who also modeled the CO and \CI\ lines.

We note that the PDR model adopted in this analysis (and other works) assumes
that the ISM of a galaxy can be described by a single PDR with a fixed input
abundane.  In reality, the ISM consists of a range of molecular clouds that all
have different properties (density, impinging UV radiation field, abundances).
Furthermore, the PDR models assume a fixed density throughout the medium,
whereas in reality the density distribution of PDRs is more complex.  Following
\cite{Valentino2018}, we also do not correct the models for the difference in
optical depth between CO, \CI\ and \LIR, and therefore restrict our relative
comparison with the literature to these data, which are consistently analysed.
Our results should therefore be treated as qualitative measures of the ISM
density and UV radiation field.  \citet{Alaghband-Zadeh2013},
\citet{Bothwell2017}, \citet{Canameras2018} and \citet{Valentino2020a} discuss
in more detail that the PDR modeling likely does not capture the full
complexity of the ISM in galaxies and should be taken as an order of magnitude
indication of the ISM properties.  Future work attempting to model the ISM
properties of galaxies should thus focus on spatially resolved observations and
multi-phase modeling of the ISM.

\section{Discussion}
\label{sec:discussion}
\subsection{Modest excitation in mid-$J$ lines at $z=1.0 - 1.6$}
\label{sec:modest-exc-high-j}
\begin{figure}[t]
\includegraphics[width=0.5\textwidth]{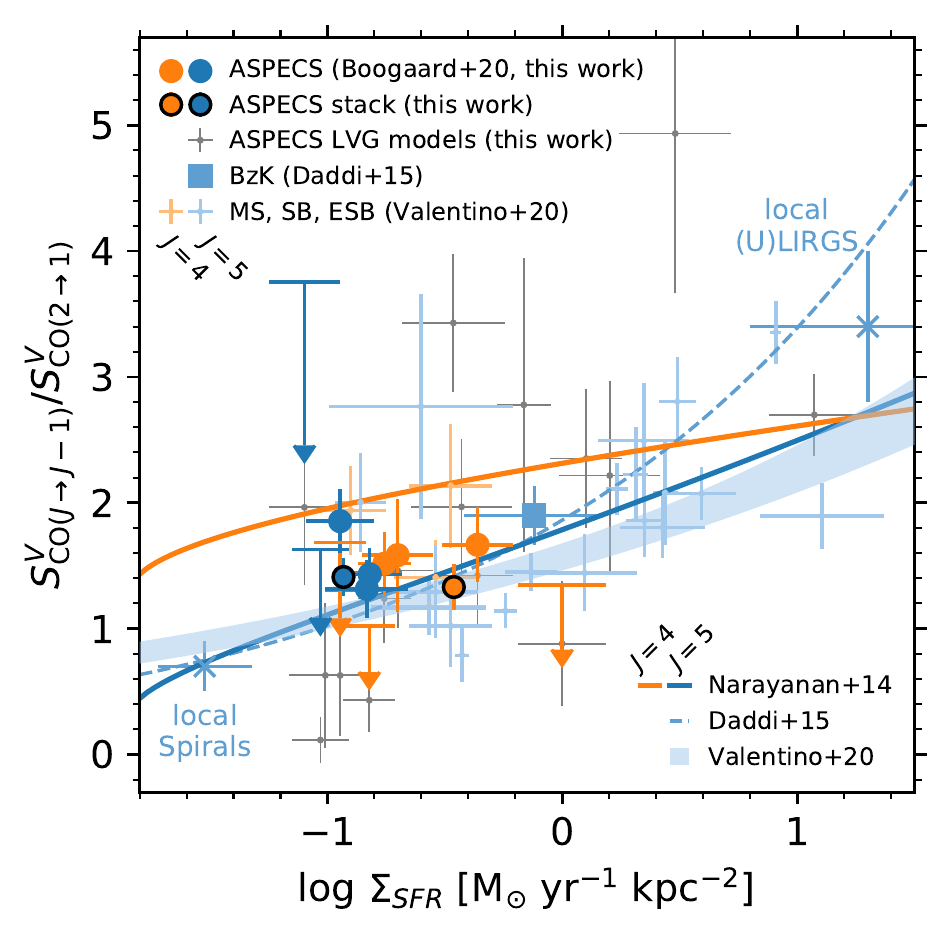}
\caption{Star formation rate surface density (\SigmaSFR) versus CO line flux
  ratio (in units of $S^{V}$), for both CO(4--3)/CO(2--1) (orange) and
  CO(5--4)/CO(2--1) (blue).  The colored points (and limits) show the observed
  line flux ratios of the ASPECS galaxies, while the gray points show the
  predict ratios from the LVG model fits for all galaxies (for
  CO(5--4)/CO(2--1) only; note the points are not visible for the galaxies in
  which we directly measure the ratio).  We also show the values from the
  stacks (\autoref{sec:average-co-ladder}) at the mean \SigmaSFR.  The blue
  square shows the average of the BzK-selected SFGs from \cite{Daddi2015} and
  the blue crosses show averages for local spirals and (U)LIRGS as reported by
  \cite{Daddi2015}.  The light shaded points show the recent data for main
  sequence and (extreme) starburst galaxies from \cite{Valentino2020b}.  The
  solid lines show the predictions from the \cite{Narayanan2014} models for
  unresolved observations, the dashed blue line shows the best-fit from
  \cite{Daddi2015} and the shaded region that from from \cite{Valentino2020b}
  (for CO(5--4) only).\label{fig:rJ1_SigmaSFR}}
\end{figure}
The ASPECS galaxies significantly expand the sample of star-forming galaxies
with CO excitation measurements at $z=1.0-1.6$.  In particular, our
observations increase the number of detections of the CO(4--3) and CO(5--4)
lines in sources at these redshifts.  A key result of our study is that the
$\avg{z}=1.2$ galaxies, selected by their CO(2--1) emission, show a range in
excitation of their $J\ge2$ lines up to CO(5--4).  In half of the sample we
find that the mid-$J$ CO lines are excited to similar (interpolated) levels as
the BzK galaxies at $\avg{z}=1.5$, suggesting the presence of a dense, warm
component in the ISM of these galaxies \citep[][]{Daddi2015}.  However, the
remaining galaxies are consistent with lower excitation, as shown by the
average stacked ladder including the individually non-detected transitions as
well (see \autoref{fig:co_individual_flux}).  This indicates that such a warm,
dense component is not as dominantly present in all galaxies.  On average, the
ASPECS galaxies at $\avg{z}=1.2$ are less excited in their mid-$J$ lines
compared to the BzK galaxies from \cite{Daddi2015}, but, the average mid-$J$
excitation is above that observed in, e.g., the Milky Way.

The lower excitation of the ASPECS galaxies can be naturally explained by their
lower surface density of star formation, as the excitation correlates with the
radiative energy input into the gas.  The CO excitation is sensitive to the gas
density and temperature and is known to correlate with the dust temperature
\citep{Rosenberg2015} and radiation field strength, star formation efficiency
and star formation rate surface density \citep{Daddi2015, Valentino2020b}.  The
excitation has also been shown to correlate, to a lesser extent, with the \LIR\
\citep[e.g.,][]{Rosenberg2015}.  The connection between \LIR\ and excitation is
less direct, however, because the total star formation rate does not correlate
with the density and temperature of the clouds as \SigmaSFR\ does
\citep[cf.][]{Narayanan2014}.  This conclusion is also reached by
\cite{Valentino2020b}, who show that the intrinsic scatter in the $r_{52}-\LIR$
relation is greater than that in the $r_{52}-\SigmaSFR$ relation.  Note that in
the case of equally-sloped $L'_{\rm CO}-\LIR$ relations for different $J$
\citep[for example, the linear relations found by ][]{Liu2015}, a correlation
between \LIR\ and excitation would also not be expected.

In \autoref{fig:rJ1_SigmaSFR}, we show the flux ratios of CO(4--3) and CO(5--4)
over CO(2--1), a proxy of the excitation in the CO ladder, as a function of
\SigmaSFR.  As anticipated, our galaxies at $\avg{z} = 1.2$ probe the low
\SigmaSFR\ regime at this redshift, compared to the sources studied in
\cite{Daddi2015} and the recent work by \cite{Valentino2020b}.  Overall, the
modest mid-$J$ excitation of the ASPECS sources appears to naturally follow
from the fact that we are probing galaxies with, on average, more moderate
surface densities of star formation.  We also compare to the models from
\cite{Narayanan2014}, who have computed theoretical CO ladders for unresolved
observations of galaxies, parameterized by \SigmaSFR.  While the models
qualitatively agree and appear to work reasonably well for $r_{52}$-ratio
transition, they seem to overpredict the $r_{42}$-ratio for the galaxies in our
sample.

\subsection{Increasing excitation with redshift}
\label{sec:evol-with-redsh}
\begin{figure*}[t]
  \centering
\includegraphics[height=0.25\textwidth]{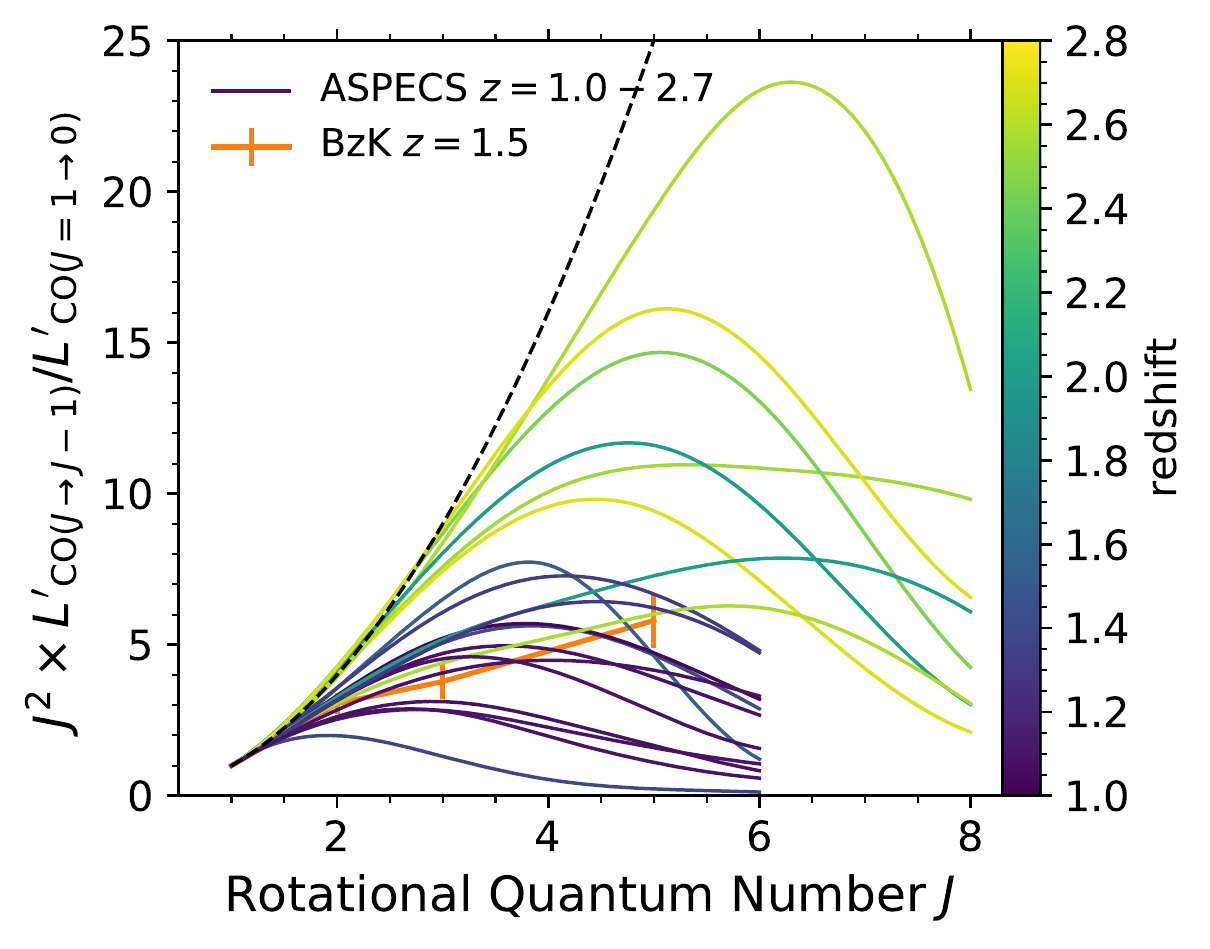}
\includegraphics[height=0.25\textwidth]{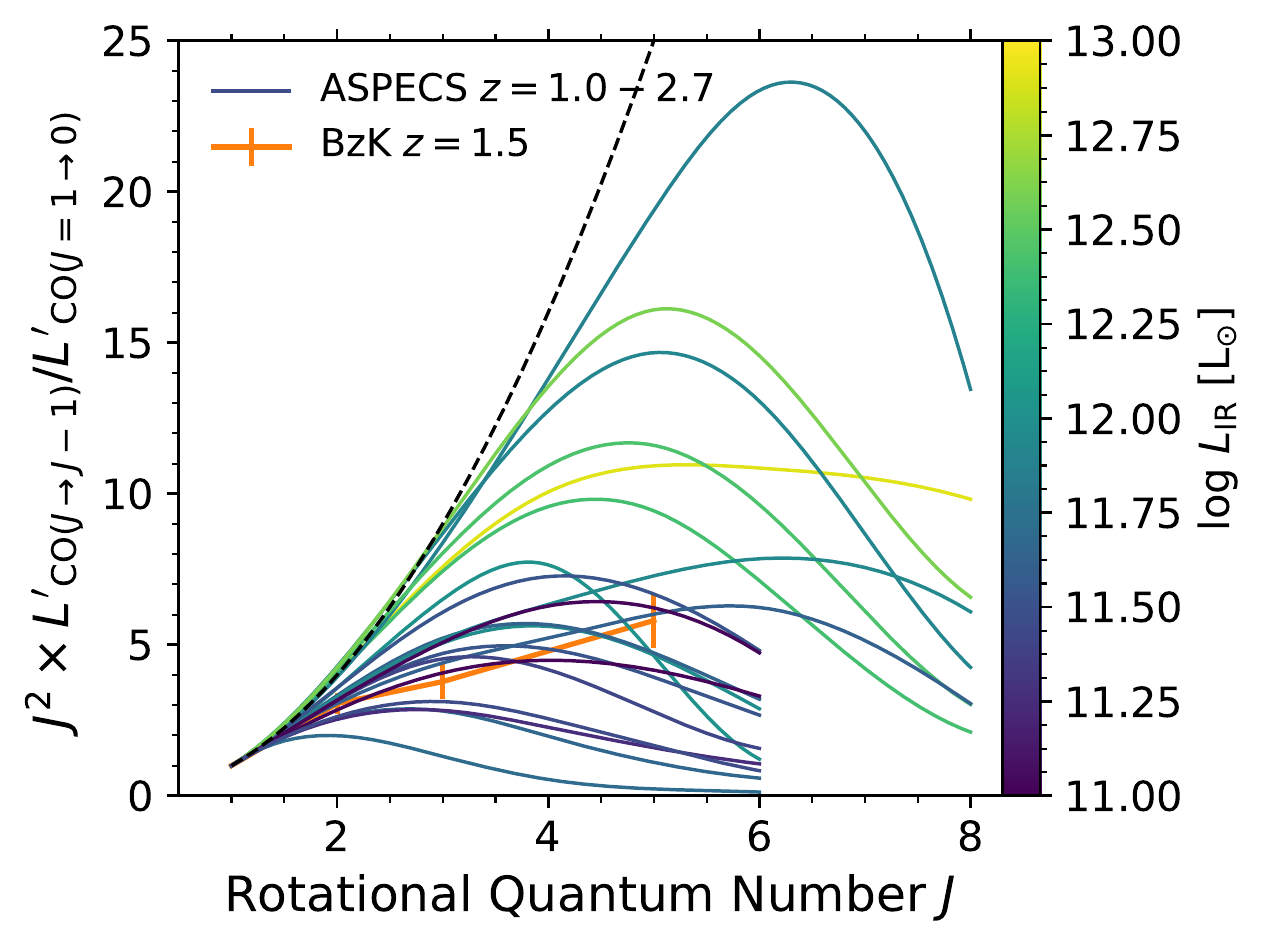}
\includegraphics[height=0.25\textwidth]{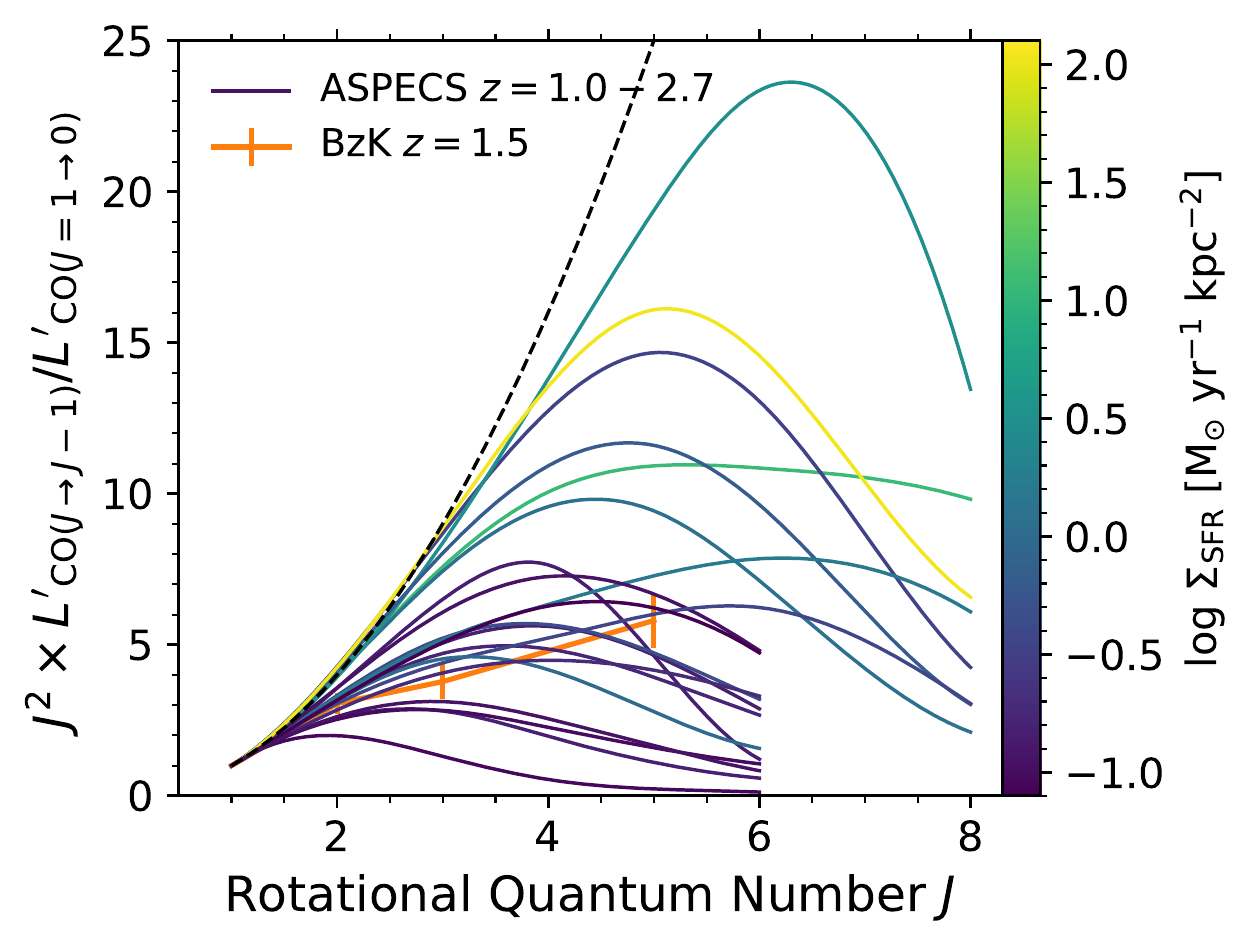}
\caption{ASPECS CO ladders from the two-component LVG models (cf.\
  \autoref{fig:CO-axel}) colored by redshift (\textbf{left}), \LIR\
  (\textbf{center}) and \SigmaSFR\ (\textbf{right}).  The ladders are now shown
  in units of $J^{2}\times L'_{\rm CO}$, normalised to $L'_{\mathrm{CO}(1-0)}$,
  for an easy comparison with the figures in units of line flux shown
  throughout the paper.  While the overall excitation increases with redshift,
  we also observe a range in excitation at fixed redshift.  The increase in
  excitation is correlated with an increase in both \LIR\ and
  \SigmaSFR. \label{fig:LVG_props}}
\end{figure*}
The CO(3--2) selected galaxies at $z\ge2$ appear to have intrinsically higher
excitation, on average, than the CO(2--1) selected galaxies at $z<2$.  This
applies not only to the high-$J$ lines, but also for the excitation in
CO(3--2).  This observation is robust against the sample being CO
flux-selected; because the volume probed in CO(2--1) at $z<2$ is merely a
factor $1.75\times$ smaller, at least some sources with a similarly high
$r_{31}$ should have been found at $z<2$, if they are equally common at both
redshifts (such a high $r_{31}$ would be indicated by an as high $r_{21}$,
which is not suggested by the LVG modeling).

The increased excitation at $z\ge2$ compared to $z<2$ suggests an intrinsic
evolution between the ISM conditions in massive main-sequence galaxies at these
redshifts.  There are several reasons why more excited CO gas may be
anticipated in star-forming galaxies going out to higher redshift.
Star-forming galaxies at fixed stellar mass are known to decrease in size
\citep[as measured in the rest-frame optical][]{vanderWel2014}, while they
increase in average star formation rate \citep[e.g.,][]{Whitaker2014,
  Schreiber2015}.  This means that the star formation rate surface density
increases with redshift for main-sequence galaxies at fixed mass
\citep[e.g.,][]{Wuyts2013}, which drives the ambient radiation field.  Indeed,
there are indications that the dust temperature increases with redshift (e.g.,
\citealt{Magdis2012, Bethermin2015, Schreiber2018a}, but see
\citealt{Dudzeviciute2020}), which is linked to the main radiation field
intensity.  As discussed in \autoref{sec:modest-exc-high-j}, the CO excitation
is expected to increase with these quantities, as they can drive the density
and temperature in the clouds\footnote{\cite{Bolatto2015} point out that a
  stronger ambient radiation field only drives the low-$J$ excitation upwards
  if this emission does not arise in a colder, more extended molecular gas
  reservoir, but is well mixed with the star formation.  This is consistent
  with our data (see \autoref{sec:similar-line-widths}), but needs to be
  verified with higher resolution observations.}.

We compare the excitation to a range of properties, finding that the galaxies
with greater excitation at higher redshift indeed have both higher \LIR\ and,
more importantly, \SigmaSFR.  This behaviour is illustrated in
\autoref{fig:LVG_props}.  To quantify the increase with \SigmaSFR, we add the
LVG model predictions for the $r_{52}$ ratio to \autoref{fig:rJ1_SigmaSFR}, now
also including galaxies at $z=2-3$ for which we do not directly measure this
line ratio.  While there is substantial scatter for the individual models, they
broadly support the scenario of increasing excitation with \SigmaSFR.

The trend in \autoref{fig:rJ1_SigmaSFR} can also be understood more
fundamentally as a trend with molecular gas surface density, as a high surface
density of gas would also drive the CO excitation upwards\footnote{From a
  radiative transfer perspective the line ratio will increase with an
  increasing CO density per velocity gradient (i.e., $N_{\HH}/dv$ for a
  constant CO abundance), because this drives the line opacity and increases
  the line trapping and thereby the excitation.  Therefore, unless the high
  column density (gas surface density) is compensated by a linearly increasing
  turbulence ($dv$) one naturally expects an increasing excitation with
  increasing gas surface density.  In addition, it is plausible that higher
  column densities correlate with higher volume densities which will again
  drive the excitation upwards}.  In that context, it is interesting to note
that, several of the galaxies at $z\approx 2.5$ are found to have a more
compact dust distribution, compared to some of the sources at $z\approx 1.5$
\citep{Rujopakarn2019, Kaasinen2020}.

The difference in excitation between the CO(2-1) and CO(3-2)-selected samples
at $z\ge2$ and $z<2$ raises the question to what extent our $r_{21}$ and
$r_{31}$ are representative of the broader population of galaxies at these
redshifts.  Whereas the higher $r_{31}$ at $z\ge2$ could in principle be the
result of the CO-flux selection (see \autoref{sec:impl-flux-limit}), it appears
that at the current sensitivity ASPECS can pick up sources with similar gas
masses but with, for example, a factor $\sim 2\times$ lower excitation in
$r_{31}$ (see Fig.~9 in \citealt{Boogaard2019}).  The conclusions here are
limited by the fact we are limited by the low number of massive sources in the
volume in the first place (cf.\ Fig.~5 in \citealt{Boogaard2019})\footnote{Note
  that while we would, in principle, pick up sources with larger gas masses but
  lower excitation, this would require ASPECS to probe a larger volume at
  similar depth.  Initial efforts are made in this direction through WIDE
  ASPECS (Decarli et al., in prep.), a survey that covers approximately seven
  times the area of the ASPECS LP, albeit at a depth that is more shallow.}.
At $z<2$, we probe well below the knee of the CO LF, while at $z\ge2$, we are
on or slightly above the knee \citep[][]{Decarli2019}.  The same appears true
in the context of the IR LF \citep[e.g.,][]{Gruppioni2013}.  This suggests we
are probing the representative part in terms of the cosmic $\rhoHH$, in
particular at $z<2$.  Indeed, the individual detections are the dominant
contributors to the total $\rhoHH(z=2.0-3.1)$ \citep{Decarli2019} and not the
corrections for sources that fall below the detection limit.

\subsection{The low-$J$ excitation}
\label{sec:low-j-exc}
\subsubsection{Constraints on $r_{21}$ at $z=1.0-1.6$}
\label{sec:constraints-r21}
From our two-component LVG model we find $r_{21} = \rTwoOneLVGzLoTwo$ for the
CO(2--1) selected galaxies at $z=1.0 -1.6$.  This is in good agreement with the
average value of $r_{21} = 0.76 \pm 0.09$ for the three massive SFGs at $z=1.5$
\citep[][which are well described by a two-component model]{Daddi2015}.

On the other hand, comparing the dust luminosity at rest-frame 850\,\micron\ to
the CO(1--0) luminosity---which is inferred from the CO(2--1) line using this
value of $r_{21}$, because we do not have direct observations of CO(1--0) at
this redshift range---we find that the dust luminosity under-predicts that of
the gas.  If such a relation holds \citep{Scoville2016}, this may suggest that
the average excitation in CO(2--1) is higher (\autoref{fig:L850_LCO}).  Looking
in detail at the galaxies from \cite{Daddi2015}, two out of three galaxies have
a $r_{21}$ consistent with unity ($0.92 \pm 0.23$ and $1.02 \pm 0.20$), while
the average subthermal excitation is driven by the third source
($r_{21}=0.48 \pm 0.08$).  For comparison, in SMGs at higher redshift, the
average $r_{21}$ is often found to be close to, or consistent with unity:
$r_{21} = 0.84 \pm 0.13$ \citep{Bothwell2013} and $r_{21} = 1.11 \pm 0.08$
\cite[][i.e., suprathermal, though we caution this sample potentially suffers
from line-dependent differences in the lensing amplification]{Spilker2014},
though the number of SMGs with direct constraints on $r_{21}$ is still small
and significantly spread in redshift \citep{Carilli2010, Riechers2013,
  Aravena2016}.  If we would assume that our sources are on average better
described by a single component model we find a higher value of
$r_{21} = \rTwoOneLVGzLoOne$.  However, given that both the mid-$J$ excitation
and \LIR\ are lower than the BzK galaxies, this appears less likely.  The fact
that the single component model is formally consistent with the two-component
solution, as well as thermalized CO, highlights that we are considering
relatively small differences in excitation in the first place, compared to the
observational uncertainties.  In any case, as observations of CO(1--0) around
$z\approx1.2$ are impossible given the atmospheric opacity, detailed
characterizations of the multi-line CO ladders are key to make progress here.

It should be noted that there are several reasons why the comparison with the
dust-luminosity as a molecular gas tracer may break down in the first place.
If the mass-weighted dust temperature in our sources is higher compared to the
sample from \cite{Scoville2016} this would increase the dust-luminosity at
fixed gas mass, relieving the need for thermalized CO.  However, even if the
luminosity-weighted dust temperature varies, the mass-weighted dust temperature
will not vary so strongly \citep{Scoville2016}.  It is also not clear that our
galaxies would have a very different dust-opacity slope ($\beta$).  A
discrepancy between the dust and CO emission could also happen if the dust
emission is distinct from the CO emission (e.g., in the case of constant
gas-to-dust-ratio, but a strong dust temperature gradient, or opacity effects).
It is not clear that our galaxies would be very distinct from the calibration
sample in this respect.  However, we are probing a fainter regime in
$L_{\nu}(850\,\micron)$, where the calibration sample is mostly local, while
the sources at comparable redshifts are generally higher luminosity.
Furthermore, our data at 1.2\,mm and 3.0\,mm probes further down the
Rayleigh-Jeans tail than some of the earlier observations.  The Scoville
relation also breaks for galaxies with a strongly sub-solar metallicity and for
that reason \cite{Scoville2016} restrict their sample to galaxies with
$M_{*} \ge 2\times 10^{10}$\,\Msun.  However, the ASPECS galaxies are generally
more massive than this and have (super-)solar metallicities
\citep{Boogaard2019}.  Finally, we do not exclude the possibility that the
apparent discrepancy (on average) is driven by low number statistics, as the
majority of the sample is consistent within the intrinsic scatter in the
relation.

\subsubsection{Measurement of $r_{31}$ at $z=2.0-2.7$}
\label{sec:measurements-r31}
Stacking the CO(3--2)-selected galaxies at $\avg{z}=2.5$, we directly derive an
$r_{31} = \rThreeOnestack$, which is supported by the average value from the
LVG modeling of all individual sources at $z=2.0 - 2.7$
($r_{31} = \rThreeOneLVGzHiTwo$).  This value is significantly higher than
found in the lower redshift BzK-selected SFGs at $z=1.5$ (\citealt{Daddi2015};
$r_{31} = 0.42 \pm 0.15$; ranging from $0.27 - 0.57$), which has implications
for the measurement of the cosmic molecular gas density (we will come back to
this in \autoref{sec:impl-cosm-molec}).  Studying two massive main-sequence
galaxies at $z=2.3$ \cite{Bolatto2015} found higher ratios, consistent with
thermalized CO: $r_{31} = 0.92\pm0.11$ and $r_{31} = 1.17\pm 0.17$ (plus two
lower limits of $\ge 0.57$ and $\ge 0.79$).  The SMGs at $z=2$ show a wide
range of excitation values, as discussed in \cite{Riechers2020b}.  Early
studies found a relatively low average \citep[$r_{31} = 0.52 \pm 0.09$
][]{Ivison2011, Bothwell2013}, while more recently, \cite{Sharon2016} finds an
average of $r_{31} = 0.78\pm0.27$.  At higher redshift, \cite{Spilker2014}
reports $r_{31} = 0.87 \pm 0.06$, although these lensed sources are arguably
more extreme.  Overall, the different samples at $z=2-3$ show a significant
spread in their $r_{31}$ ratio (see also \citealt{Riechers2020b}), driven by
different selection methods picking up galaxies with different ambient
conditions in their ISM.  In that context, ASPECS provides a well-defined
sample for further investigation---flux-limited in CO(3--2) and followed up in
CO(1--0)---which probes fainter \LIR\ than the typical samples of SMGs.  A
contribution from the AGN is not expected to dominate the low-$J$ lines and we
do not find a clear correlation between $r_{31}$ and the presence of an X-ray
AGN.

\subsubsection{Consistency with the model results}
\label{sec:link-with-pdr}
The fairly high excitation in the low-$J$ lines is generally consistent with
the densities of $\ge 10^{4}$\,cm$^{-3}$ found in the (constant density) PDR
modeling of the low-$J$ CO and \CI\ lines (though we caution that the different
types of models should not be blindly compared, given the differences in
underlying assumptions).  From a radiative transfer perspective, it is rather
easy to excite CO(2--1), even at modest densities and temperatures, and
slightly less so for CO(3--2).  Note the effective floor on the gas temperature
at each redshift is set by the Cosmic Microwave Background, which measures
$T_{\rm CMB} = 6$\,K at $z=1.2$ and 10\,K at $z=2.5$.  For comparison, the
temperatures corresponding to the energy level differences for the (dominant)
$\Delta J=1$ collisional excitations are
$T_{1 \rightarrow 2} = (E_{2} - E_{1})/k_{B} = 11.1$\,K and
$T_{2 \rightarrow 3} = 16.6$\,K, respectively, where $k_{B}$ is the Boltzmann
constant.  As such, unless galaxies harbor extended low-excitation reservoirs
(addressed in \autoref{sec:similar-line-widths}), the levels of low-$J$
excitation found in this work are not unexpected.

\subsection{Broader implications of the flux-limited survey}
\label{sec:impl-flux-limit}
Because ASPECS is a flux-limited survey, without any target preselection, it
also provides additional information on the CO excitation in the whole
population of gas mass-selected galaxies, beyond just the detected sources.
The observed CO luminosity at different redshifts depends on the product of
$r_{J1} M_{\rm mol} \aco^{-1}$ (\autoref{eq:Mmol}) and hence three selection
effects are at play.  At a given redshift we would first detect the sources
with the highest gas mass (at fixed \aco) and the highest luminosity at a given
gas mass, i.e., the sources with the highest excitation in their low-$J$ lines.
Given that we detect approximately half of the massive main-sequence galaxies
at $z=1-3$ \citep{Boogaard2019}, this implies that the galaxies that we did not
detect will have a less massive gas reservoir (and/or higher \aco) and/or lower
CO excitation in the $J=2$ and $J=3$ levels.  For that reason, in particular
for galaxies towards the lower stellar mass and SFR end of the ASPECS sample at
a given redshift (i.e., the faint end of the survey), where we are less
complete, the average excitation of the low-$J$ levels may be lower.  By the
same argument, the fact that we do not detect any galaxies in the mid-/high-$J$
CO lines alone that are in principle detectable in the low-$J$ CO(2--1) or
CO(3--2) lines, implies that the excitation in their mid-/high-$J$ levels will
not be significantly above the detected samples at the respective redshifts,
for galaxies with comparable gas masses (at fixed \aco).

\subsection{Implications for the cosmic molecular gas density}
\label{sec:impl-cosm-molec}
By measuring the CO luminosity in galaxies without any preselection over a well
defined cosmic volume, ASPECS is conducting the deepest census of the cosmic
molecular gas density, $\rhoHH(z)$, to date \citep{Decarli2016a, Decarli2019,
  Decarli2020}.  This relies on the excitation corrections from the $J\ge2$
lines back to CO(1--0).  In the initial results from ASPECS, these have been
assumed to follow a single CO ladder, as measured for BzK-selected SFGs by
\cite{Daddi2015} at $\avg{z}=1.5$, up to CO(4--3)\footnote{\cite{Daddi2015} did
  not measure the excitation in CO(4--3), but interpolating their CO ladder
  yields $r_{41}=0.31 \pm 0.06$ \citep[see][]{Decarli2016b}.}, as these were
considered to be the closest analogs of the sources observed with ASPECS at the
time\footnote{The full range of results was considered to be bracketed between
  two extreme cases: Milky Way-like low excitation conditions and thermalized
  CO, see Appendix~B in \cite{Decarli2019}}.  With our study of the CO
excitation in the ASPECS galaxies---the actual sources that defined
$\rhoHH(z)$---we can now revisit these assumptions in more detail.

Our result that the average excitation increases between $z<2$ and $z\ge2$ has
important implications for $\rhoHH(z)$.  Our results support the earlier
assumptions for the excitation corrections at $z<2$.  Adopting the new CO
ladders (\autoref{tab:lvg_model}) does not significantly alter the constraints
on $\rhoHH$ at $z<2$, with the largest change being a 25\% decrease at
$z=0.7-1.2$ (based on $r_{41}$).  In contrast, the significantly higher
excitation at $z\ge2$ implies a factor $2\times$ decrease in \rhoHH\, compared
to earlier results, for CO(3--2) at $z=2.0-3.1$ (see also
\citealt{Riechers2020b}) and CO(4--3) at $z=3.0-4.5$.  It should be noted that
we currently do lack direct constraints on the excitation for CO(4--3)-selected
samples at $z=3.0-4.5$.  However, based on the results from this paper, we do
not expect the average excitation for the sources contributing to the
measurement of \rhoHH\ to be lower than at $z=2.5$ (and certainly not as low as
in \citealt{Daddi2015}).  Note that this decrease is in line with the models
underpredicting the earlier measurements of $\rhoHH(z>2)$
\citep[e.g.,][]{Popping2019}.

In summary, we make new recommendations for the average CO ladders to be used
for the measurement of the cosmic molecular gas density (the two-component
models from \autoref{tab:lvg_model}).  The constraints on $\rhoHH(z)$ using the
new excitation corrections are presented and discussed in \cite{Decarli2020}.
Our results, combined with those of \cite{Riechers2020b}, show that direct
measurement of the CO(1--0) transition (where accessible) as well as
constructing more complete CO ladders, in order to characterize the CO
excitation and physical conditions in the cold ISM, are essential to make
progress in further constraining the cosmic molecular gas density.
\newpage
\section{Summary and Conclusions}
\label{sec:conclusions}
This paper presents a study of the carbon monoxide (CO) excitation, atomic
carbon (\CI) emission and interstellar medium (ISM) conditions in a sample of
22 star-forming galaxies at $z=0.45 - 3.60$.  These galaxies have been observed
as part of the ALMA Spectroscopic Survey in the Hubble Ultra Deep Field
(ASPECS) Large Program, designed to provide a cosmic inventory of molecular gas
by selecting galaxies purely by their CO and dust-continuum emission in ALMA
Band 3 and 6, without any preselection.  These galaxies are known to lie on,
above and below the main sequence of star-forming galaxies at their respective
redshifts \citep{Boogaard2019, Aravena2019, Aravena2020}.  We detect a total of
34 CO $J\rightarrow J-1$ lines with $J=1$ up to 8 (+ 21 upper limits, up to
$J=10$) and six $\CI$ ${^{3}P_{1}} \rightarrow {^{3}P_{0}}$ and
${^{3}P_{2}} \rightarrow {^{3}P_{1}}$ lines (+ 12 upper limits).  This includes
follow-up observations of seven sources at $z=1.99-2.70$ in CO(1--0) from
VLASPECS \citep{Riechers2020b}, that we analyze here in concert with the ASPECS
data.

The ASPECS galaxies have lower infrared luminosities (\LIR) and star formation
rate surface densities (\SigmaSFR) than earlier, targeted samples of
star-forming galaxies and sub-millimeter galaxies (including lensed samples) at
similar redshifts \citep{Daddi2015, Bothwell2013, Spilker2014}.  We study the
CO excitation of the CO(2--1) and CO(3--2) selected samples and compare them to
the average CO ladders of the targeted samples.  We focus on two redshift bins,
$\avg{z}=1.2$ and $\avg{z}=2.5$, at which we cover both a low/mid-$J$ CO
transition and a mid/high-$J$ CO transition with ASPECS.

We find that half of the galaxies at $\avg{z}=1.2$ show remarkably similar
excitation, up to CO(5--4), similar to that observed in a sample of four
BzK-color-selected star-forming galaxies at $\avg{z}=1.5$ \citep{Daddi2015},
while the remaining sources are consistent with lower excitation.  The range in
excitation suggests that a warm and/or dense component, indicated by the higher
excitation, is not omnipresent in galaxies at these redshifts.  We detect the
high-$J\ge6$ lines in several galaxies at $\avg{z}=2.5$, indicating that the
high-$J$ excitation is comparable to the levels in local starbursts and
slightly lower than SMGs at similar redshifts \citep{Bothwell2013}, although
half of the sources selected by their CO(3--2) emission are not detected in
their high-$J$ lines.

Stacking all the CO and \CI\ transitions that we cover with ASPECS (including
non-detections), we find our galaxies at $\avg{z}=1.2$ show, on average, lower
excitation than BzK-selected galaxies.  This is consistent with a picture in
which the CO excitation is driven by the star formation rate surface density
\SigmaSFR, broadly matching model predictions (although the models do not fully
reproduce our observations).  For the galaxies at $\avg{z}=2.5$, the stacking
reveals an average $r_{31} = \rThreeOnestack$ and $r_{71} = \rSevenOnestack$,
broadly comparable to SMGs at this epoch, as well as local starburst galaxies

We present the average excitation corrections for cold gas mass-selected
galaxies at $z=1.0-1.6$ and $z=2.0-2.7$, based on the interpolation of the CO
ladders using (single- and) two-component Large Velocity Gradient models.
These models predict $r_{21} = \rTwoOneLVGzLoTwo$ at $z<2$, similar to the
BzK-selected SFGs \citep{Daddi2015}.

We place our sources on the empirical correlations between
$L'_{\mathrm{CO}(1-0)}$ and dust luminosity at rest-frame 850\,\micron, probing
significantly lower $L_{\nu}(850\,\micron)$ than the earlier samples at $z>0$,
and find good agreement for the CO(3--2)-selected sources.  However, we find
that the dust luminosity on average overpredicts the CO(1--0) luminosity for
the CO(2--1)-selected sample.  This either implies that the average $r_{21}$ at
$\avg{z}=1.2$ is higher, or that the assumptions going into the correlation
break down for these sources.

Comparing our \CI(1--0) and \CI(2--1) observations to the literature, we find
that the $L_{\CI}/\LIR$ ratio of our sample is similar to main-sequence
galaxies, as observed by \cite{Valentino2018}.  We find an average neutral
atomic carbon abundance of $\CIHH = \avgCIabundCO$.  This is
comparable to the abundance measured in the main-sequence galaxies and the
Milky Way, but lower than what is measured in SMGs (although this apparent
discrepancy is degenerate with the assumption of a different \aco;
\citealt{Valentino2018}).  Modeling the CO, \CI\ and \LIR\ emission using the
\textsc{PDRToolbox} indicates densities $\ge 10^{4}$~cm$^{-3}$, generally
consistent with the (fairly high) excitation in the low-$J$ lines.

The interpolated CO ladders suggest that the intrinsic excitation is higher for
the sources at $z\ge2$ compared to $z<2$, even in the lower-$J$ lines such as
CO(3--2).  The excitation difference is robust against the ASPECS selection
function and correlated with \LIR\ and \SigmaSFR.  This implies an intrinsic
evolution in the ISM conditions of massive star-forming galaxies between these
redshifts, which we link to an increase in the surface density of star
formation (and gas) in star-forming galaxies with redshift.

Because ASPECS is a flux-limited survey, it also provides additional
information on the CO excitation in the whole population of gas mass-selected
galaxies.  Being most sensitive to galaxies with the highest excitation at a
given gas mass (at fixed \aco), this suggests that the average excitation in
sources with comparable gas masses (at fixed \aco) may be lower towards the
faint end of the survey.  At the same time, the non-detection of galaxies in
their mid-/high-$J$ alone (which are in principle detectable in their low-$J$
lines), implies that the average excitation is not much higher.

The galaxies studied in this paper are the same as those constraining the CO
luminosity function and the cosmic molecular gas density, $\rhoHH$, as measured
by ASPECS.  The increased excitation in the CO-selected galaxies at $z \ge 2$
compared to those at $z<2$ implies a decrease in the inferred $\rhoHH(z\ge2)$
compared to earlier measurements \citep{Decarli2016a, Decarli2019}.  We make
recommendations for the average CO excitation in CO-flux-limited samples of
galaxies, to be adopted in the constraints on $\rhoHH(z)$ from the complete
ASPECS survey, presented in \cite{Decarli2020}.

The observations presented here have extended the sample of star-forming
galaxies at $z=1-3$ with constraints on their CO excitation and atomic carbon
emission.  As these are the same galaxies through which the CO luminosity
function is measured, characterizing them in detail is key to further our
constraints on the cosmic molecular gas density.  Further study of such
well-defined (flux-limited) samples with multi-line observations will be
instrumental to gain a complete picture of the ISM conditions in star-forming
galaxies across cosmic time.

\acknowledgments

We would like to thank the anonymous referee for an insightful and constructive
report.  The author is also grateful to Francesco Valentino for providing early
access to the data from \cite{Valentino2020b}, that appeared on the arXiv while
this paper was under review.  This paper makes use of the following ALMA data:
ADS/JAO.ALMA\#2016.1.00324.L.  ALMA is a partnership of ESO (representing its
member states), NSF (USA) and NINS (Japan), together with NRC (Canada), NSC and
ASIAA (Taiwan), and KASI (Republic of Korea), in cooperation with the Republic
of Chile. The Joint ALMA Observatory is operated by ESO, AUI/NRAO and NAOJ.
The National Radio Astronomy Observatory is a facility of the National Science
Foundation operated under cooperative agreement by Associated Universities,
Inc.  FW acknowledges support from ERC grant 740246 (Cosmic\_Gas).  Este
trabajo cont \'o con el apoyo de CONICYT + PCI + INSTITUTO MAX PLANCK DE
ASTRONOMIA MPG190030.  DR acknowledges support from the National Science
Foundation under grant numbers AST-1614213 and AST-1910107.  DR also
acknowledges support from the Alexander von Humboldt Foundation through a
Humboldt Research Fellowship for Experienced Researchers.  IRS acknowledges
support from STFC (ST/T000244/1).  MK acknowledges support from the
International Max Planck Research School for Astronomy and Cosmic Physics at
Heidelberg University (IMPRS-HD).  T.D-S. acknowledges support from the CASSACA
and CONICYT fund CAS-CONICYT Call 2018.  HI acknowledges support from JSPS
KAKENHI Grant Number JP19K23462.

\facilities{ALMA, VLA, VLT:Yepun (MUSE)}

\software{\textsc{Topcat} \citep{Taylor2005}, \textsc{Gnuastro}
  \citep{Akhlaghi2015}, \textsc{IPython} \citep{Perez2007}, \textsc{numpy}
  \citep{VanDerWalt2011}, \textsc{Matplotlib} \citep{Hunter2007},
  \textsc{Astropy} \citep{Robitaille2013, TheAstropyCollaboration2018}.}

\appendix
\section{Similar widths for the low-$J$ and high $J$ CO lines}
\label{sec:similar-line-widths}
Previous studies have suggested that some SMGs at $z=2-4$ have line widths in
CO(1--0) that are larger than in the higher-$J$ transitions
\citep[e.g.,][although the difference is rather subtle, about $\sim 15$\%,
which is within the limits of our data]{Ivison2011}.  Together with the
observation that the excitation models to the high-$J$ CO lines underpredicted
the total molecular gas mass in these sources \citep[by a factor of
$\sim1.5 - 4.5$;][]{Riechers2011a}, this suggested the presence of extended low
excitation gas reservoirs in some SMGs, but, notably, not in all cases
\citep[e.g.,][]{Riechers2011b, Hodge2012}.  If there would be extended emission
in CO(1--0), this complicates estimates of total molecular gas mass from the
higher-$J$ lines.

We find that the line widths of the different CO and \CI\ lines are consistent
(see \autoref{fig:line-widths}), including the CO(1--0) lines.  There is one
outlier, 3mm.12, which has a potential low-S/N broad component that should be
confirmed by future observations (\citealt{Riechers2020b}; this line has
$\mathrm{S/N}<3$ in the fit where the line widths are tied together).  The
strong CO luminosity relative to the dust emission (see
\autoref{sec:dust-cont-vers}), even when assuming thermalized CO, suggests that
we are not missing a large volume of molecular gas in CO(2--1) and CO(3--2)
that would be probed by the dust.  Furthermore, from an excitation perspective
it is very unlikely to have gas that radiates purely in CO(1--0) and not at all
in CO(2--1), which is only attainable at very low \nHH\ and \Tkin.  Looking at
other SFGs at the same redshift, in the BzK-selected galaxies at $z=1.5$
\citep{Daddi2015} the line widths are also found to be very similar between
CO(2--1) \citep{Daddi2010} and CO(1--0) \citep{Aravena2014} (although the
errors on the latter are significant).  Similarly, \cite{Bolatto2015} found
consistent line widths and spatial extent between CO(3--2) and CO(1--0) in two
massive main-sequence galaxies at $z=2.3$.  In summary, while we cannot
conclusively rule out their presence with the current observations, we do not
see clear evidence of a large volumes of cold molecular gas that are not traced
by the relatively low-$J$ CO lines.  This supports the use of these transitions
in inferring the molecular gas mass.

\section{Spectral line fits}
\label{sec:spectral-line-fits}
Gaussian fits to the spectral lines of CO and \CI, performed as detailed in
\autoref{sec:spectr-line-fitt}, are shown in \autoref{fig:slfit}.  The best-fit
parameters are reported in \autoref{tab:line-fluxes}.  For each source, we fit
a single redshift and line width for all lines simultaneously.

\begin{figure*}[t]
  \centering
  \includegraphics[height=.95\textheight]{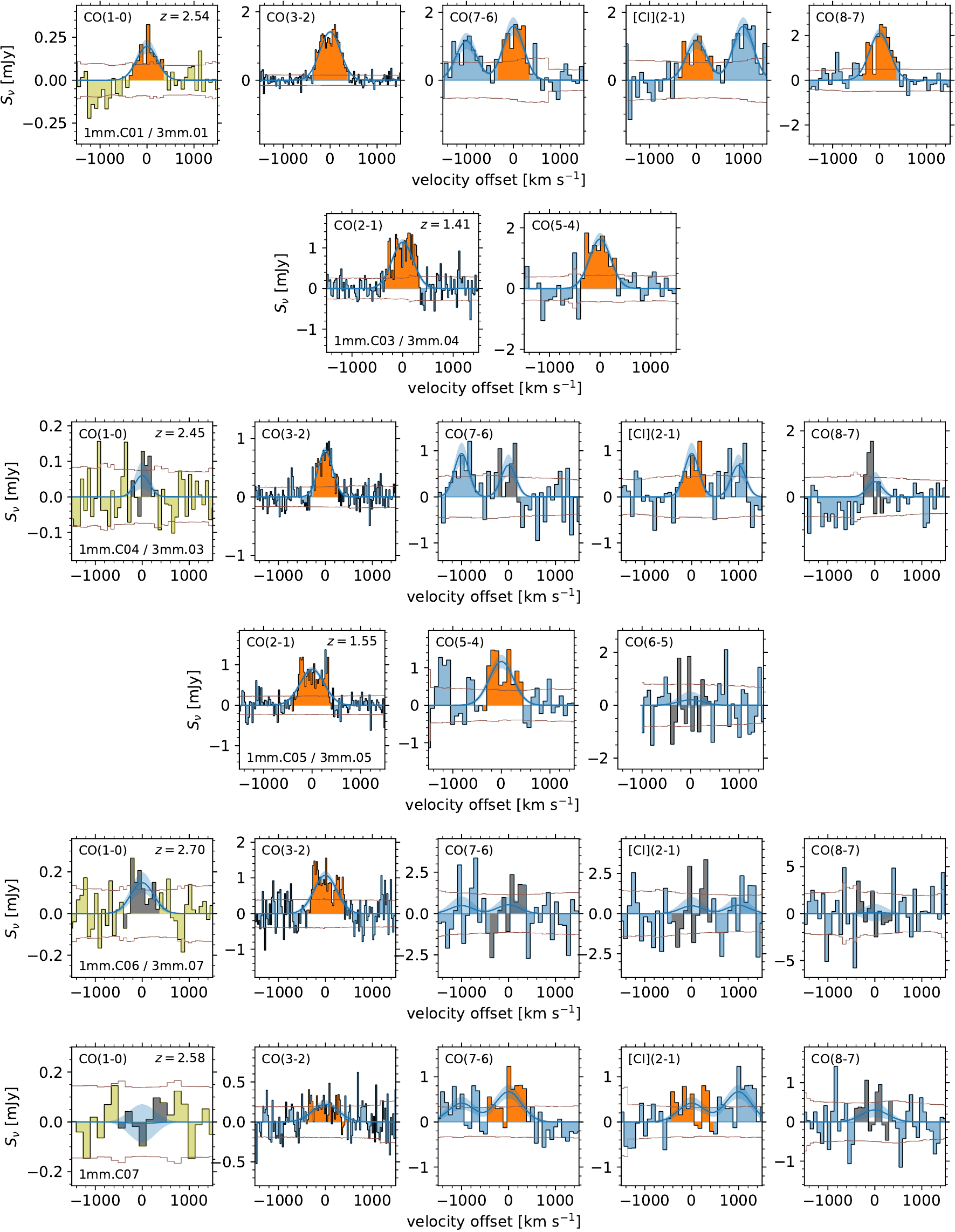}
  \caption{Gaussian fits to the $^{12}$CO and \CI\ lines in the ASPECS
    galaxies.  The groups of panels (max 2 per row) show the different
    transitions (indicated top left) in a single galaxy (identified in the
    bottom left of the leftmost panel).  The spectra are shown in blue (ASPECS)
    and yellow (VLA) and are binned in the Ka band and Band 6 for visualisation
    purposes (except for 3mm.08 and 3mm.11 with very narrow lines).  The brown
    line shows the $\pm 1 \sigma$ root-mean-square noise.  The best-fit for all
    lines (tied together in redshift and line width) and a $1 \sigma$
    confidence interval are shown by the blue line and shading.  The channels
    indicated in orange (grey) fall within $1.4 \times {\rm FWHM}$ (i.e., 90\%
    of the flux) for a detection (non-detection).
    \label{fig:slfit}}
\end{figure*}
\begin{figure*}[t]
  \figurenum{\ref{fig:slfit}}
  \centering
  \includegraphics[height=\textheight]{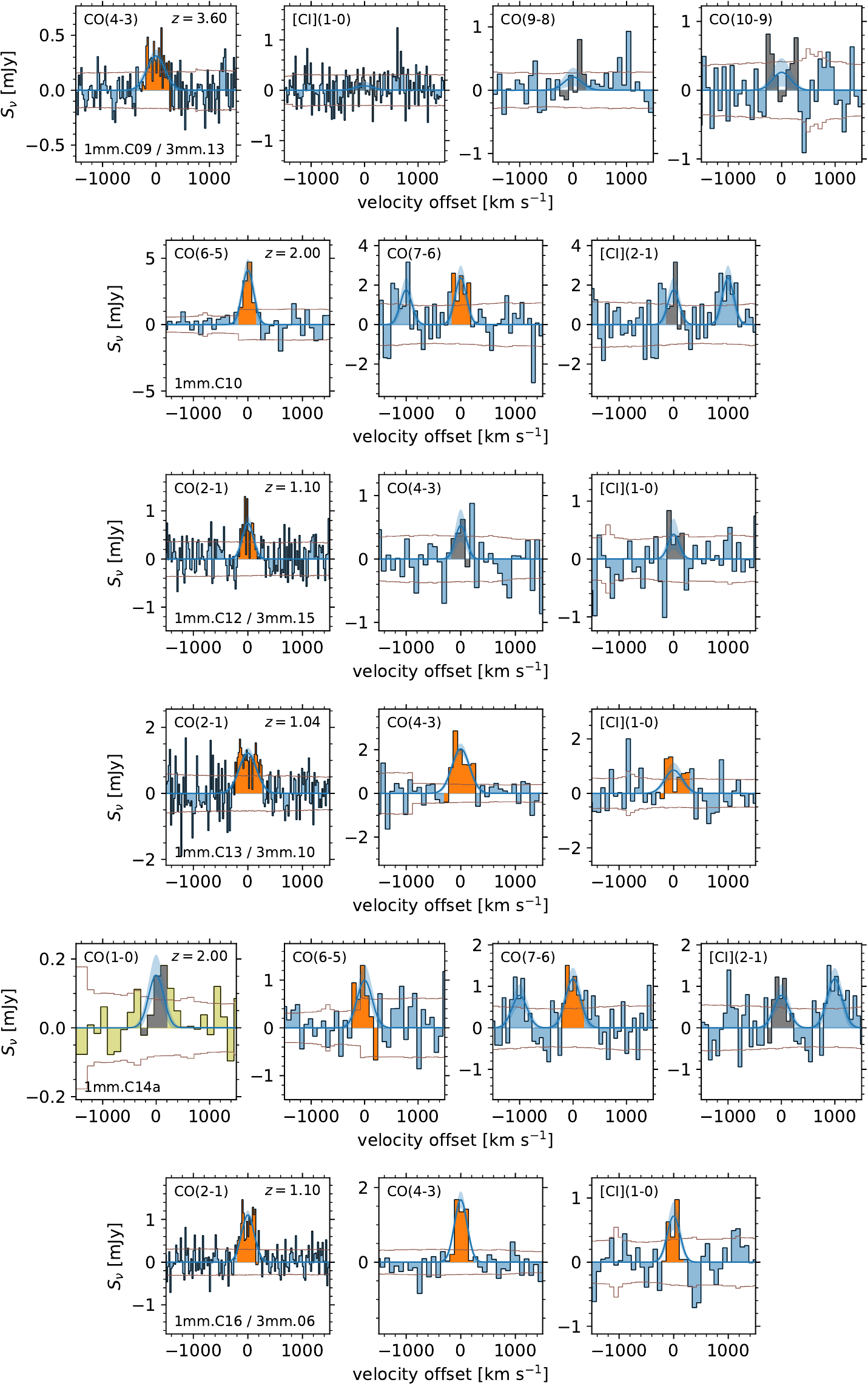}
\caption{\emph{(continued)}}
\end{figure*}
\begin{figure*}[t]
  \figurenum{\ref{fig:slfit}}
  \centering
  \includegraphics[height=\textheight]{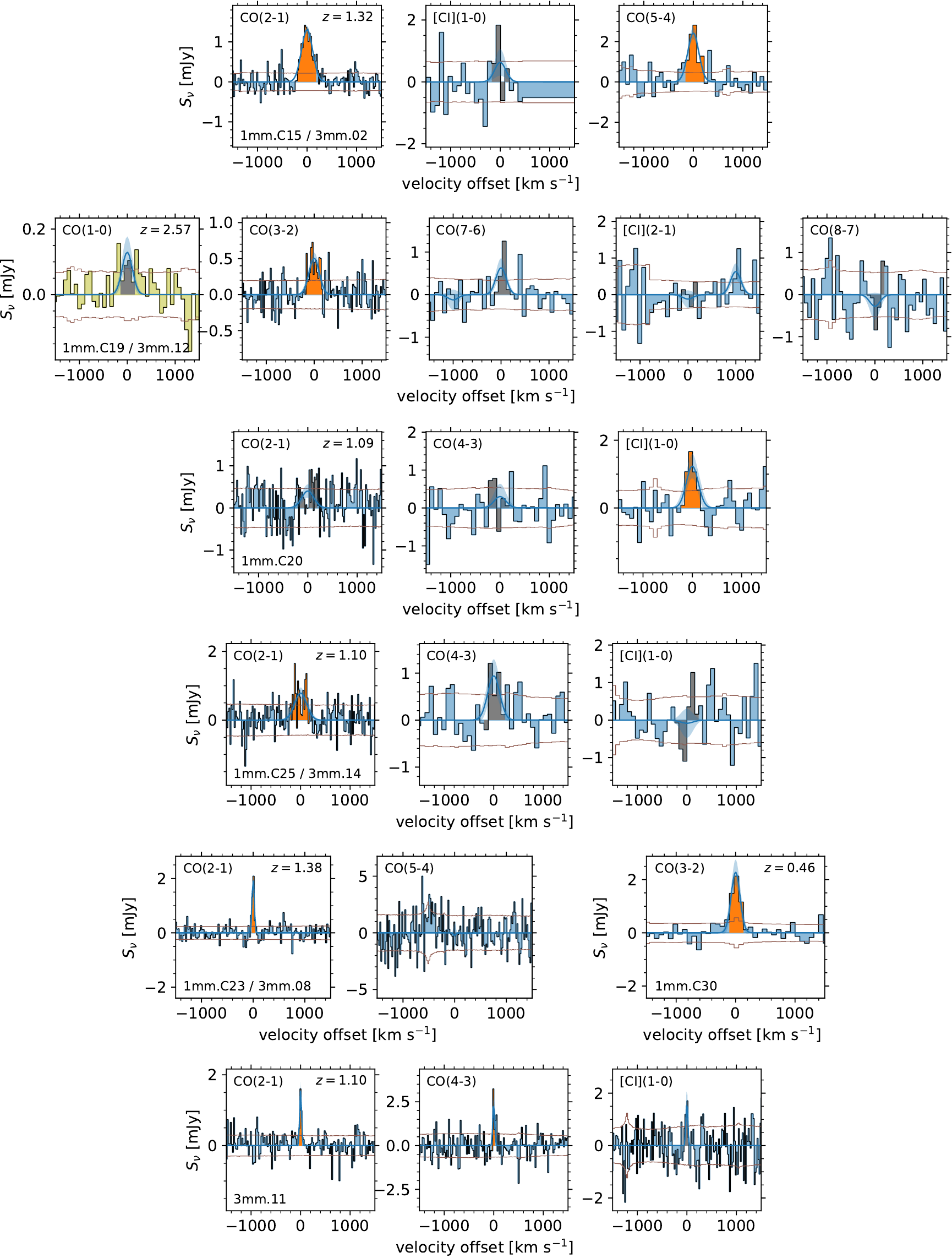}
\caption{\emph{(continued)}}
\end{figure*}
\begin{figure*}[t]
  \figurenum{\ref{fig:slfit}}
  \centering
  \includegraphics[height=0.5\textheight]{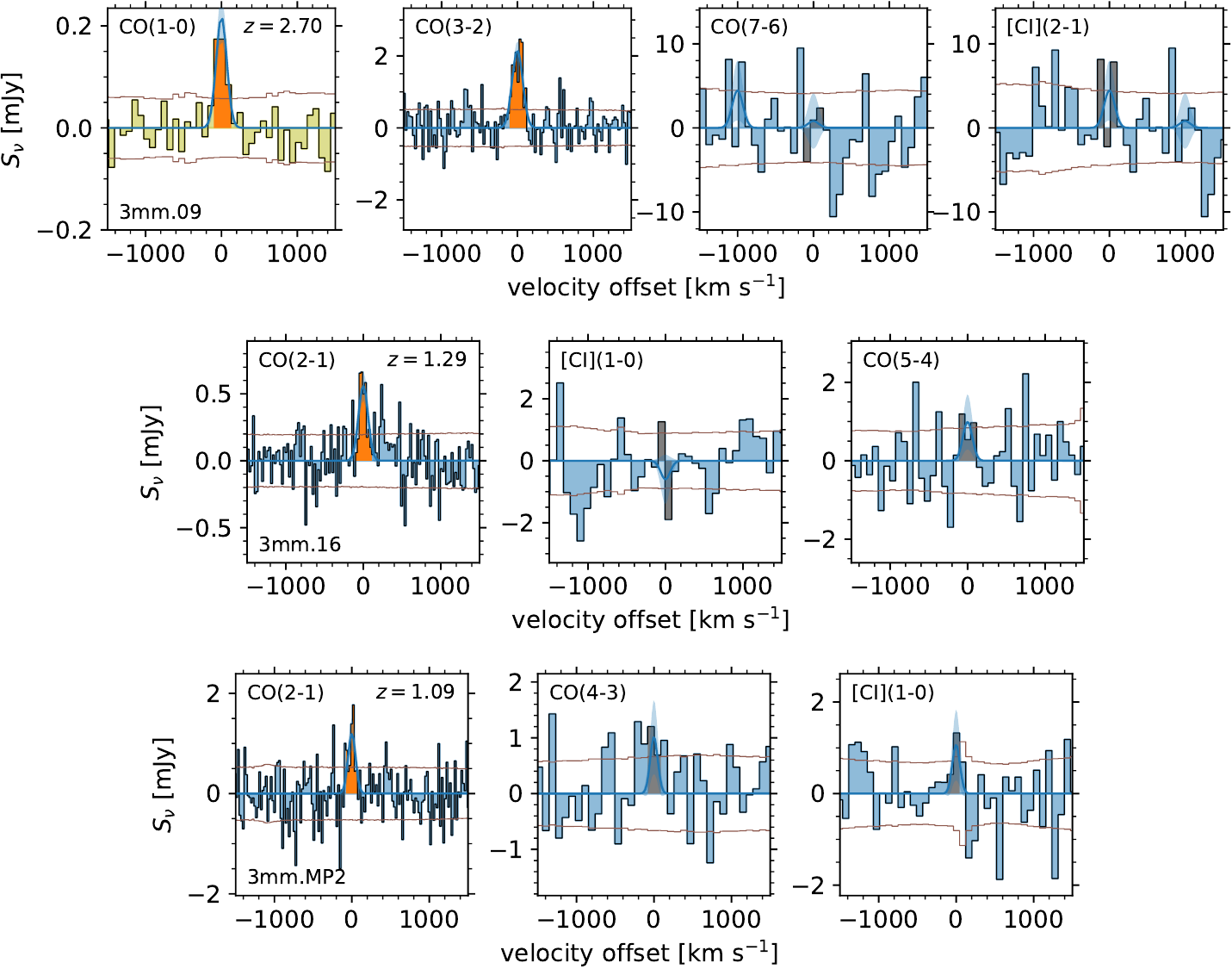}
\caption{\emph{(continued)}}
\end{figure*}

\begin{longrotatetable}
\begin{deluxetable*}{cccccccccc}
  \tablecaption{Spectral line properties determined with Gaussian fits
    \label{tab:line-fluxes}} \tablehead{ \colhead{ID 1mm} & \colhead{ID
      3mm} & \colhead{$z$} & \colhead{FWHM} & \colhead{Line} &
    \colhead{Frequency}           & \colhead{$S^{V}$} & \colhead{$L'$} &\colhead{$r_{J1}$} & \colhead{$r_{J2}$}\\
    \colhead {} & \colhead {} & \colhead{} & \colhead{(km\,s$^{-1}$)} &
    \colhead{} & \colhead{(GHz)} & \colhead{(Jy\,km\,s$^{-1}$)} &
    \colhead{($\times 10^{9}$\,K\,km\,s$^{-1}$\,pc$^{2}$)} &\colhead{} &
    \colhead{} } \colnumbers \startdata
1mm.C01 & 3mm.01 & 2.5437 $\pm$ 0.0001 & 518 $\pm$ 18 & CO(3--2) & 97.582 $\pm$ 0.002 & 0.78 $\pm$ 0.02 & 25.9 $\pm$ 0.8 & 0.79 $\pm$ 0.17 & \nodata \\
 &  &  &  & CO(7--6) & 227.633 $\pm$ 0.006 & 0.88 $\pm$ 0.15 & 5.4 $\pm$ 0.9 & 0.16 $\pm$ 0.04 & \nodata \\
 &  &  &  & \CI(2--1) & 228.392 $\pm$ 0.006 & 0.64 $\pm$ 0.13 & 3.9 $\pm$ 0.8 & \nodata & \nodata \\
 &  &  &  & CO(8--7) & 260.127 $\pm$ 0.007 & 1.16 $\pm$ 0.12 & 5.4 $\pm$ 0.5 & 0.17 $\pm$ 0.04 & \nodata \\
 &  &  &  & CO(1--0) & 32.529 $\pm$ 0.001 & 0.11 $\pm$ 0.02 & 32.7 $\pm$ 6.8 & \nodata & \nodata \\
1mm.C03 & 3mm.04 & 1.4142 $\pm$ 0.0001 & 498 $\pm$ 39 & CO(2--1) & 95.491 $\pm$ 0.005 & 0.60 $\pm$ 0.04 & 15.8 $\pm$ 1.2 & \nodata & \nodata \\
 &  &  &  & CO(5--4) & 238.697 $\pm$ 0.013 & 0.86 $\pm$ 0.11 & 3.6 $\pm$ 0.5 & \nodata & 0.23 $\pm$ 0.03 \\
1mm.C04 & 3mm.03 & 2.4535 $\pm$ 0.0002 & 367 $\pm$ 31 & CO(3--2) & 100.130 $\pm$ 0.004 & 0.31 $\pm$ 0.02 & 9.7 $\pm$ 0.7 & 1.46 $\pm$ 0.91 & \nodata \\
 &  &  &  & CO(7--6) & 233.577 $\pm$ 0.010 & 0.27 $\pm$ 0.09 & 1.6 $\pm$ 0.5 & 0.23 $\pm$ 0.16 & \nodata \\
 &  &  &  & \CI(2--1) & 234.356 $\pm$ 0.010 & 0.37 $\pm$ 0.09 & 2.1 $\pm$ 0.5 & \nodata & \nodata \\
 &  &  &  & CO(8--7) & 266.920 $\pm$ 0.012 & 0.18 $\pm$ 0.12 & 0.8 $\pm$ 0.5 & 0.12 $\pm$ 0.11 & \nodata \\
 &  &  &  & CO(1--0) & 33.378 $\pm$ 0.002 & 0.02 $\pm$ 0.01 & 6.7 $\pm$ 4.2 & \nodata & \nodata \\
1mm.C05 & 3mm.05 & 1.5503 $\pm$ 0.0002 & 587 $\pm$ 48 & CO(2--1) & 90.397 $\pm$ 0.006 & 0.55 $\pm$ 0.04 & 17.2 $\pm$ 1.3 & \nodata & \nodata \\
 &  &  &  & CO(5--4) & 225.962 $\pm$ 0.015 & 0.73 $\pm$ 0.11 & 3.6 $\pm$ 0.6 & \nodata & 0.21 $\pm$ 0.04 \\
 &  &  &  & CO(6--5) & 271.135 $\pm$ 0.018 & 0.13 $\pm$ 0.20 & 0.4 $\pm$ 0.7 & \nodata & \nodata \\
1mm.C06 & 3mm.07 & 2.6956 $\pm$ 0.0004 & 572 $\pm$ 68 & CO(3--2) & 93.570 $\pm$ 0.009 & 0.65 $\pm$ 0.07 & 24.1 $\pm$ 2.6 & 0.81 $\pm$ 0.28 & \nodata \\
 &  &  &  & CO(7--6) & 218.273 $\pm$ 0.021 & 0.32 $\pm$ 0.31 & 2.2 $\pm$ 2.1 & 0.07 $\pm$ 0.07 & \nodata \\
 &  &  &  & \CI(2--1) & 219.001 $\pm$ 0.021 & 0.29 $\pm$ 0.34 & 2.0 $\pm$ 2.3 & \nodata & \nodata \\
 &  &  &  & CO(8--7) & 249.431 $\pm$ 0.024 & 0.06 $\pm$ 0.57 & 0.3 $\pm$ 2.9 & 0.01 $\pm$ 0.10 & \nodata \\
 &  &  &  & CO(1--0) & 31.191 $\pm$ 0.003 & 0.09 $\pm$ 0.03 & 29.9 $\pm$ 10.1 & \nodata & \nodata \\
1mm.C07 & \nodata & 2.5805 $\pm$ 0.0006 & 658 $\pm$ 109 & CO(3--2) & 96.577 $\pm$ 0.017 & 0.16 $\pm$ 0.03 & 5.3 $\pm$ 1.1 & -4.19 $\pm$ 54.58 & \nodata \\
 &  &  &  & CO(7--6) & 225.288 $\pm$ 0.040 & 0.47 $\pm$ 0.10 & 2.9 $\pm$ 0.6 & -2.31 $\pm$ 30.13 & \nodata \\
 &  &  &  & \CI(2--1) & 226.040 $\pm$ 0.040 & 0.29 $\pm$ 0.10 & 1.8 $\pm$ 0.6 & \nodata & \nodata \\
 &  &  &  & CO(8--7) & 257.448 $\pm$ 0.046 & 0.22 $\pm$ 0.13 & 1.0 $\pm$ 0.6 & -0.82 $\pm$ 10.63 & \nodata \\
 &  &  &  & CO(1--0) & 32.194 $\pm$ 0.006 & -0.00 $\pm$ 0.05 & -1.3 $\pm$ 16.5 & \nodata & \nodata \\
1mm.C09 & 3mm.13 & 3.6008 $\pm$ 0.0005 & 397 $\pm$ 80 & CO(4--3) & 100.208 $\pm$ 0.011 & 0.13 $\pm$ 0.02 & 4.4 $\pm$ 0.8 & \nodata \\
 &  &  &  & \CI(1--0) & 106.972 $\pm$ 0.012 & 0.04 $\pm$ 0.04 & 1.1 $\pm$ 1.0 & \nodata & \nodata \\
 &  &  &  & CO(9--8) & 225.375 $\pm$ 0.026 & 0.09 $\pm$ 0.06 & 0.6 $\pm$ 0.4 & \nodata & \nodata \\
 &  &  &  & CO(10--9) & 250.386 $\pm$ 0.028 & 0.11 $\pm$ 0.09 & 0.6 $\pm$ 0.5 & \nodata & \nodata \\
1mm.C10 & \nodata & 1.9975 $\pm$ 0.0002 & 248 $\pm$ 44 & CO(6--5) & 230.682 $\pm$ 0.015 & 1.09 $\pm$ 0.22 & 5.9 $\pm$ 1.2 & \nodata & \nodata \\
 &  &  &  & CO(7--6) & 269.106 $\pm$ 0.018 & 0.63 $\pm$ 0.17 & 2.5 $\pm$ 0.7 & \nodata & \nodata \\
 &  &  &  & \CI(2--1) & 270.004 $\pm$ 0.018 & 0.47 $\pm$ 0.17 & 1.9 $\pm$ 0.7 & \nodata & \nodata \\
1mm.C12 & 3mm.15 & 1.0963 $\pm$ 0.0001 & 235 $\pm$ 50 & CO(2--1) & 109.975 $\pm$ 0.008 & 0.19 $\pm$ 0.04 & 3.0 $\pm$ 0.6 & \nodata & \nodata \\
 &  &  &  & CO(4--3) & 219.934 $\pm$ 0.016 & 0.13 $\pm$ 0.06 & 0.5 $\pm$ 0.3 & \nodata & 0.18 $\pm$ 0.09 \\
 &  &  &  & \CI(1--0) & 234.779 $\pm$ 0.016 & 0.11 $\pm$ 0.07 & 0.4 $\pm$ 0.2 & \nodata & \nodata \\
1mm.C13 & 3mm.10 & 1.0364 $\pm$ 0.0001 & 380 $\pm$ 42 & CO(2--1) & 113.207 $\pm$ 0.007 & 0.49 $\pm$ 0.06 & 7.1 $\pm$ 0.9 & \nodata & \nodata \\
 &  &  &  & CO(4--3) & 226.396 $\pm$ 0.013 & 0.82 $\pm$ 0.10 & 3.0 $\pm$ 0.4 & \nodata & 0.42 $\pm$ 0.07 \\
 &  &  &  & \CI(1--0) & 241.678 $\pm$ 0.014 & 0.35 $\pm$ 0.11 & 1.1 $\pm$ 0.4 & \nodata & \nodata \\
1mm.C14a & \nodata & 1.9966 $\pm$ 0.0003 & 332 $\pm$ 64 & CO(6--5) & 230.755 $\pm$ 0.021 & 0.35 $\pm$ 0.11 & 1.9 $\pm$ 0.6 & 0.18 $\pm$ 0.09 & \nodata \\
 &  &  &  & CO(7--6) & 269.191 $\pm$ 0.025 & 0.42 $\pm$ 0.10 & 1.7 $\pm$ 0.4 & 0.16 $\pm$ 0.07 & \nodata \\
 &  &  &  & \CI(2--1) & 270.089 $\pm$ 0.025 & 0.28 $\pm$ 0.09 & 1.1 $\pm$ 0.4 & \nodata & \nodata \\
 &  &  &  & CO(1--0) & 38.468 $\pm$ 0.004 & 0.05 $\pm$ 0.02 & 10.7 $\pm$ 4.2 & \nodata & \nodata \\
1mm.C16 & 3mm.06 & 1.0952 $\pm$ 0.0001 & 275 $\pm$ 27 & CO(2--1) & 110.029 $\pm$ 0.004 & 0.32 $\pm$ 0.03 & 5.2 $\pm$ 0.5 & \nodata & \nodata \\
 &  &  &  & CO(4--3) & 220.042 $\pm$ 0.008 & 0.49 $\pm$ 0.07 & 2.0 $\pm$ 0.3 & \nodata & 0.38 $\pm$ 0.06 \\
 &  &  &  & \CI(1--0) & 234.894 $\pm$ 0.009 & 0.21 $\pm$ 0.06 & 0.7 $\pm$ 0.2 & \nodata & \nodata \\
1mm.C15 & 3mm.02 & 1.3167 $\pm$ 0.0001 & 266 $\pm$ 19 & CO(2--1) & 99.512 $\pm$ 0.003 & 0.37 $\pm$ 0.03 & 8.4 $\pm$ 0.6 & \nodata & \nodata \\
 &  &  &  & \CI(1--0) & 212.442 $\pm$ 0.006 & 0.18 $\pm$ 0.12 & 0.9 $\pm$ 0.6 & \nodata & \nodata \\
 &  &  &  & CO(5--4) & 248.747 $\pm$ 0.007 & 0.69 $\pm$ 0.08 & 2.5 $\pm$ 0.3 & \nodata & 0.30 $\pm$ 0.04 \\
1mm.C19 & 3mm.12 & 2.5739 $\pm$ 0.0002 & 254 $\pm$ 43 & CO(3--2) & 96.756 $\pm$ 0.006 & 0.14 $\pm$ 0.02 & 4.6 $\pm$ 0.8 & 0.43 $\pm$ 0.18 & \nodata \\
 &  &  &  & CO(7--6) & 225.707 $\pm$ 0.014 & 0.17 $\pm$ 0.06 & 1.1 $\pm$ 0.4 & 0.10 $\pm$ 0.05 & \nodata \\
 &  &  &  & \CI(2--1) & 226.460 $\pm$ 0.014 & -0.03 $\pm$ 0.06 & -0.2 $\pm$ 0.4 & \nodata & \nodata \\
 &  &  &  & CO(8--7) & 257.926 $\pm$ 0.016 & -0.07 $\pm$ 0.09 & -0.4 $\pm$ 0.4 & -0.03 $\pm$ 0.04 & \nodata \\
 &  &  &  & CO(1--0) & 32.254 $\pm$ 0.002 & 0.04 $\pm$ 0.01 & 10.7 $\pm$ 4.0 & \nodata & \nodata \\
1mm.C20 & \nodata & 1.0931 $\pm$ 0.0002 & 287 $\pm$ 83 & CO(2--1) & 110.140 $\pm$ 0.013 & 0.12 $\pm$ 0.05 & 2.0 $\pm$ 0.8 & \nodata & \nodata \\
 &  &  &  & CO(4--3) & 220.264 $\pm$ 0.026 & 0.09 $\pm$ 0.10 & 0.4 $\pm$ 0.4 & \nodata & 0.19 $\pm$ 0.22 \\
 &  &  &  & \CI(1--0) & 235.131 $\pm$ 0.028 & 0.38 $\pm$ 0.11 & 1.3 $\pm$ 0.4 & \nodata & \nodata \\
1mm.C25 & 3mm.14 & 1.0982 $\pm$ 0.0002 & 289 $\pm$ 65 & CO(2--1) & 109.874 $\pm$ 0.010 & 0.24 $\pm$ 0.05 & 3.9 $\pm$ 0.8 & \nodata & \nodata \\
 &  &  &  & CO(4--3) & 219.732 $\pm$ 0.020 & 0.29 $\pm$ 0.11 & 1.2 $\pm$ 0.4 & \nodata & 0.30 $\pm$ 0.13 \\
 &  &  &  & \CI(1--0) & 234.564 $\pm$ 0.021 & -0.03 $\pm$ 0.12 & -0.1 $\pm$ 0.4 & \nodata & \nodata \\
1mm.C23 & 3mm.08 & 1.3821 $\pm$ 0.0000 & 50 $\pm$ 7 & CO(2--1) & 96.778 $\pm$ 0.001 & 0.11 $\pm$ 0.01 & 2.7 $\pm$ 0.3 & \nodata & \nodata \\
 &  &  &  & CO(5--4) & 241.912 $\pm$ 0.002 & -0.01 $\pm$ 0.06 & -0.0 $\pm$ 0.2 & \nodata & -0.02 $\pm$ 0.09 \\
1mm.C30 & \nodata & 0.4580 $\pm$ 0.0001 & 176 $\pm$ 36 & CO(3--2) & 237.170 $\pm$ 0.012 & 0.43 $\pm$ 0.08 & 0.5 $\pm$ 0.1 & \nodata & \nodata \\
\nodata & 3mm.11 & 1.0964 $\pm$ 0.0000 & 44 $\pm$ 7 & CO(2--1) & 109.970 $\pm$ 0.001 & 0.07 $\pm$ 0.01 & 1.2 $\pm$ 0.2 & \nodata & \nodata \\
 &  &  &  & CO(4--3) & 219.924 $\pm$ 0.002 & 0.12 $\pm$ 0.03 & 0.5 $\pm$ 0.1 & \nodata & 0.40 $\pm$ 0.11 \\
 &  &  &  & \CI(1--0) & 234.769 $\pm$ 0.002 & 0.08 $\pm$ 0.03 & 0.3 $\pm$ 0.1 & \nodata & \nodata \\
\nodata & 3mm.09 & 2.6976 $\pm$ 0.0001 & 165 $\pm$ 22 & CO(3--2) & 93.520 $\pm$ 0.003 & 0.38 $\pm$ 0.05 & 13.9 $\pm$ 1.7 & 1.09 $\pm$ 0.27 & \nodata \\
 &  &  &  & CO(7--6) & 218.158 $\pm$ 0.007 & 0.14 $\pm$ 0.58 & 1.0 $\pm$ 3.9 & 0.08 $\pm$ 0.31 & \nodata \\
 &  &  &  & \CI(2--1) & 218.885 $\pm$ 0.007 & 0.78 $\pm$ 0.63 & 5.3 $\pm$ 4.2 & \nodata & \nodata \\
 &  &  &  & CO(1--0) & 31.175 $\pm$ 0.001 & 0.04 $\pm$ 0.01 & 12.8 $\pm$ 2.7 & \nodata & \nodata \\
Faint.1mm.C20 & 3mm.16 & 1.2938 $\pm$ 0.0001 & 144 $\pm$ 32 & CO(2--1) & 100.504 $\pm$ 0.005 & 0.09 $\pm$ 0.02 & 1.9 $\pm$ 0.4 & \nodata & \nodata \\
 &  &  &  & \CI(1--0) & 214.559 $\pm$ 0.010 & -0.09 $\pm$ 0.12 & -0.4 $\pm$ 0.6 & \nodata & \nodata \\
 &  &  &  & CO(5--4) & 251.225 $\pm$ 0.012 & 0.15 $\pm$ 0.11 & 0.5 $\pm$ 0.4 & \nodata & 0.28 $\pm$ 0.21 \\
\nodata & MP.3mm.2 & 1.0873 $\pm$ 0.0001 & 115 $\pm$ 33 & CO(2--1) & 110.448 $\pm$ 0.005 & 0.15 $\pm$ 0.04 & 2.3 $\pm$ 0.6 & \nodata & \nodata \\
 &  &  &  & CO(4--3) & 220.878 $\pm$ 0.010 & 0.12 $\pm$ 0.08 & 0.5 $\pm$ 0.3 & \nodata & 0.21 $\pm$ 0.15 \\
 &  &  &  & \CI(1--0) & 235.788 $\pm$ 0.011 & 0.13 $\pm$ 0.10 & 0.5 $\pm$ 0.3 & \nodata & \nodata \\
  \enddata

  \tablecomments{(1) ASPECS-LP 1mm ID. (2) ASPECS-LP 3mm ID. (3) Redshift. (4)
    Line full width at half maximum (FWHM). (5) Line identification. (6)
    Observed line frequency. (7) Integrated line flux,
    $S^{V} = \int S_{\nu} d v$ (sometimes called $S\Delta v$). (8) Line
    luminosity (\autoref{Llineprime}). (9) Line luminosity ratio with CO(1--0)
    (\autoref{eq:rJ1}). (10) Line luminosity ratio with CO(2--1).}
 \end{deluxetable*}
 \end{longrotatetable}
\clearpage
\bibliographystyle{aasjournal}
\bibliography{library.bib}
\end{document}